\documentclass[a4paper,aps,prd,twocolumn,citeautoscript,nofootinbib,superscriptaddress,floatfix]{revtex4-1}

\usepackage{lipsum}
\usepackage{graphicx}
\usepackage{epsf}
\usepackage{epsfig}
\usepackage{amssymb,amsmath,amsfonts}
\usepackage[usenames]{color}
\usepackage{amssymb}
\usepackage{times}
\usepackage{bm}
\usepackage{dcolumn}
\usepackage{bm}
\usepackage{mathrsfs}

\usepackage{hyperref} \hypersetup{
  colorlinks=true,        
  linkcolor=blue,         
  citecolor=cyan,         
  }

\usepackage[encapsulated]{CJK}
\usepackage{ucs}
\usepackage[utf8x]{inputenc}
\newcommand{\cntext}[1]{\begin{CJK}{UTF8}{gbsn}#1\end{CJK}}

\newcommand{\ie}{i.e.,~}
\newcommand{\eg}{e.g.,~}
\newcommand{\cf}{cf.~}

\begin{document}

\title{Constraints on black-hole charges with the 2017 EHT observations
  of M87*}

\author{Prashant Kocherlakota}
\affiliation{Institut f{\"u}r Theoretische Physik,
  Goethe-Universit{\"a}t, Max-von-Laue-Str. 1, 60438 Frankfurt, Germany}

\author{Luciano Rezzolla}
\affiliation{Institut f{\"u}r Theoretische Physik,
  Goethe-Universit{\"a}t, Max-von-Laue-Str. 1, 60438 Frankfurt, Germany}
\affiliation{Frankfurt Institute for Advanced Studies, Ruth-Moufang-Str. 1, 60438 Frankfurt, Germany}
\affiliation{School of Mathematics, Trinity College, Dublin 2, Ireland}

\author{Heino Falcke}
\affiliation{Department of Astrophysics, Institute for Mathematics, Astrophysics and Particle Physics (IMAPP), Radboud University, P.O. Box 9010, 6500 GL Nijmegen, The Netherlands}

\author{Christian M. Fromm}
\affiliation{Black Hole Initiative at Harvard University, 20 Garden Street, Cambridge, MA 02138, USA}
\affiliation{Center for Astrophysics | Harvard \& Smithsonian, 60 Garden Street, Cambridge, MA 02138, USA}
\affiliation{Institut f{\"u}r Theoretische Physik, 
Goethe-Universit{\"a}t, Max-von-Laue-Str. 1, 60438 Frankfurt, Germany}

\author{Michael Kramer}
\affiliation{Max-Planck-Institut f\"ur Radioastronomie, Auf dem H\"ugel 69, D-53121 Bonn, Germany}

\author{Yosuke Mizuno}
\affiliation{Tsung-Dao Lee Institute and School of Physics and Astronomy, Shanghai Jiao Tong University, Shanghai, 200240, China}
\affiliation{Institut f\"ur Theoretische Physik, Goethe-Universit\"at Frankfurt, Max-von-Laue-Stra{\ss}e 1, D-60438 Frankfurt am Main, Germany}

\author{Antonios Nathanail}
\affiliation{Institut f\"ur Theoretische Physik, Goethe-Universit\"at Frankfurt, Max-von-Laue-Stra{\ss}e 1, D-60438 Frankfurt am Main, Germany}
\affiliation{Department of Physics, National and Kapodistrian University of Athens, Panepistimiopolis, GR 15783 Zografos, Greece}

\author{H\'ector Olivares}
\affiliation{Department of Astrophysics, Institute for Mathematics, Astrophysics and Particle Physics (IMAPP), Radboud University, P.O. Box 9010, 6500 GL Nijmegen, The Netherlands}

\author{Ziri Younsi}
\affiliation{Mullard Space Science Laboratory, University College London, Holmbury St. Mary, Dorking, Surrey, RH5 6NT, UK}
\affiliation{Institut f\"ur Theoretische Physik, Goethe-Universit\"at Frankfurt, Max-von-Laue-Stra{\ss}e 1, D-60438 Frankfurt am Main, Germany}

\author{Kazunori Akiyama}
\affiliation{Massachusetts Institute of Technology Haystack Observatory, 99 Millstone Road, Westford, MA 01886, USA}
\affiliation{National Astronomical Observatory of Japan, 2-21-1 Osawa, Mitaka, Tokyo 181-8588, Japan}
\affiliation{Black Hole Initiative at Harvard University, 20 Garden Street, Cambridge, MA 02138, USA}

% ...

\collaboration{The EHT Collaboration}
\noaffiliation

\begin{abstract}
Our understanding of strong gravity near supermassive compact objects has
recently improved thanks to the measurements made by the Event Horizon
Telescope (EHT). We use here the M87* shadow size to infer constraints on the
physical charges of a large variety of nonrotating or rotating black
holes. For example, we show that the quality of the measurements is
already sufficient to rule out that M87* is a highly charged dilaton
black hole. Similarly, when considering black holes with two physical and
independent charges, we are able to exclude considerable regions of the
space of parameters for the doubly-charged dilaton and the Sen black holes.
\end{abstract}

\maketitle

\section{Introduction} 
General relativity (GR) was
formulated to consistently account for the interaction of dynamical
gravitational fields with matter and energy, the central idea of which is
that the former manifests itself through modifications of spacetime
geometry and is fully characterized by a metric tensor. While the
physical axioms that GR is founded on are contained in the equivalence
principle \cite{Dicke1964, Will:2006LRR}, the Einstein-Hilbert action
further postulates that the associated equations of motion involve no
more than second-order derivatives of the metric tensor.

The strength of the gravitational field outside an object of mass $M$ and
characteristic size $R$, in geometrized units ($G = c = 1$), is related
to its compactness $\mathscr{C}:= M/R$, which is $\sim10^{-6}$ for the
Sun, and takes values $\sim 0.2-1$ for compact objects such as neutron
stars and black holes. Predictions from GR have been tested and validated
by various solar-system experiments to very high precision
\cite{Will:2006LRR, Collett2018}, setting it on firm footing as the
best-tested theory of classical gravity in the weak-field regime. It is
important, however, to consider whether signatures of deviations from the
Einstein-Hilbert action, \eg due to higher derivative terms
\cite{tHooft1974, *Krasnikov1987, *Lu2015}, could appear in measurements
of phenomena occurring in strong-field regimes where $\mathscr{C}$ is
large. Similarly, tests are needed to assess whether generic violations
of the equivalence principle occur in strong-fields due, \eg to the
presence of additional dynamical fields, such as scalar \cite{Scherk1979,
  *Green1988} or vector fields \cite{Barausse2011, *Barausse2013CQG,
  *Barausse2016, *Ramos2019, *Sarbach2019}, that may fall off
asymptotically. Agreement with the predictions of GR coming from
observations of binary pulsars \cite{Damour1992PRD, *Wex2014, *Wex2020},
and of the gravitational redshift \cite{Abuter2018} and geodetic
orbit-precession \cite{Abuter2020_etal} of the star S2 near our galaxy's
central supermassive compact object Sgr A$^\star$ by the GRAVITY
collaboration, all indicate the success of GR in describing strong-field
physics as well. In addition, with the gravitational-wave detections of
coalescing binaries of compact objects by the LIGO/Virgo collaboration
\cite{Abbot2016-GW-detection-prl, *Abbott2016b} and the first images of
black holes produced by EHT, it is now possible to envision testing GR at
the strongest field strengths possible.

While the inferred size of the shadow from the recently obtained
horizon-scale images of the supermassive compact object in M87 galaxy by
the EHT collaboration \cite{EHT_M87_PaperI, *EHT_M87_PaperII,
  *EHT_M87_PaperIII, *EHT_M87_PaperIV, *EHT_M87_PaperV, *EHT_M87_PaperVI}
was found to be consistent to within 17\% for a 68\% confidence interval
of the size predicted from GR for a Schwarzschild black hole using the 
\textit{a  priori} known estimates for the mass and distance of M87* 
based on stellar dynamics \cite{Gebhardt11}, this measurement admits 
other possibilities, as do various weak-field tests \cite{Will:2006LRR,
  Psaltis2020_EHT}. Since the number of alternative theories to be tested
using this measurement is large, a systematic study of the constraints
set by a strong-field measurement is naturally more tractable within a
theory-agnostic framework, and various such systems have recently been
explored \cite{Johannsen2011, *Johannsen2013PRD, *Vigeland2011,
  *Johannsen2013PRDb, *Rezzolla2014, *Younsi2016, *Konoplya2016a,
  Kocherlakota2020}. This approach allows for tests of a broad range of
possibilities that may not be captured in the limited set of known
solutions. This was exploited in Ref. \cite{Psaltis2020_EHT}, where
constraints on two deformed metrics were obtained when
determining how different M87* could be from a Kerr black hole while remaining consistent with the EHT measurements.

However, because such parametric tests cannot be connected directly to an
underlying property of the alternative theory, here we use instead the
EHT measurements to set bounds on the physical parameters, \ie angular
momentum, electric charge, scalar charge, etc., -- and which we will
generically refer to as ``charges'' (or hairs) -- that various well-known
black-hole solutions depend upon. Such analyses can be very instructive
\cite{Bambi2009, *Bambi10, *Amarilla10, *Amarilla13, *Wei2013,
  *Nedkova13, *Papnoi2014, *Wei2015, *Ghasemi-Nodehi2015,
  *Atamurotov2016, *Singh2017, *Amir2018, *Olivares2020, Mizuno2018b,
  Cunha2019} since they can shed light on which underlying theories are
promising candidates and which must be discarded or modified. At the same
time, they may provide insight into the types of additional dynamical
fields that may be necessary for a complete theoretical description of
physical phenomena, and whether associated violations of the equivalence
principle occur.

More specifically, since the bending of light in the presence of
curvature -- either in static or in stationary spacetimes -- is assured
in any metric theory of gravity, and the presence of large amounts of
mass in very small volumes can allow for the existence of a region where
null geodesics move on spherical orbits, an examination of the
characteristics of such photon regions, when they exist, is a useful
first step. The projected asymptotic collection of the photons
trajectories that are captured by the black hole -- namely, all of the photon
trajectories falling within the value of the impact parameter at the
unstable circular orbit in the case of nonrotating black holes -- will appear as
a dark area to a distant observer and thus represents the ``shadow'' of
the capturing compact object. This shadow -- which can obviously be
associated with black holes \cite{Grenzebach2016, Stuchlik2019, Kumar2019b,
  Kumar2020, Hioki2008, Abdujabbarov2016}, but also more exotic compact
objects such as gravastars \cite{Mazur2004, Chirenti2008} or naked
singularities \cite{Shaikh2019, Dey2020} -- is determined entirely by the
underlying spacetime metric. Therefore, the properties of the shadow --
and at lowest order its size -- represent valuable observables common to
all metric theories of gravity, and can be used to test them for their
agreement with EHT measurements.

While the EHT measurement contains far more information related to the
flow of magnetised plasma near M87*, we will consider only the
measurement of the size of the bright ring. Here we
consider various spherically symmetric black-hole solutions, from GR that are
either singular (see, \eg \cite{Wald84Book}) or non-singular
\cite{Bardeen68, Hayward2006PRL, Frolov2016}, and string theory
\cite{Kazakov1994, Gibbons1988, Garfinkle1991, Garcia1995,
  Kallosh1992}. Additionally, we also consider the Reissner-Nordstr{\"o}m
(RN) and the Janis-Newman-Winicour (JNW) \cite{Janis1968} naked
singularity solutions, the latter being a solution of the
Einstein-Klein-Gordon system. Many of these solutions have been recently
summarised in Ref. \cite{Kocherlakota2020}, where they were cast in a
generalised expansion of static and spherically symmetric metrics. Since
angular momentum plays a key role in astrophysical scenarios, we also
consider various rotating black-hole solutions \cite{Kerr1963, Newman1965,
  Sen1992, Bambi2013} which can be expressed in the Newman-Janis form
\cite{Newman1965JMPa} to facilitate straightforward analytical
computations. It is to be noted that this study is meant to be a proof of
principle and that while the constraints we can set here are limited, the
analytical procedure outlined below for this large class of metrics is
general, so that as future observations become available, we expect the
constraints that can be imposed following the approach proposed here to
be much stronger.

\section{Spherical Null Geodesics and Shadows} 
For all the static,
spherically symmetric spacetimes we consider here, the definition of the
shadow can be cast in rather general terms. In particular, for all the
solutions considered, the line element expressed in areal-radial polar
coordinates $(t, \tilde{r}, \theta, \phi)$ has the form%
\footnote{We use
the tilde on the radial coordinate of static spacetimes to distinguish it
from the corresponding radial coordinate of axisymmetric spacetimes.}
\begin{equation}
  \label{eq:Static_Spacetimes}
  ds^2 = g_{\mu\nu} dx^{\mu} dx^{\nu} =
  -f(\tilde{r})\text{d}t^2 +
\frac{g(\tilde{r})}{f(\tilde{r})}\text{d}\tilde{r}^2 +
\tilde{r}^2\text{d}\Omega_2^2\,,
\end{equation}
and the photon region, which degenerates into a photon sphere, is located
at $\tilde{r} =: \tilde{r}_{\text{ps}}$, which can be obtained by solving
\cite{Psaltis2020_EHT}
\begin{equation}
\tilde{r} - \frac{2f(\tilde{r})}{\partial_{\tilde{r}}f(\tilde{r})} = 0\,.
\end{equation}
The boundary of this photon sphere when observed from the frame of an
asymptotic observer, due to gravitational lensing, appears to be a circle
of size \cite{Psaltis2020_EHT}
\begin{equation}
  \label{eq:Static_Shadow_Radius}
  \tilde{r}_{\text{sh}} =
  \frac{\tilde{r}_{\text{ps}}}{\sqrt{f(\tilde{r}_{\text{ps}})}}\,.
\end{equation}
On the other hand, the class of Newman-Janis stationary,
axisymmetric spacetimes we consider here \cite{Newman1965JMPa}, which are
geodesically integrable (see, e.g., \cite{Carter68, Astorga17,
  Kumar2020}), can be expressed in Boyer-Lindquist coordinates ($t, r,
\theta, \phi$) as
\begin{align}
  \label{eq:NJA_Spacetimes}
\text{d}s^2 =&\ -f~\text{d}t^2 - 2 a \sin^2\theta(1 -
f)~\text{d}t\text{d}\phi \\ &\ + \left[\Sigma + a^2\sin^2\theta\left(2 -
  f\right)\right]\sin^2\theta~\text{d}\phi^2 +
\frac{\Sigma}{\Delta}\text{d}r^2 + \Sigma~\text{d}\theta^2 \nonumber \,,
\end{align}
where $f = f(r,\theta)$ and $\Sigma(r,\theta) := r^2 + a^2\cos^2\theta$
and $\Delta(r) := \Sigma(r,\theta)f(r,\theta) + a^2\sin^2\theta$. In
particular, these are the stationary generalisations obtained by
employing the Newman-Janis algorithm \cite{Newman1965JMPa}) for ``seed''
metrics of the form \eqref{eq:Static_Spacetimes} with $g(\tilde{r}) = 1$%
\footnote{Note that while the Sen solution can be obtained via
  the Newman-Janis algorithm \cite{Yazadjiev2000}, the starting point is
  the static EMd-1 metric written in a \textit{non}-areal-radial
  coordinate $\rho$ such that $g_{tt}g_{\rho\rho} = -1$.}.

The
Lagrangian $\mathcal{L}$ for geodesic motion in the spacetime
\eqref{eq:NJA_Spacetimes} is given as $2\mathcal{L} :=
g_{\mu\nu}\dot{x}^{\mu} \dot{x}^{\nu}$, where an overdot represents a
derivative with respect to the affine parameter, and $2\mathcal{L} = -1$
for timelike geodesics and $2\mathcal{L} = 0$ for null geodesics. The two
Killing vectors $\partial_t$ and $\partial_\phi$ yield two constants of
motion
\begin{align} \label{eq:Conserved_Quantities}
- E =&\ -f\dot{t} - a
\sin^2\theta(1 - f)\dot{\phi}, \\ 
L  =&\ - a \sin^2\theta(1 - f)\dot{t}
+ \left[\Sigma + a^2\sin^2\theta\left(2 -
  f\right)\right]\sin^2\theta~\dot{\phi}\,, \nonumber
\end{align}
in terms of which the geodesic equation for photons can be separated into
\begin{align} 
\Sigma^2\dot{r}^2 =&\ (r^2 + a^2 - a \xi)^2 - \Delta\mathcal{I} =: \mathcal{R}(r)\,, 
\label{eq:Radial_GE} \\
\Sigma^2\dot{\theta}^2 =&\ \mathcal{I} - (a\sin\theta - \xi\csc\theta)^2 =: \Theta(\theta)\,, 
\label{eq:Theta_GE}
\end{align}
where we have introduced first $\xi := L/E$, and then $\mathcal{I} :=
\eta + (a - \xi)^2$. Also, $\eta$ is the Carter constant, and the
existence of this fourth constant of motion is typically associated with
the existence of an additional Killing-Yano tensor (see for example
\cite{Hioki2009, Hioki2008}).

In particular, we are interested here in spherical null
geodesics (SNGs), which satisfy $\dot{r} = 0$ and $\ddot{r} = 0$ and are
not necessarily planar; equivalently, SNGs can exist at locations
where $\mathcal{R}(r) = 0$ and $\text{d}\mathcal{R}(r)/\text{d}r =
0$. Since these are only two equations in three variables ($r, \xi,
\eta$), it is convenient, for reasons that will become evident below, to
obtain the associated conserved quantities along such SNGs in terms of
their radii $r$ as (see also \cite{Shaikh2019PRD}),
\begin{align}
  \label{eq:Conserved_Quantities_SNGs}
\xi_{_\text{SNG}}(r) =&\ \frac{r^2 + a^2}{a} - \frac{4r\Delta}{a{\partial_{r}\Delta}} \,, \\
\eta_{_\text{SNG}}(r) =&\ \frac{r^2}{a^2(\partial_{r}\Delta)^2}\left[16
  a^2 \Delta - \left(r {\partial_{r}\Delta} - 4\Delta\right)^2\right]\,. \nonumber
\end{align}
The condition that $\Theta(\theta) \geq 0$, which must necessarily hold
as can be seen from Eq. \eqref{eq:Theta_GE}, restricts the radial range
for which SNGs exist, and it is evident that this range depends on
$\theta$. This region, which is filled by such SNGs, is called the photon
region (see, \eg Fig. 3.3 of \cite{Grenzebach2016}).

The equality $\Theta(\theta) = 0$ determines the boundaries of the photon
region, and the (disconnected) piece which lies in the exterior of the
outermost horizon is of primary interest since its image, as seen by an
asymptotic observer, is the shadow. We denote the inner and outer
surfaces of this photon region by $r_{\text{p}-}(\theta)$ and
$r_{\text{p}+}(\theta)$ respectively, with the former (smaller) SNG
corresponding to the location of a prograde photon orbit (\ie
$\xi_{_{\text{SNG}}}(r_{\text{p}-}) > 0$), and the latter to a retrograde
orbit. 

It can be shown that all of the SNGs that are admitted in the photon
region, for both the spherically symmetric and axisymmetric solutions
considered here, are unstable to radial perturbations. In particular, for
the stationary solutions, the stability of SNGs at a radius $r =
r_{\text{SNG}}$ with respect to radial perturbations is determined by the
sign of $\partial^2_{r}\mathcal{R}$, and when
$\partial^2_{r}\mathcal{R}(r_{\text{SNG}}) > 0$, SNGs at that radius are
unstable. The expression for $\partial^2_{r}\mathcal{R}$ reads
\begin{equation}
  \partial^2_{r}\mathcal{R} =
  \frac{8r}{(\partial_{r}\Delta)^2}\left[r(\partial_{r}\Delta)^2 - 2 r
    \Delta \partial^2_{r}\Delta + 2\Delta{\partial_{r}\Delta}\right]\,.
\end{equation}
To determine the appearance of the photon region and the associated shadow, as seen by asymptotic observers, we can
introduce the usual notion of celestial coordinates $(\alpha, \beta)$,
which for any photon with constants of the motion $(\xi, \eta)$ can be
obtained, for an asymptotic observer present at an inclination angle $i$
with respect to the spin-axis of the compact object as in
\cite{Bardeen72}. For photons on an SNG, we can set the conserved
quantities ($\xi, \eta$) to the values given in
Eq. \ref{eq:Conserved_Quantities_SNGs} above to obtain \cite{Hioki2009, Shaikh2019PRD}
\begin{align}
  \label{eq:beta_Celes_Coord}
  \alpha_{\text{sh}} = &\ -\xi_{_\text{SNG}}\csc{i}\,, \\
  \beta_{\text{sh}} = &\ \pm\left(\eta_{_\text{SNG}} + a^2\cos^2{i} -
  \xi_{_\text{SNG}}^2\cot^2{i}\right)^{1/2}\,. 
\end{align}
Recognizing that $\beta = \pm \sqrt{\Theta(i)}$, it becomes clear that
only the SNGs with $\Theta(i) \geq 0$ determine the apparent
shadow shape. Since the photon region is not spherically symmetric in rotating spacetimes, the associated shadow is also not circular in general. It can be shown that the band of radii for which SNGs can
exist narrows as we move away from the equatorial plane, and reduces to a
single value at the pole, i.e. in the limit $\theta
\rightarrow \pi/2$, we have $r_{\text{p}+} = r_{\text{p}-}$ (see \eg Fig. 3.3 of
\cite{Grenzebach2016}). As a result, the parametric curve of the shadow
boundary as seen by an asymptotic observer lying along the pole is
perfectly circular, $\alpha_{\text{sh}}^2 + \beta_{\text{sh}}^2 =
\eta_{_{\text{SNG}}}(r_{\text{p}\pm, \pi/2}) +
\xi^2_{_{\text{SNG}}}(r_{\text{p}\pm, \pi/2})$. 

We can now define the characteristic areal-radius of the shadow
curve as \cite{Abdujabbarov2015}
\begin{equation}
  \label{eq:Areal_Shadow_Radius}
r_{\text{sh}, A} :=
\left(\frac{2}{\pi}\int_{r_{\text{p}-}}^{r_{\text{p}+}}
  \text{d}r~\beta_{\text{sh}}(r)\partial_r\alpha_{\text{sh}}(r)\right)^{1/2}\,.
\end{equation}

\section{Shadow Size Constraints from the 2017 EHT Observations of M87*}

Measurements of stellar dynamics near M87* were previously used to
produce a posterior distribution function of the angular gravitational
radius $\theta_{\text{g}} := M/D$, where $M$ is the mass of and $D$ the
distance to M87*. The 2017 EHT observations of M87* can be similarly used
to determine such a posterior \cite{EHT_M87_PaperVI}. These observations
were used to determine the angular diameter $\hat{d}$ of the bright
emission ring that surrounds the shadow \cite{EHT_M87_PaperVI}. In
Sec. 5.3 there, using synthetic images from general-relativistic magnetohydrodynamics
(GRMHD) simulations of
accreting Kerr black holes for a wide range of physical scenarios, the
scaling factor $\alpha = \hat{d}/\theta_{\text{g}}$ was calibrated. For
emission from the outermost boundary of the photon region of a Kerr black
hole, $\alpha$ should lie in the range $\simeq 9.6 - 10.4$.

The EHT measurement picks out a class of best-fit images (``top-set'')
from the image library, with a mean value for $\alpha$ of $11.55$ (for
the ``xs-ring'' model) and $11.50$ (for the ``xs-ringauss'' model), when
using two different geometric crescent models for the images, implying
that the geometric models were accounting for emission in the top-set
GRMHD images that preferentially fell outside of the photon ring. Using
the distribution of $\alpha$ for these top-set images then enabled the
determination of the posterior in the angular gravitational radius
$P_{\text{obs}}(\theta_{\text{g}})$ for the EHT data. It is to be noted
that this posterior was also determined using direct GRMHD fitting, and
image domain feature extraction procedures, as described in Sec. 9.2
there, and a high level of consistency was found across all measurement
methods. Finally, in Sec. 9.5 of \cite{EHT_M87_PaperVI}, the fractional
deviation in the angular gravitational radius $\delta$ was introduced as
\begin{equation}
\delta := \frac{\theta_{\text{g}}}{\theta_{\text{dyn}}} - 1\,,
\end{equation}
where $\theta_{\text{g}}$ and $\theta_{\text{dyn}}$ were used to denote
the EHT and the stellar-dynamics inferences of the angular gravitational
radius, respectively. The posterior on $\delta$ -- as defined in Eq. (32)
of \cite{EHT_M87_PaperVI} -- was then obtained (see Fig. 21 there), and
its width was found to be $\delta = -0.01 \pm 0.17$, for a 68\% credible
interval. This agreement of the 2017 EHT measurement of the angular
gravitational radius for M87* with a previously existing estimate for the
same, at much larger distances, constitutes a validation of the null
hypothesis of the EHT, and in particular that M87* can be described by
the Kerr black-hole solution.

Since the stellar dynamics measurements \cite{Gebhardt11} are sensitive
only to the monopole of the metric (\ie the mass) due to negligible
spin-dependent effects at the distances involved in that analysis,
modeling M87* conservatively using the Schwarzschild solution is
reasonable with their obtained posterior. Then, using the angular
gravitational radius estimate from stellar dynamics yields a prediction
for the angular shadow radius $\theta_{\text{sh}} = r_{\text{sh}}/D$ as
being $\theta_{\text{sh}} = 3\sqrt{3}\theta_{\text{dyn}}$. The 2017 EHT
measurement, which includes spin-dependent effects as described above and
which probes near-horizon scales, then determines the allowed spread in
the angular shadow diameter as, $\theta_{\text{sh}} \approx 3\sqrt{3}(1
\pm 0.17)\,\theta_{\text{g}}$, at 68\% confidence levels
\cite{Psaltis2020_EHT}. Finally, since both angular estimates
$\theta_{\text{sh}}$ and $\theta_{\text{g}}$ make use of the same
distance estimate to M87*, it is possible to convert the $1$-$\sigma$
bounds on $\theta_{\text{sh}}$ to bounds on the allowed shadow size for
M87*.

That is, independently of whether the underlying solution be spherically
symmetric (in which case we will consider $\tilde{r}_{\text{sh}}$) or
axisymmetric $(r_{\text{sh}, A})$, the shadow size of M87* must lie in
the range $3\sqrt{3}(1 \pm 0.17) M$ \cite{Psaltis2020_EHT}, i.e., (see
gray-shaded region in Fig. \ref{fig:Spherical_Shadow_Radius_Constraints})
\begin{equation}
  \label{eq:EHT-Constraint}
  4.31 M \approx r_{\text{sh, EHT-min}} \leq \tilde{r}_{\text{sh}},
  r_{\text{sh}, A} \leq r_{\text{sh, EHT-max}} \approx 6.08 M\,,
\end{equation}
where we have introduced the maximum/minimum shadow radii $r_{\text{sh,
    EHT-max}}/r_{\text{sh, EHT-min}}$ inferred by the EHT, at 68\%
confidence levels.

Note that the bounds thus derived are consistent with compact objects
that cast shadows that are both significantly smaller and larger than the
minimum and maximum shadow sizes that a Kerr black hole could cast, which
lie in the range, $4.83\,M - 5.20\,M$ (see, e.g., \cite{Johannsen2010,
  Psaltis2020_EHT}).

An important caveat here is that the EHT posterior distribution on
$\theta_{\text{g}}$ was obtained after a comparison with a large library
of synthetic images built from GRMHD simulations of accreting Kerr black holes
\cite{EHT_M87_PaperV}. Ideally, a rigorous comparison with non-Kerr
solutions would require a similar analysis and posterior distributions
built from equivalent libraries obtained from GRMHD simulations of such
non-Kerr solutions. Besides being computationally unfeasible, this
approach is arguably not necessary in practice. For example, the recent
comparative analysis of Ref. \cite{Mizuno2018b} has shown that the image
libraries produced in this way would be very similar and essentially
indistinguishable, given the present quality of the observations. As a
result, we adopt here the working assumption that the $1$-$\sigma$
uncertainty in the shadow angular size for non-Kerr solutions is very
similar to that for Kerr black holes, and hence employ the constraints
\eqref{eq:EHT-Constraint} for all of the solutions considered here.


\section{Notable properties of various spacetimes}

As mentioned above, a rigorous comparison with non-Kerr black holes would
require constructing a series of exhaustive libraries of synthetic images
obtained from GRMHD simulations on such non-Kerr black holes. In turn,
this would provide consistent posterior distributions of angular
gravitational radii for the various black holes and hence determine how
$\delta$ varies across different non-Kerr black holes, \eg for Sen black
holes. Because this is computationally unfeasible -- the construction of
only the Kerr library has required the joint effort of several groups
with the EHTC over a good fraction of a year -- we briefly discuss below
qualitative arguments to support our use of the bounds given in
Eq. \ref{eq:EHT-Constraint} above as an approximate, yet indicative,
measure.

\begin{center}
\begin{table}
\caption{Summary of properties of spacetimes used here. For easy access,
  we show whether the spacetime contains a rotating compact object or
  not, whether it contains a spacetime singularity, and what type of
  stationary nongravitational fields are present in the
  spacetime. Starred spacetimes contain naked singularities and daggers
  indicate a violation of the equivalence principle (see, \eg
  \cite{Kocherlakota2020}); In particular, these indicate violations of
  the weak equivalence principle due to a varying fine structure
  constant, a result of the coupling of the dilaton to the EM Lagrangian
  \cite{Magueijo03, Kocherlakota2020}.}
\label{table:Solutions_Summary}
\renewcommand{\arraystretch}{1.3}
\begin{tabular}[t]{|l|c|c|c|}
\hline
Spacetime & Rotation & Singularity & Spacetime content\\
\hline
\hline
KN \cite{Newman1965} & Yes & Yes & EM fields\\
\cline{2-4}
Kerr \cite{Kerr1963} & Yes & Yes & vacuum \\ 
\cline{2-4}
RN \cite{Wald84Book} & No & Yes & EM fields\\
\cline{2-4}
RN* \cite{Wald84Book} & No & Yes & EM fields\\
\cline{2-4}
Schwarzschild \cite{Wald84Book} & No & Yes & vacuum \\
\hline
Rot. Bardeen \cite{Bambi2013} & Yes & No & matter \\
\cline{2-4}
Bardeen \cite{Bardeen68} & No & No & matter \\
\hline
Rot. Hayward \cite{Bambi2013} & Yes & No & matter \\
\cline{2-4}
Frolov \cite{Frolov2016} & No & No & EM fields, matter \\
\cline{2-4}
Hayward \cite{Hayward2006PRL} & No & No & matter \\
\hline 
JNW* \cite{Janis1968} & No & Yes & scalar field \\
\hline 
KS \cite{Kazakov1994} & No & Yes & vacuum \\
\hline
Sen$^\dagger$ \cite{Sen1992} & Yes & Yes & EM, dilaton, axion fields\\ 
\cline{2-4}
EMd-1$^\dagger$ \cite{Gibbons1988, Garfinkle1991} & No & Yes & EM,
dilaton fields \\
\hline
EMd-2$^\dagger$ \cite{Kallosh1992} & No & Yes & EM, EM, dilaton fields\\
\hline
\hline
\end{tabular}
\end{table}
\end{center}

\begin{figure*}
\includegraphics[width=0.49\textwidth]{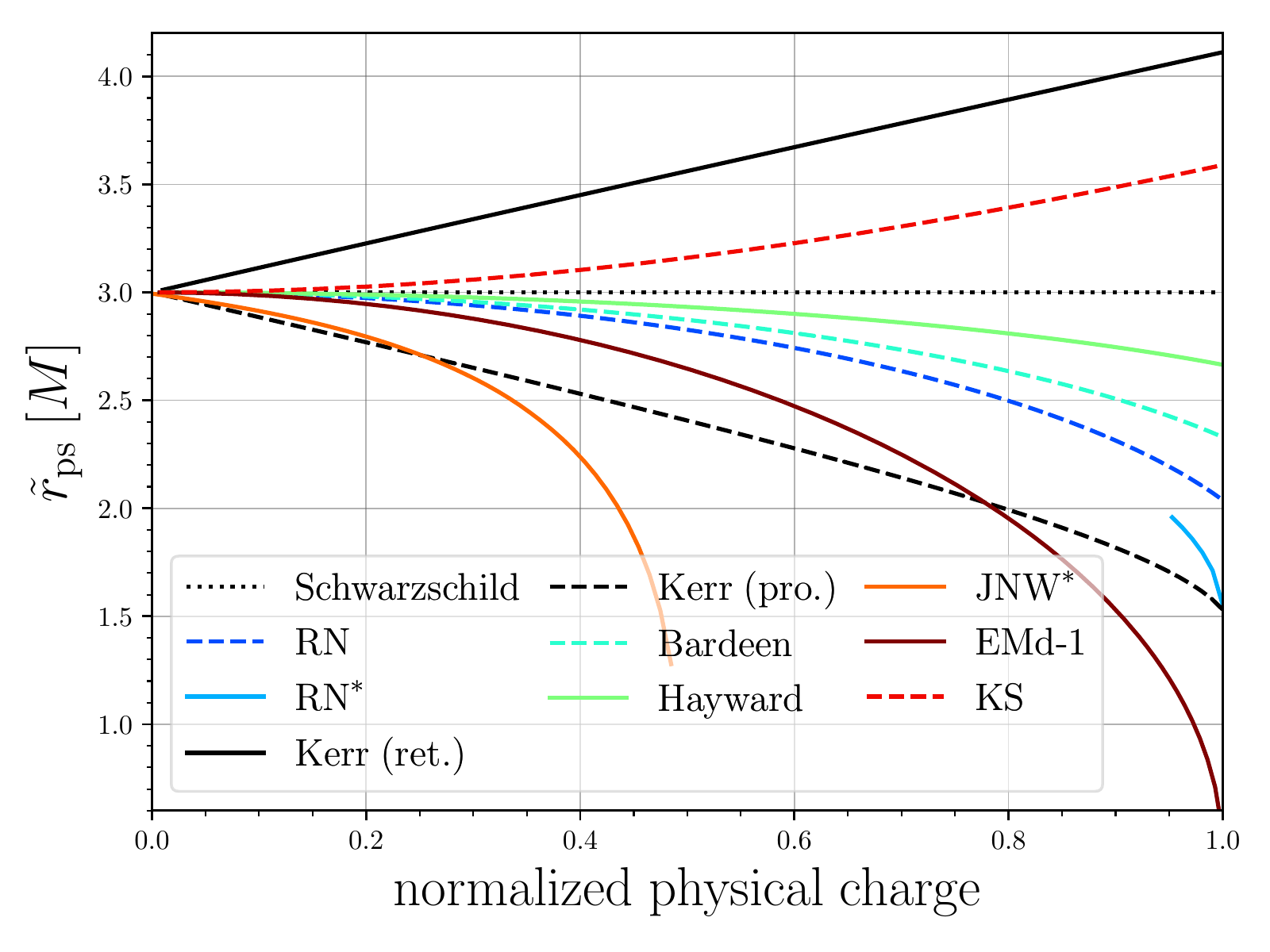}
\includegraphics[width=0.49\textwidth]{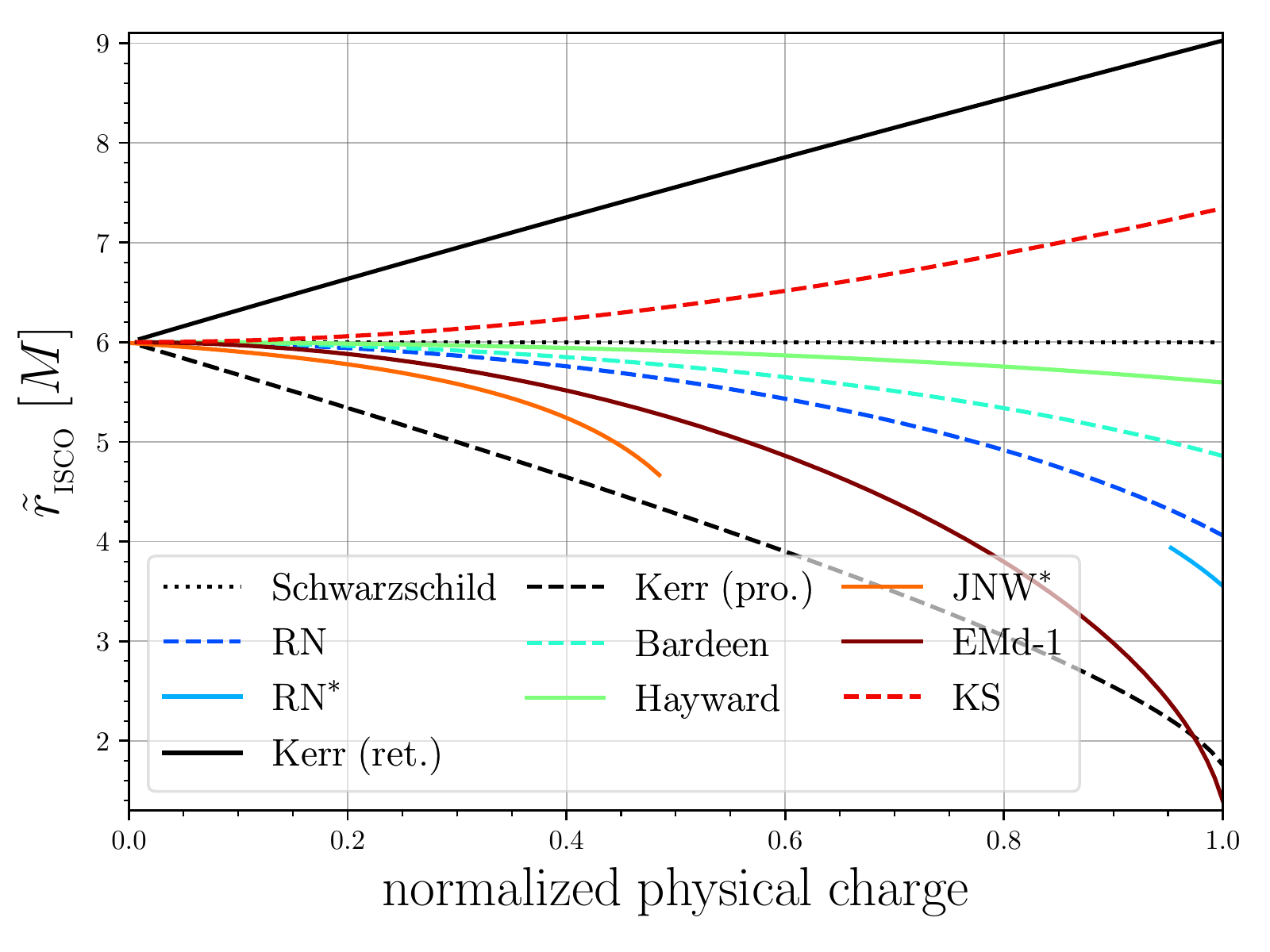}
\caption{\textit{Left:} variation in the photon sphere radii for the
  single-charge nonrotating solutions as a function of the normalized
  physical charge. \textit{Right:} The same as in the left panel but for
  the ISCO radii. We include also, for comparison, the variation in the
  Kerr equatorial prograde and retrograde photon sphere and ISCO radii in
  the left and right panels respectively.}
\label{fig:Photon_Spheres}
\end{figure*}

To this end, we summarize in Table \ref{table:Solutions_Summary} the
relevant properties of the various solutions used here. First, we have
considered here solutions from three types of theories, i.e., the
underlying actions are either (a) Einstein-Hilbert-Maxwell-matter
\cite{Janis1968, Wald84Book, Kazakov1994, Kerr1963, Bardeen68,
  Hayward2006PRL, Bambi2013, Frolov2016}, (b)
Einstein-Hilbert-Maxwell-dilaton-axion \cite{Gibbons1988, Garfinkle1991,
  Sen1992}, or (c) Einstein-Hilbert-Maxwell-Maxwell-dilaton
\cite{Kallosh1992}. This careful choice implies that the gravitational
piece of the action is always given by the Einstein-Hilbert term and that
matter is minimally coupled to gravity. As a result, the dynamical
evolution of the accreting plasma is expected to be very similar to that
in GR, as indeed found in Ref. \cite{Mizuno2018b}. Second, since a
microphysical description that allows one to describe the interaction of
the exotic matter present in some of the regular black-hole spacetimes
used here \cite{Bardeen68, Hayward2006PRL} -- which typically do not
satisfy some form of the energy conditions \cite{Bambi2013,
  Rodrigues2018} -- with the ordinary matter is thus far lacking, it is
reasonable to assume that the interaction between these two types of
fluids is gravitational only. This is indeed what is done in standard
numerical simulations, either in dynamical spacetimes (see, \eg
\cite{Liebling2012}), or in fixed ones \cite{Meliani2016,
  Olivares2020}. Third, since the mass-energy in the matter and
electromagnetic fields for the non-vacuum spacetimes used here is of the
order of the mass of the central compact object $M$, while the total mass
of the accreting plasma in the GRMHD simulations is only a tiny fraction
of the same, it is reasonable to treat the spacetime geometry and the
stationary fields as unaffected by the plasma. Fourth, we have also been
careful not to use solutions from theories with modified electrodynamics
(such as nonlinear electrodynamics). As a result, the electromagnetic
Lagrangian in all of the theories considered here is the Maxwell
Lagrangian (see, \eg the discussion in \cite{Kocherlakota2020} and
compare with \cite{Stuchlik2019}). This ensures that in these spacetimes
light moves along the null geodesics of the metric tensor (see, \eg
Sec. 4.3 of \cite{Wald84Book} and compare against Sec. 2 of
\cite{Hirschmann2018}). Therefore, we are also assured that ray-tracing
the radiation emitted from the accreting matter in these spacetimes can
be handled similarly as in the Kerr spacetime.

Finally, under the assumption that the dominant effects in determining
the angular gravitational radii come from variations in the location of
the photon region and in location of the inner edge of the accretion disk
in these spacetimes, it is instructive to learn how these two physical
quantities vary when changing physical charges, and, in particular, to
demonstrate that they are quantitatively comparable to the corresponding
values for the Kerr spacetime.

For this purpose, we study the single-charge solutions used here and
report in Fig. \ref{fig:Photon_Spheres} the variation in the location of
the photon spheres (left panel) and innermost stable circular orbit
(ISCO) radii (right panel) as a function of the relevant physical charge
(\cf left panel of Fig. 1 in the main text).  Note that both the
photon-sphere and the ISCO radii depend exclusively on the $g_{tt}$
component of the metric when expressed using an areal radial coordinate
$\tilde{r}$ (see, \eg \cite{Psaltis2020_EHT, Kocherlakota2020}). To gauge
the effect of spin, we also show the variation in the locations of the
equatorial prograde and retrograde circular photon orbits and the ISCOs
in the Kerr black-hole spacetime, expressed in terms of the Cartesian
Kerr-Schild radial coordinate $r_{\text{CKS}}$, which, in the equatorial
plane, is related to the Boyer-Lindquist radial coordinate used elsewhere
in this work $r$ simply via \cite{Wiltshire2009}
\begin{equation}
r_{\text{CKS}} = \sqrt{r^2 + a^2}\,.
\end{equation} 

It is apparent from Fig. \ref{fig:Photon_Spheres} that the maximum
deviation in the photon-sphere size from the Schwarzschild solution
occurs for the EMd-1 black hole and is $\approx 75\%$, while the size of
the prograde equatorial circular photon orbit for Kerr deviates by at
most $\approx 50\%$. Similarly, the maximum variation in the ISCO size
also occurs for the EMd-1 solution and is $\approx 73\%$, while the
prograde equatorial ISCO for Kerr can differ by $\approx 66\%$.

\section{Charge Constraints from the EHT M87* Observations}

\begin{figure*}
\centering
\includegraphics[width=0.49\textwidth]{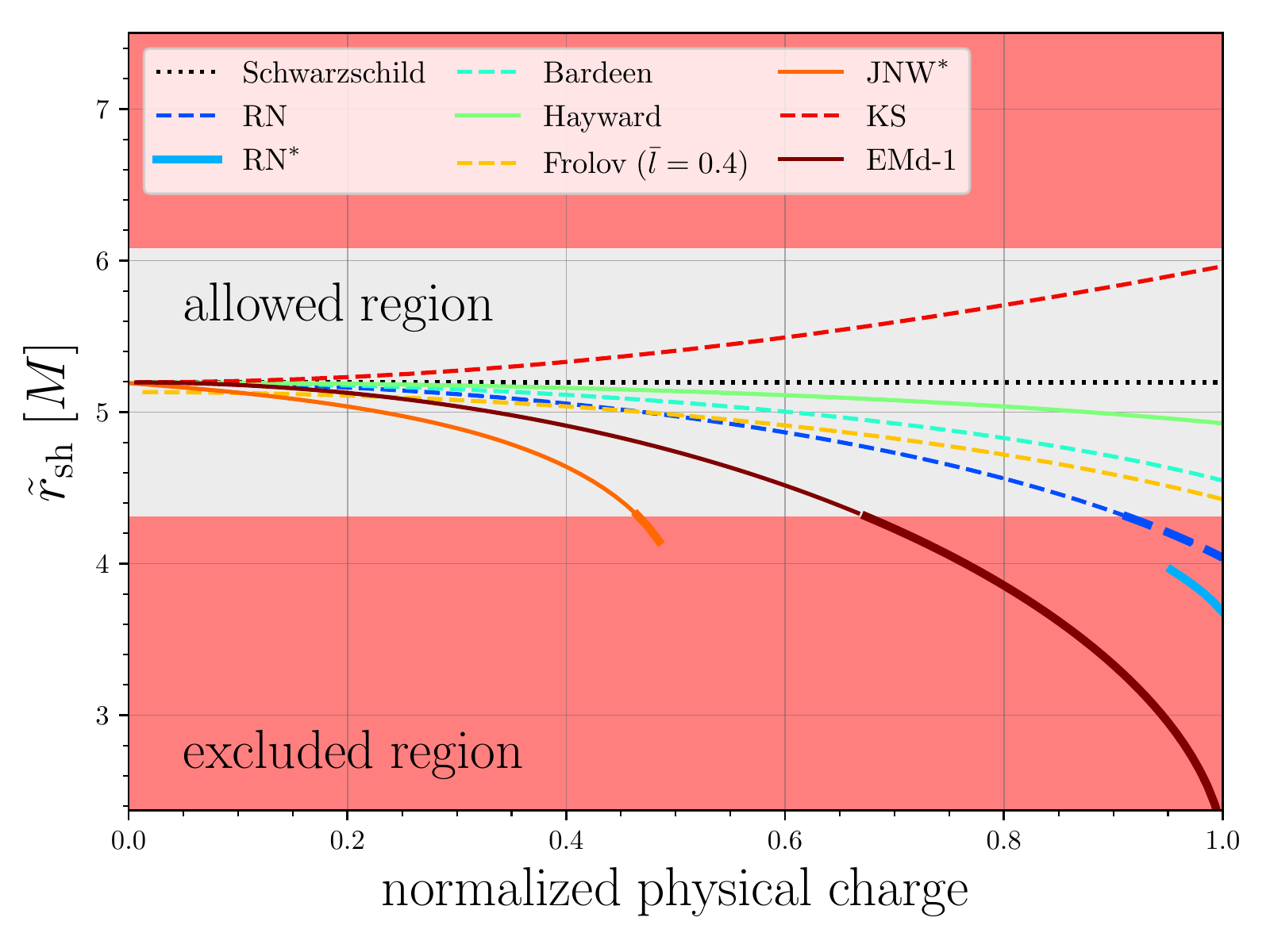}
\includegraphics[width=0.49\textwidth]{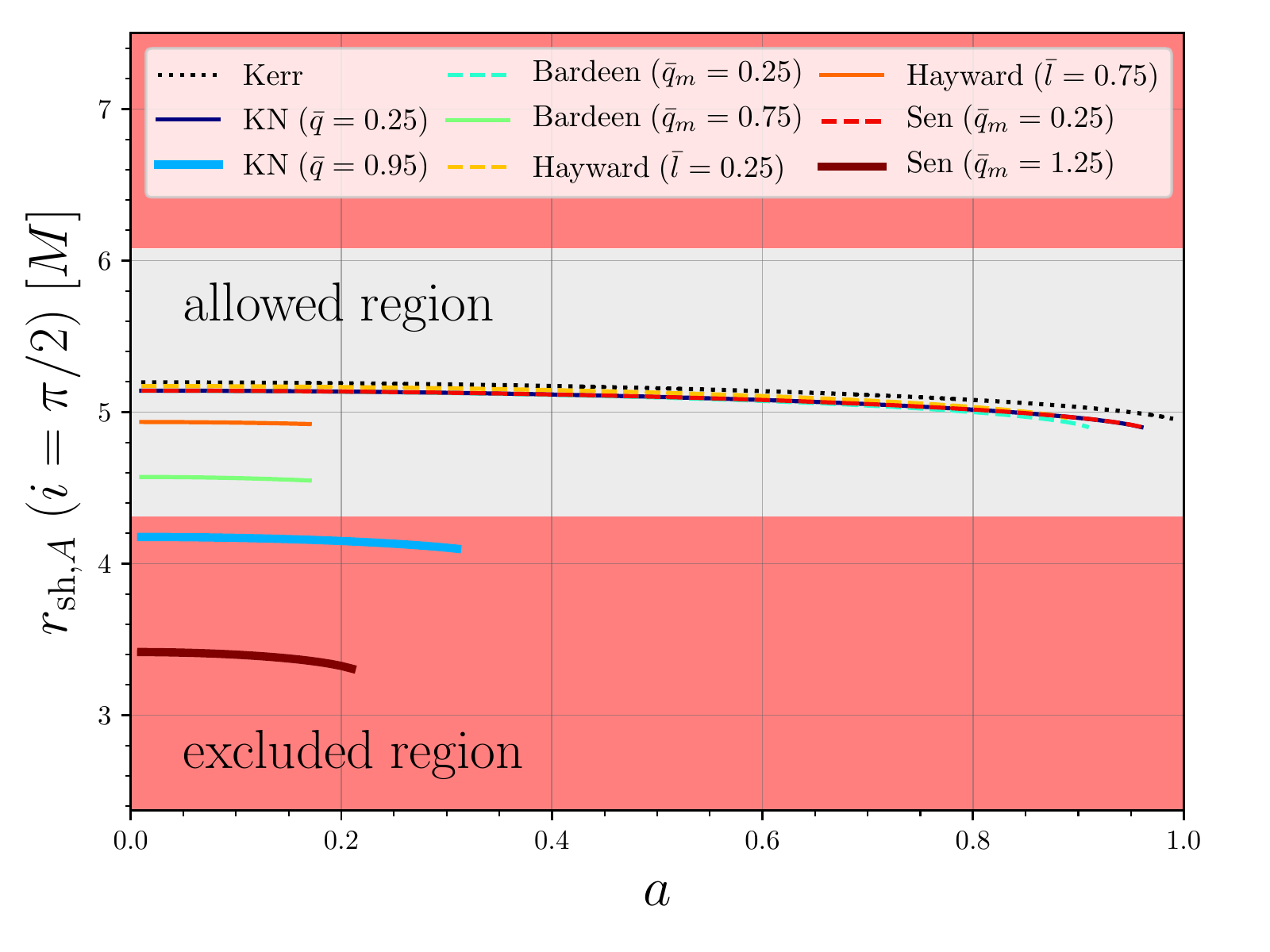}
\caption{\textit{Left:} Shadow radii $\tilde{r}_{\text{sh}}$ for various
  spherically symmetric black-hole solutions, as well as for the JNW and
  RN naked singularities (marked with an asterisk), as a function of the
  physical charge normalized to its maximum value. The gray/red shaded
  regions refer to the areas that are $1$-$\sigma$
  consistent/inconsistent with the 2017 EHT observations and highlight
  that the latter set constraints on the physical charges (see also
  Fig. \ref{fig:EMd-2_Sen_Constraints} for the EMd-2 black
  hole). \textit{Right:} Shadow areal radii $r_{\text{sh}, A}$ as a
  function of the dimensionless spin $a$ for four families of black-hole
  solutions when viewed on the equatorial plane ($i=\pi/2$). Also in this
  case, the observations restrict the ranges of the physical charges of
  the Kerr-Newman and the Sen black holes (see also
  Fig. \ref{fig:EMd-2_Sen_Constraints}).}
\label{fig:Spherical_Shadow_Radius_Constraints}
\end{figure*} 

We first consider compact objects with a single ``charge,'' and report in
the left panel of Fig. \ref{fig:Spherical_Shadow_Radius_Constraints} the
variation in the shadow radius for various spherically symmetric black
hole solutions, as well as for the RN and JNW naked
singularities\footnote{While the electromagnetic and scalar charge
parameters for the RN and JNW spacetimes are allowed to take values
$\bar{q} > 1$ and $0 < \hat{\bar{\nu}} := 1 - \bar{\nu} < 1$
respectively, they do not cast shadows for $\bar{q} > \sqrt{9/8}$ and
$0.5 \leq \hat{\bar{\nu}} < 1$ (see, e.g., Sec. IV D of
\cite{Kocherlakota2020} and references therein).}.  More specifically, we
consider the black-hole solutions given by Reissner-Nordstr\"om (RN)
\cite{Wald84Book}, Bardeen \cite{Bardeen68, Bambi2013}, Hayward
\cite{Hayward2006PRL, Held2019}, Kazakov-Solodhukin (KS)
\cite{Kazakov1994}, and also the asymptotically-flat
Einstein-Maxwell-dilaton (EMd-1) with $\phi_\infty = 0$ and $\alpha_1 =
1$ \cite{Gibbons1988, Garfinkle1991, Hirschmann2018} solution (see
Sec. IV of \cite{Kocherlakota2020} for further details on these
solutions). For each of these solutions we vary the corresponding charge
(in units of $M$) in the allowed range, \ie RN: $0 < \bar{q} \leq 1$;
Bardeen: $0 < \bar{q}_{m} \leq \sqrt{16/27}$; Hayward: $0 < \bar{l} \leq
\sqrt{16/27}$; Frolov: $0 < \bar{l} \leq \sqrt{16/27}$, $0 < \bar{q} \leq
1$; KS: $0 < \bar{l}$; EMd-1: $0 < \bar{q} < \sqrt{2}$, but report the
normalised value in the figure so that all curves are in a range between
0 and 1. The figure shows the variation in the shadow size of KS
  black holes over the parameter range $0 < \bar{l} < \sqrt{2}$. Note
that the shadow radii tend to become smaller with increasing physical
charge, but also that this is not universal behaviour, since the KS black
holes have increasing shadow radii (the singularity is smeared out on a
surface for this solution, which increases in size with increasing
$\bar{l}$).

\begin{figure}
\includegraphics[width=\columnwidth]{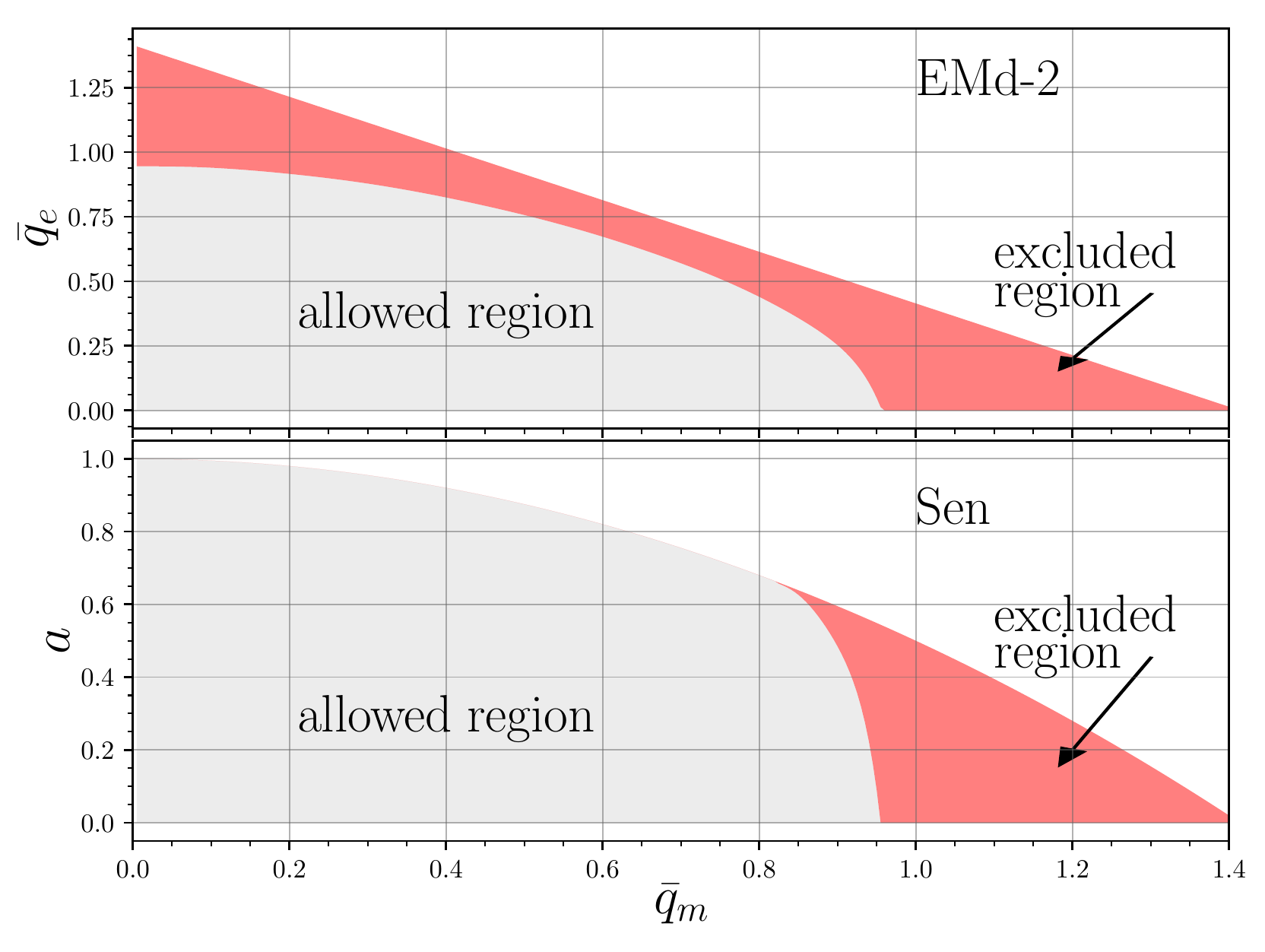}
\caption{Constraints set by the 2017 EHT observations on the nonrotating
  Einstein-Maxwell-dilaton 2 and on the rotating Sen black holes. Also in
  this case, the gray/red shaded regions refer to the areas that are
  $1$-$\sigma$ consistent/inconsistent with the 2017 EHT observations).}
\label{fig:EMd-2_Sen_Constraints}
\end{figure}

Overall, it is apparent that the regular Bardeen, Hayward, and Frolov
black-hole solutions are compatible with the present constraints. At the
same time, the Reissner-Nordstr\"om and Einstein-Maxwell-dilaton 1
black-hole solutions, for certain values of the physical charge, produce
shadow radii that lie outside the $1$-$\sigma$ region allowed by the 2017
EHT observations, and we find that these solutions are now constrained to
take values in, $0 < \bar{q} \lesssim 0.90$ and $0 < \bar{q} \lesssim
0.95$ respectively. Furthermore, the Reissner-Nordstr\"om naked
singularity is entirely eliminated as a viable model for M87* and the
Janis-Newman-Winicour naked singularity parameter space is restricted
further by this measurement to $0 < \hat{\bar{\nu}} \lesssim
0.47$. Finally, we also find that the KS black hole is also restricted to
have charges in the range $\bar{l} < 1.53$. In addition, note that the
nonrotating Einstein-Maxwell-dilaton 2 (EMd-2) solution
\cite{Kallosh1992} -- which depends on two independent charges -- can
also produce shadow radii that are incompatible with the EHT
observations; we will discuss this further below. The two EMd black-hole
solutions (1 and 2) correspond to fundamentally different field contents,
as discussed in \cite{Kallosh1992}.

We report in the right panel of
Fig. \ref{fig:Spherical_Shadow_Radius_Constraints} the shadow areal
radius $r_{{\rm sh}, A}$ for a number of stationary black holes, such as
Kerr \cite{Kerr1963}, Kerr-Newman (KN) \cite{Newman1965}, Sen
\cite{Sen1992}, and the rotating versions of the Bardeen and Hayward
black holes \cite{Bambi2013}. The data refers to an observer inclination
angle of $i=\pi/2$, and we find that the variation in the shadow size
with spin at higher inclinations (of up to $i = \pi/100$) is at most
about $7.1\%$ (for $i = \pi/2$, this is $5\%$); of course, at zero-spin
the shadow size does not change with inclination. The shadow areal radii
are shown as a function of the dimensionless spin of the black hole
$a:=J/M^2$, where $J$ is its angular momentum, and for representative
values of the additional parameters that characterize the solutions. Note
that -- similar to the angular momentum for a Kerr black hole -- the role
of an electric charge or the presence of a de Sitter core (as in the case
of the Hayward black holes) is to reduce the apparent size of the
shadow. Furthermore, on increasing the spin parameter, we recover the
typical trend that the shadow becomes increasingly noncircular, as
encoded, e.g., in the distortion parameter $\delta_{\text{sh}}$ defined
in \cite{Abdujabbarov2015, Abdujabbarov2016} (see Appendix
  \ref{sec:Distortion_Parameters}). Also in this case, while the regular
rotating Bardeen and Hayward solutions are compatible with the present
constraints set by the 2017 EHT observations, the Kerr-Newman and Sen
families of black holes can produce shadow areal radii that lie outside
of the $1$-$\sigma$ region allowed by the observations.

To further explore the constraints on the excluded regions for the
Einstein-Maxwell-dilaton 2 and the Sen black holes, we report in
Fig. \ref{fig:EMd-2_Sen_Constraints} the relevant ranges for these two
solutions. The Einstein-Maxwell-dilaton 2 black holes are nonrotating but
have two physical charges expressed by the coefficients $0 < \bar{q}_{e}
< \sqrt{2}$ and $0 < \bar{q}_{m} < \sqrt{2}$, while the Sen black holes
spin ($a$) and have an additional electromagnetic charge
$\bar{q}_{m}$. Also in this case, the gray/red shaded regions refer to
the areas that are consistent/inconsistent with the 2017 EHT
observations. The figure shows rather easily that for these two
black-hole families there are large areas of the space of parameters that
are excluded at the $1$-$\sigma$ level. Not surprisingly, these areas are
those where the physical charges take their largest values and hence the
corresponding black-hole solutions are furthest away from the
corresponding Schwarzschild or Kerr solutions. The obvious prospect is of
course that as the EHT increases the precision of it's measurements,
increasingly larger portions of the space of parameters of these black
holes will be excluded. Furthermore, other solutions that are presently
still compatible with the observations may see their corresponding
physical charges restricted.

\section{Conclusions} 

As our understanding of gravity under extreme regimes improves, and as
physical measurements of these regimes are now becoming available --
either through the imaging of supermassive black holes or the detection
of gravitational waves from stellar-mass black holes -- we are finally in
the position of setting some constraints to the large landscape of
non-Kerr black holes that have been proposed over the years. We have used
here the recent 2017 EHT observations of M87* to set constraints, at the
$1$-$\sigma$-level, on the physical charges -- either electric, scalar,
or angular momentum -- of a large variety of static (nonrotating) or
stationary (rotating) black holes.

In this way, when considering nonrotating black holes with a single
physical charge, we have been able to rule out, at 68\% confidence
levels, the possibility that M87* is a near-extremal Reissner-Nordstr\"om
or Einstein-Maxwell-dilaton 1 black hole and that the corresponding
physical charge must be in the range, RN: $0 < \bar{q} \lesssim 0.90$ and
EMd-1: $0 < \bar{q} \lesssim 0.95$. We also find that it cannot be a
Reissner-Nordstr\"om naked singularity or a JNW naked singularity with
large scalar charge, \ie only $0 < \hat{\bar{\nu}} \lesssim 0.47$ is
allowed. Similarly, when considering black holes with two physical
charges (either nonrotating or rotating), we have been able to exclude,
with 68\% confidence, considerable regions of the space of parameters in
the case of the Einstein-Maxwell-dilaton 2, Kerr-Newman and Sen black
holes. Although the idea of setting such constraints is an old one (see,
\eg \cite{Johannsen2011, *Johannsen2013PRD, *Vigeland2011,
  *Johannsen2013PRDb, *Rezzolla2014, *Younsi2016, *Konoplya2016a,
  Kocherlakota2020, Cunha2019, Kumar2019b, Kumar2020}), and while there
have been recent important developments in the study of other possible
observational signatures of such alternative solutions, such as in X-ray
spectra of accreting black holes (see, e.g., \cite{Bambi2014}) and in
gravitational waves \cite{Barausse2016b, *Konoplya2016c,
  *Felix-Louis2018, *Felix-Louis2018b, *Siahaan2020, Hirschmann2018}, to
the best of our knowledge, constraints of this type have not been set
before for the spacetimes considered here.

As a final remark, we note that while we have chosen only a few solutions
that can be seen as deviations from the Schwarzschild/Kerr solutions
since they share the same basic Einstein-Hilbert-Maxwell action of GR,
the work presented here is meant largely as a proof-of-concept
investigation and a methodological example of how to exploit observations
and measurements that impact the photon region.  While a certain
degeneracy in the shadow size induced by mass and spin remains and is
inevitable, when in the future the relative difference in the posterior
for the angular gravitational radius for M87* can be pushed to $\lesssim
5\%$, we should be able to constrain its spin, when modeling it as a Kerr
black hole. Furthermore, since this posterior implies a spread in the
estimated mass, one can expect small changes in the exact values of the
maximum allowed charges reported here. Hence, as future observations --
either in terms of black-hole imaging or of gravitational-wave detection
-- will become more precise and notwithstanding a poor measurement of the
black-hole spin, the methodology presented here can be readily applied to
set even tighter constraints on the physical charges of non-Einsteinian
black holes.


\medskip\noindent\textit{Acknowledgements.} It is a pleasure to thank
Enrico Barausse, Sebastian V\"olkel and Nicola Franchini for insightful
discussions on alternative black holes. 
During the completion of this work we have become aware of a related work
 by S. V\"olkel et al. \cite{Voelkel2020}, which deals with topics that 
 partly overlap with those of this manuscript (\ie EHT tests of the 
 strong-field regime of GR).


The  authors  of  the  present paper thank the following organizations 
and programs: the Academy of Finland (projects 274477, 284495, 312496, 
315721); the Alexander von Humboldt Stiftung; Agencia Nacional de 
Investigaci{\'o}n y Desarrollo (ANID), Chile via NCN19\_058 (TITANs), 
and Fondecyt 3190878; an Alfred P. Sloan Research Fellowship; Allegro, 
the European ALMA Regional Centre node in the Netherlands, the NL 
astronomy research network NOVA and the astronomy institutes of the 
University of Amsterdam, Leiden University and Radboud University; the 
Black Hole Initiative at Harvard University, through a grant (60477) 
from the John Templeton Foundation; the China Scholarship Council; 
Consejo Nacional de Ciencia y Tecnolog{\'i}a (CONACYT, Mexico, projects 
U0004-246083, U0004-259839, F0003-272050, M0037-279006, F0003-281692, 
104497, 275201, 263356); the Delaney Family via the Delaney Family John 
A.Wheeler Chair at Perimeter Institute; Direcci{\'o}n General de Asuntos 
del Personal Acad{\'e}mico-Universidad Nacional Autónomade M{\'e}xico 
(DGAPA-UNAM, projects IN112417 and IN112820); the EACOA Fellowship of 
the East Asia Core Observatories Association; the European Research 
Council Synergy Grant ``BlackHoleCam: Imaging the Event Horizon of Black 
Holes'' (grant 610058); the Generalitat Valenciana postdoctoral grant 
APOSTD/2018/177 and GenT Program (project CIDEGENT/2018/021); MICINN 
Research Project PID2019-108995GB-C22; the Gordon and Betty Moore 
Foundation (grants GBMF-3561, GBMF-5278); the Istituto Nazionale di 
Fisica Nucleare (INFN) sezione  di  Napoli, iniziative  specifiche 
TEONGRAV;  the International Max Planck Research School for Astronomy 
and Astrophysics at the Universities of Bonn and Cologne; Joint 
Princeton/Flatiron and  Joint Columbia/Flatiron Postdoctoral Fellowships, 
research at the Flatiron Institute is supported by the Simons Foundation; 
the Japanese Government (Monbukagakusho: MEXT) Scholarship; the Japan 
Society for the Promotion of Science (JSPS) Grant-in-Aid for JSPS 
Research Fellowship (JP17J08829); the Key Research Program of Frontier 
Sciences, Chinese Academy of Sciences (CAS, grants QYZDJ-SSW-SLH057, 
QYZDJSSW- SYS008, ZDBS-LY-SLH011); the Lever-hulme Trust Early Career 
Research Fellowship; the Max-Planck-Gesellschaft (MPG); the Max Planck 
Partner Group of the MPG and the CAS; the MEXT/JSPS KAKENHI (grants 
18KK0090, JP18K13594, JP18K03656, JP18H03721, 18K03709, 18H01245, 
25120007); the Malaysian Fundamental Research Grant Scheme (FRGS) 
FRGS/1/2019/STG02/UM/02/6; the MIT International Science and Technology 
Initiatives (MISTI) Funds; the Ministry of Science and Technology (MOST) 
of Taiwan (105-2112-M-001-025-MY3, 106-2112-M-001-011, 
106-2119-M-001-027, 107-2119-M-001-017,  107-2119-M-001-020,  
107-2119-M-110-005,108-2112-M-001-048, and 109-2124-M-001-005); the 
National Aeronautics and Space Administration (NASA, Fermi Guest 
Investigator grant 80NSSC20K1567 and 80NSSC20K1567, NASA Astrophysics 
Theory Program grant 80NSSC20K0527, NASA NuSTAR award  80NSSC20K0645, 
NASA  grant NNX17AL82G, and Hubble Fellowship grant HST-HF2-51431.001-A 
awarded by the Space Telescope Science Institute, which is operated by 
the Association of Universities for Research in Astronomy, Inc., for 
NASA, under contract NAS5-26555); the National Institute of Natural 
Sciences (NINS) of Japan; the National Key Research and Development 
Program of China (grant 2016YFA0400704, 2016YFA0400702); the National 
ScienceFoundation (NSF, grants AST-0096454, AST-0352953, AST-0521233, 
AST-0705062, AST-0905844, AST-0922984, AST-1126433, AST-1140030, 
DGE-1144085, AST-1207704, AST-1207730, AST-1207752, MRI-1228509, 
OPP-1248097, AST-1310896, AST-1337663, AST-1440254, AST-1555365, 
AST-1615796, AST-1715061, AST-1716327, AST-1716536, OISE-1743747, 
AST-1816420, AST-1903847, AST-1935980, AST-2034306); the Natural Science 
Foundation of China (grants 11573051, 11633006,  11650110427,  10625314,  
11721303, 11725312, 11933007, 11991052, 11991053); a fellowship of China 
Postdoctoral Science Foundation (2020M671266); the Natural Sciences and 
Engineering Research Council of Canada (NSERC, including a Discovery 
Grant and the NSERC Alexander Graham Bell Canada Graduate 
Scholarships-Doctoral Program); the  National  Research Foundation of 
Korea (the  Global PhD Fellowship  Grant: grants 2014H1A2A1018695,  
NRF-2015H1A2A1033752, 2015- R1D1A1A01056807, the Korea Research 
Fellowship Program: NRF-2015H1D3A1066561, Basic Research Support grant 
2019R1F1A1059721); the Netherlands Organization for 
Scientific Research (NWO) VICI award (grant 639.043.513) and Spinoza 
Prize SPI 78-409; the New Scientific Frontiers with Precision Radio 
Interferometry Fellowship awarded by the South African Radio Astronomy 
Observatory (SARAO), which is a facility of the National Research 
Foundation (NRF), an agency of the Department of Science and Innovation 
(DSI) of South Africa; the South African Research Chairs Initiative of 
the Department of Science and Innovation and National Research 
Foundation;  the  Onsala  Space  Observatory (OSO) national 
infrastructure, for the provisioning of its facilities/observational 
support (OSO receives funding through the Swedish Research Council under 
grant 2017-00648) the Perimeter Institute for Theoretical Physics 
(research at Perimeter Institute is supported by the Government of 
Canada through the Department of Innovation, Science and Economic 
Development and by the Province of Ontario through the Ministry of 
Research, Innovation and Science);  the  Spanish  Ministerio  de  
Ciencia  e  Innovaci{\'o}n (grants PGC2018-098915-B-C21, 
AYA2016-80889-P; PID2019-108995GB-C21, PGC2018-098915-B-C21); the State 
Agency for Research of the Spanish MCIU through the ``Center of 
Excellence Severo Ochoa'' award for the Instituto de Astrof{\'i}sica de 
Andaluc{\'i}a (SEV-2017-0709); the Toray Science Foundation; the 
Consejer{\'i}a de Econom{\'i}a, Conocimiento, Empresas y Universidad of 
the Junta de Andaluc{\'i}a (grant P18-FR-1769), the Consejo Superior de 
Investigaciones Cient{\'i}ficas (grant 2019AEP112); the US Department of 
Energy (USDOE) through the Los Alamos National Laboratory (operated by 
Triad National Security, LLC, for the National Nuclear Security 
Administration of the USDOE (Contract 89233218CNA000001); the European 
Union's Horizon 2020 research and innovation program under grant 
agreement No 730562 RadioNet; ALMA North America Development Fund; the 
Academia Sinica; Chandra TM6-17006X; Chandra award DD7-18089X. This work 
used the Extreme  Science  and  Engineering  Discovery  Environment 
(XSEDE), supported by NSF grant ACI-1548562, and CyVerse, supported by 
NSF grants DBI-0735191, DBI-1265383, and DBI-1743442. XSEDE Stampede2 
resource at TACC was allocated through TG-AST170024 and TG-AST080026N. 
XSEDE Jet-Stream resource at PTI and TACC was allocated through 
AST170028. The simulations were performed in part on the SuperMUC 
cluster at the LRZ in Garching, on the LOEWE cluster in CSC in 
Frankfurt, and on the HazelHen cluster at the HLRS inStuttgart. This 
research was enabled in part by support provided by Compute  Ontario 
(\url{http://computeontario.ca}),  Calcul  Quebec 
(\url{http://www.calculquebec.ca}) and  Compute  Canada 
(\url{http://www.computecanada.ca}). We thank the staff at the 
participating observatories, correlation centers, and institutions for 
their enthusiastic support. This paper makes use of the following ALMA 
data: ADS/JAO.ALMA\#2016.1.01154.V. ALMA is  a partnership of the 
European Southern Observatory (ESO; Europe, representing its member 
states), NSF, and National Institutes of Natural Sciences of Japan, 
together with National Research Council (Canada), Ministry of Science 
and Technology (MOST; Taiwan), Academia Sinica Institute of Astronomy 
and Astro-physics(ASIAA; Taiwan), and Korea Astronomy and Space Science 
Institute (KASI; Republic of Korea), in cooperation with the Republic 
of Chile. The Joint ALMA Observatory is operated by ESO, Associated 
Universities, Inc. (AUI)/NRAO, and the National Astronomical Observatory 
of Japan (NAOJ). The NRAO is a facility of the NSF operated under 
cooperative agreement by AUI. This paper has made use of the following 
APEX data: Project ID T-091.F-0006-2013. APEX is a collaboration between 
the Max-Planck-Institut f{\"u}r Radioastronomie (Germany), ESO, and the 
Onsala Space Observatory (Sweden). The SMA is a joint project between 
the SAO and ASIAA and is funded by the Smithsonian Institution and the 
Academia Sinica. The JCMT is operated by the East Asian Observatory on 
behalf of the NAOJ, ASIAA, and KASI, as well as the Ministry of Finance 
of China, Chinese Academy of Sciences, and the National Key R\&D Program 
(No. 2017YFA0402700) of China. Additional funding support for the JCMT 
is provided by the Science and Technologies Facility Council (UK) and 
participating universities in the UK and Canada. The LMT is a project 
operated by the Instituto Nacionalde Astrof{\'i}sica, {\'O}ptica, y 
Electr{\'o}nica (Mexico) and the University of Massachusetts at 
Amherst (USA), with financial support from the Consejo Nacional de 
Ciencia y Tecnolog{\'i}a and the National Science Foundation. The IRAM 
30-m telescope on Pico Veleta, Spain is operated by IRAM and supported 
by CNRS (Centre National de la Recherche Scientifique, France), MPG 
(Max-Planck- Gesellschaft, Germany) and IGN (Instituto Geogr{\'a}fico 
Nacional, Spain). The SMT is operated by the Arizona Radio Observatory, 
a part of the Steward Observatory of the University of Arizona, with 
financial support of operations from the State of Arizona and financial 
support for instrumentation development from the NSF. The SPT is 
supported by the National Science Foundation through grant PLR-1248097. 
Partial support is also provided by the NSF Physics Frontier Center 
grant PHY-1125897 to the Kavli Institute of Cosmological Physics at the 
University of Chicago, the Kavli Foundation and the Gordon and Betty 
Moore Foundation grant GBMF 947. The SPT hydrogen maser was provided on 
loan from the GLT, courtesy of ASIAA. The EHTC has received generous 
donations of FPGA chips from Xilinx Inc., under the Xilinx University 
Program. The EHTC has benefited from  technology  shared  under  
open-source  license  by  the Collaboration for Astronomy Signal 
Processing and Electronics Research (CASPER). The EHT project is 
grateful to T4Science and Microsemi for their assistance with Hydrogen 
Masers. This research has made use of NASA's Astrophysics Data System. 
We gratefully acknowledge the support provided by the extended staff of 
the ALMA, both from the inception of the ALMA Phasing Project through 
the observational campaigns of 2017 and 2018. We would like to thank 
A. Deller and W. Brisken for EHT-specific support with the use of DiFX. 
We acknowledge the significance that Maunakea, where the SMA and JCMT 
EHT stations are located, has for the indigenous Hawaiian people. 

\textit{Facilities}: EHT,  ALMA,  APEX,  IRAM:30 m, JCMT, LMT, SMA, 
ARO:SMT, SPT.

\textit{Software}: AIPS \cite{Greisen2003}, ParselTongue 
\cite{Kettenis2006}, GNU Parallel \cite{Tange2011}, 
$\mathtt{eht}$-$\mathtt{imaging}$ \cite{Chael2016}, Difmap 
\cite{Shepherd2011}, Numpy \cite{vanderWalt2011}, Scipy 
\cite{Jones2001}, Pandas \cite{McKinney2010Pandas}, Astropy 
\cite{Astropy2013, Astropy2018}, Jupyter \cite{Kluyver2016}, Matplotlib 
\cite{Hunter2007}, THEMIS \cite{Broderick2020}, DMC \cite{Pesce2021}, 
$\mathtt{polsolve}$ \cite{MartiVidal2021}, GPCAL \cite{Park2021}.


\bibliographystyle{apsrev4-2}
\bibliography{aeireferences}

\begin{thebibliography}{114}%
\makeatletter
\providecommand \@ifxundefined [1]{%
 \@ifx{#1\undefined}
}%
\providecommand \@ifnum [1]{%
 \ifnum #1\expandafter \@firstoftwo
 \else \expandafter \@secondoftwo
 \fi
}%
\providecommand \@ifx [1]{%
 \ifx #1\expandafter \@firstoftwo
 \else \expandafter \@secondoftwo
 \fi
}%
\providecommand \natexlab [1]{#1}%
\providecommand \enquote  [1]{``#1''}%
\providecommand \bibnamefont  [1]{#1}%
\providecommand \bibfnamefont [1]{#1}%
\providecommand \citenamefont [1]{#1}%
\providecommand \href@noop [0]{\@secondoftwo}%
\providecommand \href [0]{\begingroup \@sanitize@url \@href}%
\providecommand \@href[1]{\@@startlink{#1}\@@href}%
\providecommand \@@href[1]{\endgroup#1\@@endlink}%
\providecommand \@sanitize@url [0]{\catcode `\\12\catcode `\$12\catcode
  `\&12\catcode `\#12\catcode `\^12\catcode `\_12\catcode `\%12\relax}%
\providecommand \@@startlink[1]{}%
\providecommand \@@endlink[0]{}%
\providecommand \url  [0]{\begingroup\@sanitize@url \@url }%
\providecommand \@url [1]{\endgroup\@href {#1}{\urlprefix }}%
\providecommand \urlprefix  [0]{URL }%
\providecommand \Eprint [0]{\href }%
\providecommand \doibase [0]{https://doi.org/}%
\providecommand \selectlanguage [0]{\@gobble}%
\providecommand \bibinfo  [0]{\@secondoftwo}%
\providecommand \bibfield  [0]{\@secondoftwo}%
\providecommand \translation [1]{[#1]}%
\providecommand \BibitemOpen [0]{}%
\providecommand \bibitemStop [0]{}%
\providecommand \bibitemNoStop [0]{.\EOS\space}%
\providecommand \EOS [0]{\spacefactor3000\relax}%
\providecommand \BibitemShut  [1]{\csname bibitem#1\endcsname}%
\let\auto@bib@innerbib\@empty
\bibitem [{\citenamefont {{Dicke}}(2019)}]{Dicke1964}%
  \BibitemOpen
  \bibfield  {author} {\bibinfo {author} {\bibfnamefont {R.~H.}\ \bibnamefont
  {{Dicke}}},\ }\href {https://doi.org/10.1007/s10714-019-2509-2} {\bibfield
  {journal} {\bibinfo  {journal} {General Relativity and Gravitation}\ }\textbf
  {\bibinfo {volume} {51}},\ \bibinfo {eid} {57} (\bibinfo {year}
  {2019})}\BibitemShut {NoStop}%
\bibitem [{\citenamefont {{Will}}(2006)}]{Will:2006LRR}%
  \BibitemOpen
  \bibfield  {author} {\bibinfo {author} {\bibfnamefont {C.~M.}\ \bibnamefont
  {{Will}}},\ }\href {https://doi.org/10.12942/lrr-2006-3} {\bibfield
  {journal} {\bibinfo  {journal} {Living Rev. Relativity}\ }\textbf {\bibinfo
  {volume} {9}},\ \bibinfo {pages} {3} (\bibinfo {year} {2006})},\ \Eprint
  {https://arxiv.org/abs/arXiv:gr-qc/0510072} {arXiv:gr-qc/0510072}
  \BibitemShut {NoStop}%
\bibitem [{\citenamefont {{Collett}}\ \emph {et~al.}(2018)\citenamefont
  {{Collett}}, \citenamefont {{Oldham}}, \citenamefont {{Smith}}, \citenamefont
  {{Auger}}, \citenamefont {{Westfall}}, \citenamefont {{Bacon}}, \citenamefont
  {{Nichol}}, \citenamefont {{Masters}}, \citenamefont {{Koyama}},\ and\
  \citenamefont {{van den Bosch}}}]{Collett2018}%
  \BibitemOpen
  \bibfield  {author} {\bibinfo {author} {\bibfnamefont {T.~E.}\ \bibnamefont
  {{Collett}}}, \bibinfo {author} {\bibfnamefont {L.~J.}\ \bibnamefont
  {{Oldham}}}, \bibinfo {author} {\bibfnamefont {R.~J.}\ \bibnamefont
  {{Smith}}}, \bibinfo {author} {\bibfnamefont {M.~W.}\ \bibnamefont
  {{Auger}}}, \bibinfo {author} {\bibfnamefont {K.~B.}\ \bibnamefont
  {{Westfall}}}, \bibinfo {author} {\bibfnamefont {D.}~\bibnamefont {{Bacon}}},
  \bibinfo {author} {\bibfnamefont {R.~C.}\ \bibnamefont {{Nichol}}}, \bibinfo
  {author} {\bibfnamefont {K.~L.}\ \bibnamefont {{Masters}}}, \bibinfo {author}
  {\bibfnamefont {K.}~\bibnamefont {{Koyama}}},\ and\ \bibinfo {author}
  {\bibfnamefont {R.}~\bibnamefont {{van den Bosch}}},\ }\href
  {https://doi.org/10.1126/science.aao2469} {\bibfield  {journal} {\bibinfo
  {journal} {Science}\ }\textbf {\bibinfo {volume} {360}},\ \bibinfo {pages}
  {134} (\bibinfo {year} {2018})},\ \Eprint {https://arxiv.org/abs/1806.08300}
  {arXiv:1806.08300 [astro-ph.CO]} \BibitemShut {NoStop}%
\bibitem [{\citenamefont {{'t Hooft}}\ and\ \citenamefont
  {{Veltman}}(1974)}]{tHooft1974}%
  \BibitemOpen
  \bibfield  {author} {\bibinfo {author} {\bibfnamefont {G.}~\bibnamefont {{'t
  Hooft}}}\ and\ \bibinfo {author} {\bibfnamefont {M.}~\bibnamefont
  {{Veltman}}},\ }\href@noop {} {\bibfield  {journal} {\bibinfo  {journal}
  {Annales de L'Institut Henri Poincare Section (A) Physique Theorique}\
  }\textbf {\bibinfo {volume} {20}},\ \bibinfo {pages} {69} (\bibinfo {year}
  {1974})}\BibitemShut {NoStop}%
\bibitem [{\citenamefont {{Krasnikov}}(1987)}]{Krasnikov1987}%
  \BibitemOpen
  \bibfield  {author} {\bibinfo {author} {\bibfnamefont {N.~V.}\ \bibnamefont
  {{Krasnikov}}},\ }\href {https://doi.org/10.1007/BF01017588} {\bibfield
  {journal} {\bibinfo  {journal} {Theoretical and Mathematical Physics}\
  }\textbf {\bibinfo {volume} {73}},\ \bibinfo {pages} {1184} (\bibinfo {year}
  {1987})}\BibitemShut {NoStop}%
\bibitem [{\citenamefont {{L{\"u}}}\ \emph {et~al.}(2015)\citenamefont
  {{L{\"u}}}, \citenamefont {{Perkins}}, \citenamefont {{Pope}},\ and\
  \citenamefont {{Stelle}}}]{Lu2015}%
  \BibitemOpen
  \bibfield  {author} {\bibinfo {author} {\bibfnamefont {H.}~\bibnamefont
  {{L{\"u}}}}, \bibinfo {author} {\bibfnamefont {A.}~\bibnamefont {{Perkins}}},
  \bibinfo {author} {\bibfnamefont {C.~N.}\ \bibnamefont {{Pope}}},\ and\
  \bibinfo {author} {\bibfnamefont {K.~S.}\ \bibnamefont {{Stelle}}},\ }\href
  {https://doi.org/10.1103/PhysRevLett.114.171601} {\bibfield  {journal}
  {\bibinfo  {journal} {Phys. Rev. Lett.}\ }\textbf {\bibinfo {volume} {114}},\
  \bibinfo {eid} {171601} (\bibinfo {year} {2015})},\ \Eprint
  {https://arxiv.org/abs/1502.01028} {arXiv:1502.01028 [hep-th]} \BibitemShut
  {NoStop}%
\bibitem [{\citenamefont {{Scherk}}\ and\ \citenamefont
  {{Schwarz}}(1979)}]{Scherk1979}%
  \BibitemOpen
  \bibfield  {author} {\bibinfo {author} {\bibfnamefont {J.}~\bibnamefont
  {{Scherk}}}\ and\ \bibinfo {author} {\bibfnamefont {J.~H.}\ \bibnamefont
  {{Schwarz}}},\ }\href {https://doi.org/10.1016/0550-3213(79)90592-3}
  {\bibfield  {journal} {\bibinfo  {journal} {Nuclear Physics B}\ }\textbf
  {\bibinfo {volume} {153}},\ \bibinfo {pages} {61} (\bibinfo {year}
  {1979})}\BibitemShut {NoStop}%
\bibitem [{\citenamefont {{Green}}\ \emph {et~al.}(1988)\citenamefont
  {{Green}}, \citenamefont {{Schwarz}},\ and\ \citenamefont
  {{Witten}}}]{Green1988}%
  \BibitemOpen
  \bibfield  {author} {\bibinfo {author} {\bibfnamefont {M.~B.}\ \bibnamefont
  {{Green}}}, \bibinfo {author} {\bibfnamefont {J.~H.}\ \bibnamefont
  {{Schwarz}}},\ and\ \bibinfo {author} {\bibfnamefont {E.}~\bibnamefont
  {{Witten}}},\ }\href@noop {} {\emph {\bibinfo {title} {{Superstring
  Theory}}}}\ (\bibinfo {year} {1988})\BibitemShut {NoStop}%
\bibitem [{\citenamefont {{Barausse}}\ \emph {et~al.}(2011)\citenamefont
  {{Barausse}}, \citenamefont {{Jacobson}},\ and\ \citenamefont
  {{Sotiriou}}}]{Barausse2011}%
  \BibitemOpen
  \bibfield  {author} {\bibinfo {author} {\bibfnamefont {E.}~\bibnamefont
  {{Barausse}}}, \bibinfo {author} {\bibfnamefont {T.}~\bibnamefont
  {{Jacobson}}},\ and\ \bibinfo {author} {\bibfnamefont {T.~P.}\ \bibnamefont
  {{Sotiriou}}},\ }\href {https://doi.org/10.1103/PhysRevD.83.124043}
  {\bibfield  {journal} {\bibinfo  {journal} {Phys. Rev. D}\ }\textbf {\bibinfo
  {volume} {83}},\ \bibinfo {eid} {124043} (\bibinfo {year} {2011})},\ \Eprint
  {https://arxiv.org/abs/1104.2889} {arXiv:1104.2889 [gr-qc]} \BibitemShut
  {NoStop}%
\bibitem [{\citenamefont {{Barausse}}\ and\ \citenamefont
  {{Sotiriou}}(2013)}]{Barausse2013CQG}%
  \BibitemOpen
  \bibfield  {author} {\bibinfo {author} {\bibfnamefont {E.}~\bibnamefont
  {{Barausse}}}\ and\ \bibinfo {author} {\bibfnamefont {T.~P.}\ \bibnamefont
  {{Sotiriou}}},\ }\href {https://doi.org/10.1088/0264-9381/30/24/244010}
  {\bibfield  {journal} {\bibinfo  {journal} {Classical and Quantum Gravity}\
  }\textbf {\bibinfo {volume} {30}},\ \bibinfo {eid} {244010} (\bibinfo {year}
  {2013})},\ \Eprint {https://arxiv.org/abs/1307.3359} {arXiv:1307.3359
  [gr-qc]} \BibitemShut {NoStop}%
\bibitem [{\citenamefont {{Barausse}}\ \emph
  {et~al.}(2016{\natexlab{a}})\citenamefont {{Barausse}}, \citenamefont
  {{Sotiriou}},\ and\ \citenamefont {{Vega}}}]{Barausse2016}%
  \BibitemOpen
  \bibfield  {author} {\bibinfo {author} {\bibfnamefont {E.}~\bibnamefont
  {{Barausse}}}, \bibinfo {author} {\bibfnamefont {T.~P.}\ \bibnamefont
  {{Sotiriou}}},\ and\ \bibinfo {author} {\bibfnamefont {I.}~\bibnamefont
  {{Vega}}},\ }\href {https://doi.org/10.1103/PhysRevD.93.044044} {\bibfield
  {journal} {\bibinfo  {journal} {Phys. Rev. D}\ }\textbf {\bibinfo {volume}
  {93}},\ \bibinfo {eid} {044044} (\bibinfo {year} {2016}{\natexlab{a}})},\
  \Eprint {https://arxiv.org/abs/1512.05894} {arXiv:1512.05894 [gr-qc]}
  \BibitemShut {NoStop}%
\bibitem [{\citenamefont {{Ramos}}\ and\ \citenamefont
  {{Barausse}}(2019)}]{Ramos2019}%
  \BibitemOpen
  \bibfield  {author} {\bibinfo {author} {\bibfnamefont {O.}~\bibnamefont
  {{Ramos}}}\ and\ \bibinfo {author} {\bibfnamefont {E.}~\bibnamefont
  {{Barausse}}},\ }\href {https://doi.org/10.1103/PhysRevD.99.024034}
  {\bibfield  {journal} {\bibinfo  {journal} {Phys. Rev. D}\ }\textbf {\bibinfo
  {volume} {99}},\ \bibinfo {eid} {024034} (\bibinfo {year} {2019})},\ \Eprint
  {https://arxiv.org/abs/1811.07786} {arXiv:1811.07786 [gr-qc]} \BibitemShut
  {NoStop}%
\bibitem [{\citenamefont {{Sarbach}}\ \emph {et~al.}(2019)\citenamefont
  {{Sarbach}}, \citenamefont {{Barausse}},\ and\ \citenamefont
  {{Preciado-L{\'o}pez}}}]{Sarbach2019}%
  \BibitemOpen
  \bibfield  {author} {\bibinfo {author} {\bibfnamefont {O.}~\bibnamefont
  {{Sarbach}}}, \bibinfo {author} {\bibfnamefont {E.}~\bibnamefont
  {{Barausse}}},\ and\ \bibinfo {author} {\bibfnamefont {J.~A.}\ \bibnamefont
  {{Preciado-L{\'o}pez}}},\ }\href {https://doi.org/10.1088/1361-6382/ab2e13}
  {\bibfield  {journal} {\bibinfo  {journal} {Classical and Quantum Gravity}\
  }\textbf {\bibinfo {volume} {36}},\ \bibinfo {eid} {165007} (\bibinfo {year}
  {2019})},\ \Eprint {https://arxiv.org/abs/1902.05130} {arXiv:1902.05130
  [gr-qc]} \BibitemShut {NoStop}%
\bibitem [{\citenamefont {{Damour}}\ and\ \citenamefont
  {{Taylor}}(1992)}]{Damour1992PRD}%
  \BibitemOpen
  \bibfield  {author} {\bibinfo {author} {\bibfnamefont {T.}~\bibnamefont
  {{Damour}}}\ and\ \bibinfo {author} {\bibfnamefont {J.~H.}\ \bibnamefont
  {{Taylor}}},\ }\href {https://doi.org/10.1103/PhysRevD.45.1840} {\bibfield
  {journal} {\bibinfo  {journal} {Phys. Rev. D}\ }\textbf {\bibinfo {volume}
  {45}},\ \bibinfo {pages} {1840} (\bibinfo {year} {1992})}\BibitemShut
  {NoStop}%
\bibitem [{\citenamefont {{Wex}}(2014)}]{Wex2014}%
  \BibitemOpen
  \bibfield  {author} {\bibinfo {author} {\bibfnamefont {N.}~\bibnamefont
  {{Wex}}},\ }\href@noop {} {\bibfield  {journal} {\bibinfo  {journal} {arXiv
  e-prints}\ } (\bibinfo {year} {2014})},\ \Eprint
  {https://arxiv.org/abs/1402.5594} {arXiv:1402.5594 [gr-qc]} \BibitemShut
  {NoStop}%
\bibitem [{\citenamefont {{Wex}}\ and\ \citenamefont
  {{Kramer}}(2020)}]{Wex2020}%
  \BibitemOpen
  \bibfield  {author} {\bibinfo {author} {\bibfnamefont {N.}~\bibnamefont
  {{Wex}}}\ and\ \bibinfo {author} {\bibfnamefont {M.}~\bibnamefont
  {{Kramer}}},\ }\href {https://doi.org/10.3390/universe6090156} {\bibfield
  {journal} {\bibinfo  {journal} {Universe}\ }\textbf {\bibinfo {volume} {6}},\
  \bibinfo {pages} {156} (\bibinfo {year} {2020})}\BibitemShut {NoStop}%
\bibitem [{\citenamefont {Abuter}\ \emph {et~al.}(2018)\citenamefont {Abuter}
  \emph {et~al.}}]{Abuter2018}%
  \BibitemOpen
  \bibfield  {author} {\bibinfo {author} {\bibfnamefont {R.}~\bibnamefont
  {Abuter}} \emph {et~al.} (\bibinfo {collaboration} {GRAVITY}),\ }\href
  {https://doi.org/10.1051/0004-6361/201833718} {\bibfield  {journal} {\bibinfo
   {journal} {Astron. Astrophys.}\ }\textbf {\bibinfo {volume} {615}},\
  \bibinfo {pages} {L15} (\bibinfo {year} {2018})},\ \Eprint
  {https://arxiv.org/abs/1807.09409} {arXiv:1807.09409 [astro-ph.GA]}
  \BibitemShut {NoStop}%
\bibitem [{\citenamefont {Abuter}\ \emph {et~al.}(2020)\citenamefont {Abuter},
  \citenamefont {Amorim}, \citenamefont {Baub{\"{o}}ck}, \citenamefont
  {Berger}, \citenamefont {Bonnet}, \citenamefont {Brandner}, \citenamefont
  {Cardoso}, \citenamefont {Cl{\'{e}}net}, \citenamefont {de~Zeeuw},\ and\
  \citenamefont {Dexter}}]{Abuter2020_etal}%
  \BibitemOpen
  \bibfield  {author} {\bibinfo {author} {\bibfnamefont {R.}~\bibnamefont
  {Abuter}}, \bibinfo {author} {\bibfnamefont {A.}~\bibnamefont {Amorim}},
  \bibinfo {author} {\bibfnamefont {M.}~\bibnamefont {Baub{\"{o}}ck}}, \bibinfo
  {author} {\bibfnamefont {J.~P.}\ \bibnamefont {Berger}}, \bibinfo {author}
  {\bibfnamefont {H.}~\bibnamefont {Bonnet}}, \bibinfo {author} {\bibfnamefont
  {W.}~\bibnamefont {Brandner}}, \bibinfo {author} {\bibfnamefont
  {V.}~\bibnamefont {Cardoso}}, \bibinfo {author} {\bibfnamefont
  {Y.}~\bibnamefont {Cl{\'{e}}net}}, \bibinfo {author} {\bibfnamefont {P.~T.}\
  \bibnamefont {de~Zeeuw}},\ and\ \bibinfo {author} {\bibfnamefont
  {J.}~\bibnamefont {Dexter}},\ }\href
  {https://doi.org/10.1051/0004-6361/202037813} {\bibfield  {journal} {\bibinfo
   {journal} {Astronomy {\&} Astrophysics}\ }\textbf {\bibinfo {volume}
  {636}},\ \bibinfo {pages} {L5} (\bibinfo {year} {2020})},\ \Eprint
  {https://arxiv.org/abs/2004.07187} {arXiv:2004.07187 [astro-ph.GA]}
  \BibitemShut {NoStop}%
\bibitem [{\citenamefont {{Abbott}}\ \emph
  {et~al.}(2016{\natexlab{a}})\citenamefont {{Abbott}}, \citenamefont
  {{Abbott}}, \citenamefont {{Abbott}}, \citenamefont {{Abernathy}},
  \citenamefont {{Acernese}}, \citenamefont {{Ackley}}, \citenamefont
  {{Adams}}, \citenamefont {{Adams}}, \citenamefont {{Addesso}}, \citenamefont
  {{Adhikari}},\ and\ \citenamefont {et~al.}}]{Abbot2016-GW-detection-prl}%
  \BibitemOpen
  \bibfield  {author} {\bibinfo {author} {\bibfnamefont {B.~P.}\ \bibnamefont
  {{Abbott}}}, \bibinfo {author} {\bibfnamefont {R.}~\bibnamefont {{Abbott}}},
  \bibinfo {author} {\bibfnamefont {T.~D.}\ \bibnamefont {{Abbott}}}, \bibinfo
  {author} {\bibfnamefont {M.~R.}\ \bibnamefont {{Abernathy}}}, \bibinfo
  {author} {\bibfnamefont {F.}~\bibnamefont {{Acernese}}}, \bibinfo {author}
  {\bibfnamefont {K.}~\bibnamefont {{Ackley}}}, \bibinfo {author}
  {\bibfnamefont {C.}~\bibnamefont {{Adams}}}, \bibinfo {author} {\bibfnamefont
  {T.}~\bibnamefont {{Adams}}}, \bibinfo {author} {\bibfnamefont
  {P.}~\bibnamefont {{Addesso}}}, \bibinfo {author} {\bibfnamefont {R.~X.}\
  \bibnamefont {{Adhikari}}},\ and\ \bibinfo {author} {\bibnamefont {et~al.}},\
  }\href {https://doi.org/10.1103/PhysRevLett.116.061102} {\bibfield  {journal}
  {\bibinfo  {journal} {Phys. Rev. Lett.}\ }\textbf {\bibinfo {volume} {116}},\
  \bibinfo {eid} {061102} (\bibinfo {year} {2016}{\natexlab{a}})},\ \Eprint
  {https://arxiv.org/abs/1602.03837} {arXiv:1602.03837 [gr-qc]} \BibitemShut
  {NoStop}%
\bibitem [{\citenamefont {{Abbott}}\ \emph
  {et~al.}(2016{\natexlab{b}})\citenamefont {{Abbott}}, \citenamefont
  {{Abbott}}, \citenamefont {{Abbott}}, \citenamefont {{Abernathy}},
  \citenamefont {{Acernese}}, \citenamefont {{Ackley}}, \citenamefont
  {{Adams}}, \citenamefont {{Adams}}, \citenamefont {{Addesso}}, \citenamefont
  {{Adhikari}},\ and\ \citenamefont {et~al.}}]{Abbott2016b}%
  \BibitemOpen
  \bibfield  {author} {\bibinfo {author} {\bibfnamefont {B.~P.}\ \bibnamefont
  {{Abbott}}}, \bibinfo {author} {\bibfnamefont {R.}~\bibnamefont {{Abbott}}},
  \bibinfo {author} {\bibfnamefont {T.~D.}\ \bibnamefont {{Abbott}}}, \bibinfo
  {author} {\bibfnamefont {M.~R.}\ \bibnamefont {{Abernathy}}}, \bibinfo
  {author} {\bibfnamefont {F.}~\bibnamefont {{Acernese}}}, \bibinfo {author}
  {\bibfnamefont {K.}~\bibnamefont {{Ackley}}}, \bibinfo {author}
  {\bibfnamefont {C.}~\bibnamefont {{Adams}}}, \bibinfo {author} {\bibfnamefont
  {T.}~\bibnamefont {{Adams}}}, \bibinfo {author} {\bibfnamefont
  {P.}~\bibnamefont {{Addesso}}}, \bibinfo {author} {\bibfnamefont {R.~X.}\
  \bibnamefont {{Adhikari}}},\ and\ \bibinfo {author} {\bibnamefont {et~al.}},\
  }\href {https://doi.org/10.3847/2041-8205/818/2/L22} {\bibfield  {journal}
  {\bibinfo  {journal} {Astrophys. J. Lett.}\ }\textbf {\bibinfo {volume}
  {818}},\ \bibinfo {eid} {L22} (\bibinfo {year} {2016}{\natexlab{b}})},\
  \Eprint {https://arxiv.org/abs/1602.03846} {arXiv:1602.03846 [astro-ph.HE]}
  \BibitemShut {NoStop}%
\bibitem [{\citenamefont {{Event Horizon Telescope Collaboration}}\ \emph
  {et~al.}(2019{\natexlab{a}})\citenamefont {{Event Horizon Telescope
  Collaboration}}, \citenamefont {{Akiyama}}, \citenamefont {{Alberdi}},
  \citenamefont {{Alef}}, \citenamefont {{Asada}}, \citenamefont {{Azulay}},
  \citenamefont {{Baczko}}, \citenamefont {{Ball}}, \citenamefont
  {{Balokovi{\'c}}}, \citenamefont {{Barrett}} \emph
  {et~al.}}]{EHT_M87_PaperI}%
  \BibitemOpen
  \bibfield  {author} {\bibinfo {author} {\bibnamefont {{Event Horizon
  Telescope Collaboration}}}, \bibinfo {author} {\bibfnamefont
  {K.}~\bibnamefont {{Akiyama}}}, \bibinfo {author} {\bibfnamefont
  {A.}~\bibnamefont {{Alberdi}}}, \bibinfo {author} {\bibfnamefont
  {W.}~\bibnamefont {{Alef}}}, \bibinfo {author} {\bibfnamefont
  {K.}~\bibnamefont {{Asada}}}, \bibinfo {author} {\bibfnamefont
  {R.}~\bibnamefont {{Azulay}}}, \bibinfo {author} {\bibfnamefont {A.-K.}\
  \bibnamefont {{Baczko}}}, \bibinfo {author} {\bibfnamefont {D.}~\bibnamefont
  {{Ball}}}, \bibinfo {author} {\bibfnamefont {M.}~\bibnamefont
  {{Balokovi{\'c}}}}, \bibinfo {author} {\bibfnamefont {J.}~\bibnamefont
  {{Barrett}}}, \emph {et~al.},\ }\href
  {https://doi.org/10.3847/2041-8213/ab0ec7} {\bibfield  {journal} {\bibinfo
  {journal} {Astrophys. J. Lett.}\ }\textbf {\bibinfo {volume} {875}},\
  \bibinfo {eid} {L1} (\bibinfo {year} {2019}{\natexlab{a}})}\BibitemShut
  {NoStop}%
\bibitem [{\citenamefont {{Event Horizon Telescope Collaboration}}\ \emph
  {et~al.}(2019{\natexlab{b}})\citenamefont {{Event Horizon Telescope
  Collaboration}}, \citenamefont {{Akiyama}}, \citenamefont {{Alberdi}},
  \citenamefont {{Alef}}, \citenamefont {{Asada}}, \citenamefont {{Azulay}},
  \citenamefont {{Baczko}}, \citenamefont {{Ball}}, \citenamefont
  {{Balokovi{\'c}}}, \citenamefont {{Barrett}} \emph
  {et~al.}}]{EHT_M87_PaperII}%
  \BibitemOpen
  \bibfield  {author} {\bibinfo {author} {\bibnamefont {{Event Horizon
  Telescope Collaboration}}}, \bibinfo {author} {\bibfnamefont
  {K.}~\bibnamefont {{Akiyama}}}, \bibinfo {author} {\bibfnamefont
  {A.}~\bibnamefont {{Alberdi}}}, \bibinfo {author} {\bibfnamefont
  {W.}~\bibnamefont {{Alef}}}, \bibinfo {author} {\bibfnamefont
  {K.}~\bibnamefont {{Asada}}}, \bibinfo {author} {\bibfnamefont
  {R.}~\bibnamefont {{Azulay}}}, \bibinfo {author} {\bibfnamefont {A.-K.}\
  \bibnamefont {{Baczko}}}, \bibinfo {author} {\bibfnamefont {D.}~\bibnamefont
  {{Ball}}}, \bibinfo {author} {\bibfnamefont {M.}~\bibnamefont
  {{Balokovi{\'c}}}}, \bibinfo {author} {\bibfnamefont {J.}~\bibnamefont
  {{Barrett}}}, \emph {et~al.},\ }\href
  {https://doi.org/10.3847/2041-8213/ab0c96} {\bibfield  {journal} {\bibinfo
  {journal} {Astrophys. J. Lett.}\ }\textbf {\bibinfo {volume} {875}},\
  \bibinfo {eid} {L2} (\bibinfo {year} {2019}{\natexlab{b}})}\BibitemShut
  {NoStop}%
\bibitem [{\citenamefont {{Event Horizon Telescope Collaboration}}\ \emph
  {et~al.}(2019{\natexlab{c}})\citenamefont {{Event Horizon Telescope
  Collaboration}}, \citenamefont {{Akiyama}}, \citenamefont {{Alberdi}},
  \citenamefont {{Alef}}, \citenamefont {{Asada}}, \citenamefont {{Azulay}},
  \citenamefont {{Baczko}}, \citenamefont {{Ball}}, \citenamefont
  {{Balokovi{\'c}}}, \citenamefont {{Barrett}} \emph
  {et~al.}}]{EHT_M87_PaperIII}%
  \BibitemOpen
  \bibfield  {author} {\bibinfo {author} {\bibnamefont {{Event Horizon
  Telescope Collaboration}}}, \bibinfo {author} {\bibfnamefont
  {K.}~\bibnamefont {{Akiyama}}}, \bibinfo {author} {\bibfnamefont
  {A.}~\bibnamefont {{Alberdi}}}, \bibinfo {author} {\bibfnamefont
  {W.}~\bibnamefont {{Alef}}}, \bibinfo {author} {\bibfnamefont
  {K.}~\bibnamefont {{Asada}}}, \bibinfo {author} {\bibfnamefont
  {R.}~\bibnamefont {{Azulay}}}, \bibinfo {author} {\bibfnamefont {A.-K.}\
  \bibnamefont {{Baczko}}}, \bibinfo {author} {\bibfnamefont {D.}~\bibnamefont
  {{Ball}}}, \bibinfo {author} {\bibfnamefont {M.}~\bibnamefont
  {{Balokovi{\'c}}}}, \bibinfo {author} {\bibfnamefont {J.}~\bibnamefont
  {{Barrett}}}, \emph {et~al.},\ }\href
  {https://doi.org/10.3847/2041-8213/ab0c57} {\bibfield  {journal} {\bibinfo
  {journal} {Astrophys. J. Lett.}\ }\textbf {\bibinfo {volume} {875}},\
  \bibinfo {eid} {L3} (\bibinfo {year} {2019}{\natexlab{c}})}\BibitemShut
  {NoStop}%
\bibitem [{\citenamefont {{Event Horizon Telescope Collaboration}}\ \emph
  {et~al.}(2019{\natexlab{d}})\citenamefont {{Event Horizon Telescope
  Collaboration}}, \citenamefont {{Akiyama}}, \citenamefont {{Alberdi}},
  \citenamefont {{Alef}}, \citenamefont {{Asada}}, \citenamefont {{Azulay}},
  \citenamefont {{Baczko}}, \citenamefont {{Ball}}, \citenamefont
  {{Balokovi{\'c}}}, \citenamefont {{Barrett}} \emph
  {et~al.}}]{EHT_M87_PaperIV}%
  \BibitemOpen
  \bibfield  {author} {\bibinfo {author} {\bibnamefont {{Event Horizon
  Telescope Collaboration}}}, \bibinfo {author} {\bibfnamefont
  {K.}~\bibnamefont {{Akiyama}}}, \bibinfo {author} {\bibfnamefont
  {A.}~\bibnamefont {{Alberdi}}}, \bibinfo {author} {\bibfnamefont
  {W.}~\bibnamefont {{Alef}}}, \bibinfo {author} {\bibfnamefont
  {K.}~\bibnamefont {{Asada}}}, \bibinfo {author} {\bibfnamefont
  {R.}~\bibnamefont {{Azulay}}}, \bibinfo {author} {\bibfnamefont {A.-K.}\
  \bibnamefont {{Baczko}}}, \bibinfo {author} {\bibfnamefont {D.}~\bibnamefont
  {{Ball}}}, \bibinfo {author} {\bibfnamefont {M.}~\bibnamefont
  {{Balokovi{\'c}}}}, \bibinfo {author} {\bibfnamefont {J.}~\bibnamefont
  {{Barrett}}}, \emph {et~al.},\ }\href
  {https://doi.org/10.3847/2041-8213/ab0e85} {\bibfield  {journal} {\bibinfo
  {journal} {Astrophys. J. Lett.}\ }\textbf {\bibinfo {volume} {875}},\
  \bibinfo {eid} {L4} (\bibinfo {year} {2019}{\natexlab{d}})}\BibitemShut
  {NoStop}%
\bibitem [{\citenamefont {{Event Horizon Telescope Collaboration}}\ \emph
  {et~al.}(2019{\natexlab{e}})\citenamefont {{Event Horizon Telescope
  Collaboration}}, \citenamefont {{Akiyama}}, \citenamefont {{Alberdi}},
  \citenamefont {{Alef}}, \citenamefont {{Asada}}, \citenamefont {{Azulay}},
  \citenamefont {{Baczko}}, \citenamefont {{Ball}}, \citenamefont
  {{Balokovi{\'c}}}, \citenamefont {{Barrett}} \emph
  {et~al.}}]{EHT_M87_PaperV}%
  \BibitemOpen
  \bibfield  {author} {\bibinfo {author} {\bibnamefont {{Event Horizon
  Telescope Collaboration}}}, \bibinfo {author} {\bibfnamefont
  {K.}~\bibnamefont {{Akiyama}}}, \bibinfo {author} {\bibfnamefont
  {A.}~\bibnamefont {{Alberdi}}}, \bibinfo {author} {\bibfnamefont
  {W.}~\bibnamefont {{Alef}}}, \bibinfo {author} {\bibfnamefont
  {K.}~\bibnamefont {{Asada}}}, \bibinfo {author} {\bibfnamefont
  {R.}~\bibnamefont {{Azulay}}}, \bibinfo {author} {\bibfnamefont {A.-K.}\
  \bibnamefont {{Baczko}}}, \bibinfo {author} {\bibfnamefont {D.}~\bibnamefont
  {{Ball}}}, \bibinfo {author} {\bibfnamefont {M.}~\bibnamefont
  {{Balokovi{\'c}}}}, \bibinfo {author} {\bibfnamefont {J.}~\bibnamefont
  {{Barrett}}}, \emph {et~al.},\ }\href
  {https://doi.org/10.3847/2041-8213/ab0f43} {\bibfield  {journal} {\bibinfo
  {journal} {Astrophys. J. Lett.}\ }\textbf {\bibinfo {volume} {875}},\
  \bibinfo {eid} {L5} (\bibinfo {year} {2019}{\natexlab{e}})}\BibitemShut
  {NoStop}%
\bibitem [{\citenamefont {{Event Horizon Telescope Collaboration}}\ \emph
  {et~al.}(2019{\natexlab{f}})\citenamefont {{Event Horizon Telescope
  Collaboration}}, \citenamefont {{Akiyama}}, \citenamefont {{Alberdi}},
  \citenamefont {{Alef}}, \citenamefont {{Asada}}, \citenamefont {{Azulay}},
  \citenamefont {{Baczko}}, \citenamefont {{Ball}}, \citenamefont
  {{Balokovi{\'c}}}, \citenamefont {{Barrett}} \emph
  {et~al.}}]{EHT_M87_PaperVI}%
  \BibitemOpen
  \bibfield  {author} {\bibinfo {author} {\bibnamefont {{Event Horizon
  Telescope Collaboration}}}, \bibinfo {author} {\bibfnamefont
  {K.}~\bibnamefont {{Akiyama}}}, \bibinfo {author} {\bibfnamefont
  {A.}~\bibnamefont {{Alberdi}}}, \bibinfo {author} {\bibfnamefont
  {W.}~\bibnamefont {{Alef}}}, \bibinfo {author} {\bibfnamefont
  {K.}~\bibnamefont {{Asada}}}, \bibinfo {author} {\bibfnamefont
  {R.}~\bibnamefont {{Azulay}}}, \bibinfo {author} {\bibfnamefont {A.-K.}\
  \bibnamefont {{Baczko}}}, \bibinfo {author} {\bibfnamefont {D.}~\bibnamefont
  {{Ball}}}, \bibinfo {author} {\bibfnamefont {M.}~\bibnamefont
  {{Balokovi{\'c}}}}, \bibinfo {author} {\bibfnamefont {J.}~\bibnamefont
  {{Barrett}}}, \emph {et~al.},\ }\href
  {https://doi.org/10.3847/2041-8213/ab1141} {\bibfield  {journal} {\bibinfo
  {journal} {Astrophys. J. Lett.}\ }\textbf {\bibinfo {volume} {875}},\
  \bibinfo {eid} {L6} (\bibinfo {year} {2019}{\natexlab{f}})}\BibitemShut
  {NoStop}%
\bibitem [{\citenamefont {{Gebhardt}}\ \emph {et~al.}(2011)\citenamefont
  {{Gebhardt}}, \citenamefont {{Adams}}, \citenamefont {{Richstone}},
  \citenamefont {{Lauer}}, \citenamefont {{Faber}}, \citenamefont
  {{G{\"u}ltekin}}, \citenamefont {{Murphy}},\ and\ \citenamefont
  {{Tremaine}}}]{Gebhardt11}%
  \BibitemOpen
  \bibfield  {author} {\bibinfo {author} {\bibfnamefont {K.}~\bibnamefont
  {{Gebhardt}}}, \bibinfo {author} {\bibfnamefont {J.}~\bibnamefont {{Adams}}},
  \bibinfo {author} {\bibfnamefont {D.}~\bibnamefont {{Richstone}}}, \bibinfo
  {author} {\bibfnamefont {T.~R.}\ \bibnamefont {{Lauer}}}, \bibinfo {author}
  {\bibfnamefont {S.~M.}\ \bibnamefont {{Faber}}}, \bibinfo {author}
  {\bibfnamefont {K.}~\bibnamefont {{G{\"u}ltekin}}}, \bibinfo {author}
  {\bibfnamefont {J.}~\bibnamefont {{Murphy}}},\ and\ \bibinfo {author}
  {\bibfnamefont {S.}~\bibnamefont {{Tremaine}}},\ }\href
  {https://doi.org/10.1088/0004-637X/729/2/119} {\bibfield  {journal} {\bibinfo
   {journal} {Astrophysical Journal}\ }\textbf {\bibinfo {volume} {729}},\
  \bibinfo {eid} {119} (\bibinfo {year} {2011})},\ \Eprint
  {https://arxiv.org/abs/1101.1954} {arXiv:1101.1954 [astro-ph.CO]}
  \BibitemShut {NoStop}%
\bibitem [{\citenamefont {{Psaltis}}\ \emph {et~al.}(2020)\citenamefont
  {{Psaltis}}, \citenamefont {{Medeiros}}, \citenamefont {{Christian}},
  \citenamefont {{Ozel}},\ and\ \citenamefont {{the EHT
  Collaboration}}}]{Psaltis2020_EHT}%
  \BibitemOpen
  \bibfield  {author} {\bibinfo {author} {\bibfnamefont {D.}~\bibnamefont
  {{Psaltis}}}, \bibinfo {author} {\bibfnamefont {L.}~\bibnamefont
  {{Medeiros}}}, \bibinfo {author} {\bibfnamefont {P.}~\bibnamefont
  {{Christian}}}, \bibinfo {author} {\bibfnamefont {F.}~\bibnamefont
  {{Ozel}}},\ and\ \bibinfo {author} {\bibnamefont {{the EHT Collaboration}}},\
  }\href {https://doi.org/10.1103/PhysRevLett.125.141104} {\bibfield  {journal}
  {\bibinfo  {journal} {Phys. Rev. Lett.}\ }\textbf {\bibinfo {volume} {125}},\
  \bibinfo {eid} {141104} (\bibinfo {year} {2020})},\ \Eprint
  {https://arxiv.org/abs/2010.01055} {arXiv:2010.01055 [gr-qc]} \BibitemShut
  {NoStop}%
\bibitem [{\citenamefont {{Johannsen}}\ and\ \citenamefont
  {{Psaltis}}(2011)}]{Johannsen2011}%
  \BibitemOpen
  \bibfield  {author} {\bibinfo {author} {\bibfnamefont {T.}~\bibnamefont
  {{Johannsen}}}\ and\ \bibinfo {author} {\bibfnamefont {D.}~\bibnamefont
  {{Psaltis}}},\ }\href {https://doi.org/10.1103/PhysRevD.83.124015} {\bibfield
   {journal} {\bibinfo  {journal} {Phys. Rev. D}\ }\textbf {\bibinfo {volume}
  {83}},\ \bibinfo {eid} {124015} (\bibinfo {year} {2011})},\ \Eprint
  {https://arxiv.org/abs/1105.3191} {arXiv:1105.3191 [gr-qc]} \BibitemShut
  {NoStop}%
\bibitem [{\citenamefont {{Johannsen}}(2013{\natexlab{a}})}]{Johannsen2013PRD}%
  \BibitemOpen
  \bibfield  {author} {\bibinfo {author} {\bibfnamefont {T.}~\bibnamefont
  {{Johannsen}}},\ }\href {https://doi.org/10.1103/PhysRevD.87.124017}
  {\bibfield  {journal} {\bibinfo  {journal} {Phys. Rev. D}\ }\textbf {\bibinfo
  {volume} {87}},\ \bibinfo {eid} {124017} (\bibinfo {year}
  {2013}{\natexlab{a}})},\ \Eprint {https://arxiv.org/abs/1304.7786}
  {arXiv:1304.7786 [gr-qc]} \BibitemShut {NoStop}%
\bibitem [{\citenamefont {{Vigeland}}\ \emph {et~al.}(2011)\citenamefont
  {{Vigeland}}, \citenamefont {{Yunes}},\ and\ \citenamefont
  {{Stein}}}]{Vigeland2011}%
  \BibitemOpen
  \bibfield  {author} {\bibinfo {author} {\bibfnamefont {S.}~\bibnamefont
  {{Vigeland}}}, \bibinfo {author} {\bibfnamefont {N.}~\bibnamefont
  {{Yunes}}},\ and\ \bibinfo {author} {\bibfnamefont {L.~C.}\ \bibnamefont
  {{Stein}}},\ }\href {https://doi.org/10.1103/PhysRevD.83.104027} {\bibfield
  {journal} {\bibinfo  {journal} {Phys. Rev. D}\ }\textbf {\bibinfo {volume}
  {83}},\ \bibinfo {eid} {104027} (\bibinfo {year} {2011})},\ \Eprint
  {https://arxiv.org/abs/1102.3706} {arXiv:1102.3706 [gr-qc]} \BibitemShut
  {NoStop}%
\bibitem [{\citenamefont
  {{Johannsen}}(2013{\natexlab{b}})}]{Johannsen2013PRDb}%
  \BibitemOpen
  \bibfield  {author} {\bibinfo {author} {\bibfnamefont {T.}~\bibnamefont
  {{Johannsen}}},\ }\href {https://doi.org/10.1103/PhysRevD.88.044002}
  {\bibfield  {journal} {\bibinfo  {journal} {Phys. Rev. D}\ }\textbf {\bibinfo
  {volume} {88}},\ \bibinfo {eid} {044002} (\bibinfo {year}
  {2013}{\natexlab{b}})},\ \Eprint {https://arxiv.org/abs/1501.02809}
  {arXiv:1501.02809 [gr-qc]} \BibitemShut {NoStop}%
\bibitem [{\citenamefont {{Rezzolla}}\ and\ \citenamefont
  {{Zhidenko}}(2014)}]{Rezzolla2014}%
  \BibitemOpen
  \bibfield  {author} {\bibinfo {author} {\bibfnamefont {L.}~\bibnamefont
  {{Rezzolla}}}\ and\ \bibinfo {author} {\bibfnamefont {A.}~\bibnamefont
  {{Zhidenko}}},\ }\href {https://doi.org/10.1103/PhysRevD.90.084009}
  {\bibfield  {journal} {\bibinfo  {journal} {Phys. Rev. D}\ }\textbf {\bibinfo
  {volume} {90}},\ \bibinfo {eid} {084009} (\bibinfo {year} {2014})},\ \Eprint
  {https://arxiv.org/abs/1407.3086} {arXiv:1407.3086 [gr-qc]} \BibitemShut
  {NoStop}%
\bibitem [{\citenamefont {{Younsi}}\ \emph {et~al.}(2016)\citenamefont
  {{Younsi}}, \citenamefont {{Zhidenko}}, \citenamefont {{Rezzolla}},
  \citenamefont {{Konoplya}},\ and\ \citenamefont {{Mizuno}}}]{Younsi2016}%
  \BibitemOpen
  \bibfield  {author} {\bibinfo {author} {\bibfnamefont {Z.}~\bibnamefont
  {{Younsi}}}, \bibinfo {author} {\bibfnamefont {A.}~\bibnamefont
  {{Zhidenko}}}, \bibinfo {author} {\bibfnamefont {L.}~\bibnamefont
  {{Rezzolla}}}, \bibinfo {author} {\bibfnamefont {R.}~\bibnamefont
  {{Konoplya}}},\ and\ \bibinfo {author} {\bibfnamefont {Y.}~\bibnamefont
  {{Mizuno}}},\ }\href {https://doi.org/10.1103/PhysRevD.94.084025} {\bibfield
  {journal} {\bibinfo  {journal} {Phys. Rev. D}\ }\textbf {\bibinfo {volume}
  {94}},\ \bibinfo {eid} {084025} (\bibinfo {year} {2016})},\ \Eprint
  {https://arxiv.org/abs/1607.05767} {arXiv:1607.05767 [gr-qc]} \BibitemShut
  {NoStop}%
\bibitem [{\citenamefont {{Konoplya}}\ \emph {et~al.}(2016)\citenamefont
  {{Konoplya}}, \citenamefont {{Rezzolla}},\ and\ \citenamefont
  {{Zhidenko}}}]{Konoplya2016a}%
  \BibitemOpen
  \bibfield  {author} {\bibinfo {author} {\bibfnamefont {R.}~\bibnamefont
  {{Konoplya}}}, \bibinfo {author} {\bibfnamefont {L.}~\bibnamefont
  {{Rezzolla}}},\ and\ \bibinfo {author} {\bibfnamefont {A.}~\bibnamefont
  {{Zhidenko}}},\ }\href {https://doi.org/10.1103/PhysRevD.93.064015}
  {\bibfield  {journal} {\bibinfo  {journal} {Phys. Rev. D}\ }\textbf {\bibinfo
  {volume} {93}},\ \bibinfo {eid} {064015} (\bibinfo {year} {2016})},\ \Eprint
  {https://arxiv.org/abs/1602.02378} {arXiv:1602.02378 [gr-qc]} \BibitemShut
  {NoStop}%
\bibitem [{\citenamefont {{Kocherlakota}}\ and\ \citenamefont
  {{Rezzolla}}(2020)}]{Kocherlakota2020}%
  \BibitemOpen
  \bibfield  {author} {\bibinfo {author} {\bibfnamefont {P.}~\bibnamefont
  {{Kocherlakota}}}\ and\ \bibinfo {author} {\bibfnamefont {L.}~\bibnamefont
  {{Rezzolla}}},\ }\href {https://doi.org/10.1103/PhysRevD.102.064058}
  {\bibfield  {journal} {\bibinfo  {journal} {Phys. Rev. D}\ }\textbf {\bibinfo
  {volume} {102}},\ \bibinfo {eid} {064058} (\bibinfo {year} {2020})},\ \Eprint
  {https://arxiv.org/abs/2007.15593} {arXiv:2007.15593 [gr-qc]} \BibitemShut
  {NoStop}%
\bibitem [{\citenamefont {{Bambi}}\ and\ \citenamefont
  {{Freese}}(2009)}]{Bambi2009}%
  \BibitemOpen
  \bibfield  {author} {\bibinfo {author} {\bibfnamefont {C.}~\bibnamefont
  {{Bambi}}}\ and\ \bibinfo {author} {\bibfnamefont {K.}~\bibnamefont
  {{Freese}}},\ }\href {https://doi.org/10.1103/PhysRevD.79.043002} {\bibfield
  {journal} {\bibinfo  {journal} {Phys. Rev. D}\ }\textbf {\bibinfo {volume}
  {79}},\ \bibinfo {eid} {043002} (\bibinfo {year} {2009})},\ \Eprint
  {https://arxiv.org/abs/0812.1328} {arXiv:0812.1328 [astro-ph]} \BibitemShut
  {NoStop}%
\bibitem [{\citenamefont {{Bambi}}\ and\ \citenamefont
  {{Yoshida}}(2010)}]{Bambi10}%
  \BibitemOpen
  \bibfield  {author} {\bibinfo {author} {\bibfnamefont {C.}~\bibnamefont
  {{Bambi}}}\ and\ \bibinfo {author} {\bibfnamefont {N.}~\bibnamefont
  {{Yoshida}}},\ }\href {https://doi.org/10.1088/0264-9381/27/20/205006}
  {\bibfield  {journal} {\bibinfo  {journal} {Class. Quantum Grav.}\ }\textbf
  {\bibinfo {volume} {27}},\ \bibinfo {eid} {205006} (\bibinfo {year}
  {2010})},\ \Eprint {https://arxiv.org/abs/1004.3149} {arXiv:1004.3149
  [gr-qc]} \BibitemShut {NoStop}%
\bibitem [{\citenamefont {{Amarilla}}\ \emph {et~al.}(2010)\citenamefont
  {{Amarilla}}, \citenamefont {{Eiroa}},\ and\ \citenamefont
  {{Giribet}}}]{Amarilla10}%
  \BibitemOpen
  \bibfield  {author} {\bibinfo {author} {\bibfnamefont {L.}~\bibnamefont
  {{Amarilla}}}, \bibinfo {author} {\bibfnamefont {E.~F.}\ \bibnamefont
  {{Eiroa}}},\ and\ \bibinfo {author} {\bibfnamefont {G.}~\bibnamefont
  {{Giribet}}},\ }\href {https://doi.org/10.1103/PhysRevD.81.124045} {\bibfield
   {journal} {\bibinfo  {journal} {Phys. Rev. D}\ }\textbf {\bibinfo {volume}
  {81}},\ \bibinfo {eid} {124045} (\bibinfo {year} {2010})},\ \Eprint
  {https://arxiv.org/abs/1005.0607} {arXiv:1005.0607 [gr-qc]} \BibitemShut
  {NoStop}%
\bibitem [{\citenamefont {{Amarilla}}\ and\ \citenamefont
  {{Eiroa}}(2013)}]{Amarilla13}%
  \BibitemOpen
  \bibfield  {author} {\bibinfo {author} {\bibfnamefont {L.}~\bibnamefont
  {{Amarilla}}}\ and\ \bibinfo {author} {\bibfnamefont {E.~F.}\ \bibnamefont
  {{Eiroa}}},\ }\href {https://doi.org/10.1103/PhysRevD.87.044057} {\bibfield
  {journal} {\bibinfo  {journal} {Phys. Rev. D}\ }\textbf {\bibinfo {volume}
  {87}},\ \bibinfo {eid} {044057} (\bibinfo {year} {2013})},\ \Eprint
  {https://arxiv.org/abs/1301.0532} {arXiv:1301.0532 [gr-qc]} \BibitemShut
  {NoStop}%
\bibitem [{\citenamefont {{Wei}}\ and\ \citenamefont {{Liu}}(2013)}]{Wei2013}%
  \BibitemOpen
  \bibfield  {author} {\bibinfo {author} {\bibfnamefont {S.-W.}\ \bibnamefont
  {{Wei}}}\ and\ \bibinfo {author} {\bibfnamefont {Y.-X.}\ \bibnamefont
  {{Liu}}},\ }\href {https://doi.org/10.1088/1475-7516/2013/11/063} {\bibfield
  {journal} {\bibinfo  {journal} {JCAP}\ }\textbf {\bibinfo {volume}
  {2013}}\bibfield  {number} {\bibinfo  {number} { (11)},\ \bibinfo {eid}
  {063}},\ }\Eprint {https://arxiv.org/abs/1311.4251} {arXiv:1311.4251 [gr-qc]}
  \BibitemShut {NoStop}%
\bibitem [{\citenamefont {{Nedkova}}\ \emph {et~al.}(2013)\citenamefont
  {{Nedkova}}, \citenamefont {{Tinchev}},\ and\ \citenamefont
  {{Yazadjiev}}}]{Nedkova13}%
  \BibitemOpen
  \bibfield  {author} {\bibinfo {author} {\bibfnamefont {P.~G.}\ \bibnamefont
  {{Nedkova}}}, \bibinfo {author} {\bibfnamefont {V.~K.}\ \bibnamefont
  {{Tinchev}}},\ and\ \bibinfo {author} {\bibfnamefont {S.~S.}\ \bibnamefont
  {{Yazadjiev}}},\ }\href {https://doi.org/10.1103/PhysRevD.88.124019}
  {\bibfield  {journal} {\bibinfo  {journal} {Phys. Rev. D}\ }\textbf {\bibinfo
  {volume} {88}},\ \bibinfo {eid} {124019} (\bibinfo {year} {2013})},\ \Eprint
  {https://arxiv.org/abs/1307.7647} {arXiv:1307.7647 [gr-qc]} \BibitemShut
  {NoStop}%
\bibitem [{\citenamefont {{Papnoi}}\ \emph {et~al.}(2014)\citenamefont
  {{Papnoi}}, \citenamefont {{Atamurotov}}, \citenamefont {{Ghosh}},\ and\
  \citenamefont {{Ahmedov}}}]{Papnoi2014}%
  \BibitemOpen
  \bibfield  {author} {\bibinfo {author} {\bibfnamefont {U.}~\bibnamefont
  {{Papnoi}}}, \bibinfo {author} {\bibfnamefont {F.}~\bibnamefont
  {{Atamurotov}}}, \bibinfo {author} {\bibfnamefont {S.~G.}\ \bibnamefont
  {{Ghosh}}},\ and\ \bibinfo {author} {\bibfnamefont {B.}~\bibnamefont
  {{Ahmedov}}},\ }\href {https://doi.org/10.1103/PhysRevD.90.024073} {\bibfield
   {journal} {\bibinfo  {journal} {Phys. Rev. D}\ }\textbf {\bibinfo {volume}
  {90}},\ \bibinfo {eid} {024073} (\bibinfo {year} {2014})},\ \Eprint
  {https://arxiv.org/abs/1407.0834} {arXiv:1407.0834 [gr-qc]} \BibitemShut
  {NoStop}%
\bibitem [{\citenamefont {{Wei}}\ \emph {et~al.}(2015)\citenamefont {{Wei}},
  \citenamefont {{Cheng}}, \citenamefont {{Zhong}},\ and\ \citenamefont
  {{Zhou}}}]{Wei2015}%
  \BibitemOpen
  \bibfield  {author} {\bibinfo {author} {\bibfnamefont {S.-W.}\ \bibnamefont
  {{Wei}}}, \bibinfo {author} {\bibfnamefont {P.}~\bibnamefont {{Cheng}}},
  \bibinfo {author} {\bibfnamefont {Y.}~\bibnamefont {{Zhong}}},\ and\ \bibinfo
  {author} {\bibfnamefont {X.-N.}\ \bibnamefont {{Zhou}}},\ }\href
  {https://doi.org/10.1088/1475-7516/2015/08/004} {\bibfield  {journal}
  {\bibinfo  {journal} {JCAP}\ }\textbf {\bibinfo {volume} {2015}}\bibfield
  {number} {\bibinfo  {number} { (8)},\ \bibinfo {eid} {004}},\ }\Eprint
  {https://arxiv.org/abs/1501.06298} {arXiv:1501.06298 [gr-qc]} \BibitemShut
  {NoStop}%
\bibitem [{\citenamefont {{Ghasemi-Nodehi}}\ \emph {et~al.}(2015)\citenamefont
  {{Ghasemi-Nodehi}}, \citenamefont {{Li}},\ and\ \citenamefont
  {{Bambi}}}]{Ghasemi-Nodehi2015}%
  \BibitemOpen
  \bibfield  {author} {\bibinfo {author} {\bibfnamefont {M.}~\bibnamefont
  {{Ghasemi-Nodehi}}}, \bibinfo {author} {\bibfnamefont {Z.}~\bibnamefont
  {{Li}}},\ and\ \bibinfo {author} {\bibfnamefont {C.}~\bibnamefont
  {{Bambi}}},\ }\href {https://doi.org/10.1140/epjc/s10052-015-3539-x}
  {\bibfield  {journal} {\bibinfo  {journal} {European Physical Journal C}\
  }\textbf {\bibinfo {volume} {75}},\ \bibinfo {eid} {315} (\bibinfo {year}
  {2015})},\ \Eprint {https://arxiv.org/abs/1506.02627} {arXiv:1506.02627
  [gr-qc]} \BibitemShut {NoStop}%
\bibitem [{\citenamefont {{Atamurotov}}\ \emph {et~al.}(2016)\citenamefont
  {{Atamurotov}}, \citenamefont {{Ghosh}},\ and\ \citenamefont
  {{Ahmedov}}}]{Atamurotov2016}%
  \BibitemOpen
  \bibfield  {author} {\bibinfo {author} {\bibfnamefont {F.}~\bibnamefont
  {{Atamurotov}}}, \bibinfo {author} {\bibfnamefont {S.~G.}\ \bibnamefont
  {{Ghosh}}},\ and\ \bibinfo {author} {\bibfnamefont {B.}~\bibnamefont
  {{Ahmedov}}},\ }\href {https://doi.org/10.1140/epjc/s10052-016-4122-9}
  {\bibfield  {journal} {\bibinfo  {journal} {European Physical Journal C}\
  }\textbf {\bibinfo {volume} {76}},\ \bibinfo {eid} {273} (\bibinfo {year}
  {2016})},\ \Eprint {https://arxiv.org/abs/1506.03690} {arXiv:1506.03690
  [gr-qc]} \BibitemShut {NoStop}%
\bibitem [{\citenamefont {{Pratap Singh}}\ and\ \citenamefont
  {{Ghosh}}(2017)}]{Singh2017}%
  \BibitemOpen
  \bibfield  {author} {\bibinfo {author} {\bibfnamefont {B.}~\bibnamefont
  {{Pratap Singh}}}\ and\ \bibinfo {author} {\bibfnamefont {S.~G.}\
  \bibnamefont {{Ghosh}}},\ }\href@noop {} {\bibfield  {journal} {\bibinfo
  {journal} {arXiv e-prints}\ ,\ \bibinfo {eid} {arXiv:1707.07125}} (\bibinfo
  {year} {2017})},\ \Eprint {https://arxiv.org/abs/1707.07125}
  {arXiv:1707.07125 [gr-qc]} \BibitemShut {NoStop}%
\bibitem [{\citenamefont {{Amir}}\ \emph {et~al.}(2018)\citenamefont {{Amir}},
  \citenamefont {{Singh}},\ and\ \citenamefont {{Ghosh}}}]{Amir2018}%
  \BibitemOpen
  \bibfield  {author} {\bibinfo {author} {\bibfnamefont {M.}~\bibnamefont
  {{Amir}}}, \bibinfo {author} {\bibfnamefont {B.~P.}\ \bibnamefont
  {{Singh}}},\ and\ \bibinfo {author} {\bibfnamefont {S.~G.}\ \bibnamefont
  {{Ghosh}}},\ }\href {https://doi.org/10.1140/epjc/s10052-018-5872-3}
  {\bibfield  {journal} {\bibinfo  {journal} {European Physical Journal C}\
  }\textbf {\bibinfo {volume} {78}},\ \bibinfo {eid} {399} (\bibinfo {year}
  {2018})},\ \Eprint {https://arxiv.org/abs/1707.09521} {arXiv:1707.09521
  [gr-qc]} \BibitemShut {NoStop}%
\bibitem [{\citenamefont {{Olivares}}\ \emph {et~al.}(2020)\citenamefont
  {{Olivares}}, \citenamefont {{Younsi}}, \citenamefont {{Fromm}},
  \citenamefont {{De Laurentis}}, \citenamefont {{Porth}}, \citenamefont
  {{Mizuno}}, \citenamefont {{Falcke}}, \citenamefont {{Kramer}},\ and\
  \citenamefont {{Rezzolla}}}]{Olivares2020}%
  \BibitemOpen
  \bibfield  {author} {\bibinfo {author} {\bibfnamefont {H.}~\bibnamefont
  {{Olivares}}}, \bibinfo {author} {\bibfnamefont {Z.}~\bibnamefont
  {{Younsi}}}, \bibinfo {author} {\bibfnamefont {C.~M.}\ \bibnamefont
  {{Fromm}}}, \bibinfo {author} {\bibfnamefont {M.}~\bibnamefont {{De
  Laurentis}}}, \bibinfo {author} {\bibfnamefont {O.}~\bibnamefont {{Porth}}},
  \bibinfo {author} {\bibfnamefont {Y.}~\bibnamefont {{Mizuno}}}, \bibinfo
  {author} {\bibfnamefont {H.}~\bibnamefont {{Falcke}}}, \bibinfo {author}
  {\bibfnamefont {M.}~\bibnamefont {{Kramer}}},\ and\ \bibinfo {author}
  {\bibfnamefont {L.}~\bibnamefont {{Rezzolla}}},\ }\href
  {https://doi.org/10.1093/mnras/staa1878} {\bibfield  {journal} {\bibinfo
  {journal} {MNRAS}\ }\textbf {\bibinfo {volume} {497}},\ \bibinfo {pages}
  {521} (\bibinfo {year} {2020})},\ \Eprint {https://arxiv.org/abs/1809.08682}
  {arXiv:1809.08682 [gr-qc]} \BibitemShut {NoStop}%
\bibitem [{\citenamefont {{Mizuno}}\ \emph {et~al.}(2018)\citenamefont
  {{Mizuno}}, \citenamefont {{Younsi}}, \citenamefont {{Fromm}}, \citenamefont
  {{Porth}}, \citenamefont {{De Laurentis}}, \citenamefont {{Olivares}},
  \citenamefont {{Falcke}}, \citenamefont {{Kramer}},\ and\ \citenamefont
  {{Rezzolla}}}]{Mizuno2018b}%
  \BibitemOpen
  \bibfield  {author} {\bibinfo {author} {\bibfnamefont {Y.}~\bibnamefont
  {{Mizuno}}}, \bibinfo {author} {\bibfnamefont {Z.}~\bibnamefont {{Younsi}}},
  \bibinfo {author} {\bibfnamefont {C.~M.}\ \bibnamefont {{Fromm}}}, \bibinfo
  {author} {\bibfnamefont {O.}~\bibnamefont {{Porth}}}, \bibinfo {author}
  {\bibfnamefont {M.}~\bibnamefont {{De Laurentis}}}, \bibinfo {author}
  {\bibfnamefont {H.}~\bibnamefont {{Olivares}}}, \bibinfo {author}
  {\bibfnamefont {H.}~\bibnamefont {{Falcke}}}, \bibinfo {author}
  {\bibfnamefont {M.}~\bibnamefont {{Kramer}}},\ and\ \bibinfo {author}
  {\bibfnamefont {L.}~\bibnamefont {{Rezzolla}}},\ }\href
  {https://doi.org/10.1038/s41550-018-0449-5} {\bibfield  {journal} {\bibinfo
  {journal} {Nature Astronomy}\ }\textbf {\bibinfo {volume} {2}},\ \bibinfo
  {eid} {585} (\bibinfo {year} {2018})},\ \Eprint
  {https://arxiv.org/abs/1804.05812} {arXiv:1804.05812 [astro-ph.GA]}
  \BibitemShut {NoStop}%
\bibitem [{\citenamefont {{Cunha}}\ \emph {et~al.}(2019)\citenamefont
  {{Cunha}}, \citenamefont {{Herdeiro}},\ and\ \citenamefont
  {{Radu}}}]{Cunha2019}%
  \BibitemOpen
  \bibfield  {author} {\bibinfo {author} {\bibfnamefont {P.~V.~P.}\
  \bibnamefont {{Cunha}}}, \bibinfo {author} {\bibfnamefont {C.~A.~R.}\
  \bibnamefont {{Herdeiro}}},\ and\ \bibinfo {author} {\bibfnamefont
  {E.}~\bibnamefont {{Radu}}},\ }\href
  {https://doi.org/10.1103/PhysRevLett.123.011101} {\bibfield  {journal}
  {\bibinfo  {journal} {Phys. Rev. Lett.}\ }\textbf {\bibinfo {volume} {123}},\
  \bibinfo {eid} {011101} (\bibinfo {year} {2019})},\ \Eprint
  {https://arxiv.org/abs/1904.09997} {arXiv:1904.09997 [gr-qc]} \BibitemShut
  {NoStop}%
\bibitem [{\citenamefont {{Grenzebach}}(2016)}]{Grenzebach2016}%
  \BibitemOpen
  \bibfield  {author} {\bibinfo {author} {\bibfnamefont {A.}~\bibnamefont
  {{Grenzebach}}},\ }\href {https://doi.org/10.1007/978-3-319-30066-5} {\emph
  {\bibinfo {title} {{The Shadow of Black Holes}}}}\ (\bibinfo  {publisher}
  {Springer International Publishing},\ \bibinfo {year} {2016})\BibitemShut
  {NoStop}%
\bibitem [{\citenamefont {{Stuchl{\'\i}k}}\ and\ \citenamefont
  {{Schee}}(2019)}]{Stuchlik2019}%
  \BibitemOpen
  \bibfield  {author} {\bibinfo {author} {\bibfnamefont {Z.}~\bibnamefont
  {{Stuchl{\'\i}k}}}\ and\ \bibinfo {author} {\bibfnamefont {J.}~\bibnamefont
  {{Schee}}},\ }\href {https://doi.org/10.1140/epjc/s10052-019-6543-8}
  {\bibfield  {journal} {\bibinfo  {journal} {European Physical Journal C}\
  }\textbf {\bibinfo {volume} {79}},\ \bibinfo {eid} {44} (\bibinfo {year}
  {2019})}\BibitemShut {NoStop}%
\bibitem [{\citenamefont {{Kumar}}\ \emph {et~al.}(2019)\citenamefont
  {{Kumar}}, \citenamefont {{Ghosh}},\ and\ \citenamefont
  {{Wang}}}]{Kumar2019b}%
  \BibitemOpen
  \bibfield  {author} {\bibinfo {author} {\bibfnamefont {R.}~\bibnamefont
  {{Kumar}}}, \bibinfo {author} {\bibfnamefont {S.~G.}\ \bibnamefont
  {{Ghosh}}},\ and\ \bibinfo {author} {\bibfnamefont {A.}~\bibnamefont
  {{Wang}}},\ }\href {https://doi.org/10.1103/PhysRevD.100.124024} {\bibfield
  {journal} {\bibinfo  {journal} {Phys. Rev. D}\ }\textbf {\bibinfo {volume}
  {100}},\ \bibinfo {eid} {124024} (\bibinfo {year} {2019})},\ \Eprint
  {https://arxiv.org/abs/1912.05154} {arXiv:1912.05154 [gr-qc]} \BibitemShut
  {NoStop}%
\bibitem [{\citenamefont {{Kumar}}\ \emph {et~al.}(2020)\citenamefont
  {{Kumar}}, \citenamefont {{Kumar}},\ and\ \citenamefont
  {{Ghosh}}}]{Kumar2020}%
  \BibitemOpen
  \bibfield  {author} {\bibinfo {author} {\bibfnamefont {R.}~\bibnamefont
  {{Kumar}}}, \bibinfo {author} {\bibfnamefont {A.}~\bibnamefont {{Kumar}}},\
  and\ \bibinfo {author} {\bibfnamefont {S.~G.}\ \bibnamefont {{Ghosh}}},\
  }\href {https://doi.org/10.3847/1538-4357/ab8c4a} {\bibfield  {journal}
  {\bibinfo  {journal} {Astrophys. J.}\ }\textbf {\bibinfo {volume} {896}},\
  \bibinfo {eid} {89} (\bibinfo {year} {2020})},\ \Eprint
  {https://arxiv.org/abs/2006.09869} {arXiv:2006.09869 [gr-qc]} \BibitemShut
  {NoStop}%
\bibitem [{\citenamefont {{Hioki}}\ and\ \citenamefont
  {{Miyamoto}}(2008)}]{Hioki2008}%
  \BibitemOpen
  \bibfield  {author} {\bibinfo {author} {\bibfnamefont {K.}~\bibnamefont
  {{Hioki}}}\ and\ \bibinfo {author} {\bibfnamefont {U.}~\bibnamefont
  {{Miyamoto}}},\ }\href {https://doi.org/10.1103/PhysRevD.78.044007}
  {\bibfield  {journal} {\bibinfo  {journal} {Phys. Rev. D}\ }\textbf {\bibinfo
  {volume} {78}},\ \bibinfo {eid} {044007} (\bibinfo {year} {2008})},\ \Eprint
  {https://arxiv.org/abs/0805.3146} {arXiv:0805.3146 [gr-qc]} \BibitemShut
  {NoStop}%
\bibitem [{\citenamefont {{Abdujabbarov}}\ \emph {et~al.}(2016)\citenamefont
  {{Abdujabbarov}}, \citenamefont {{Amir}}, \citenamefont {{Ahmedov}},\ and\
  \citenamefont {{Ghosh}}}]{Abdujabbarov2016}%
  \BibitemOpen
  \bibfield  {author} {\bibinfo {author} {\bibfnamefont {A.}~\bibnamefont
  {{Abdujabbarov}}}, \bibinfo {author} {\bibfnamefont {M.}~\bibnamefont
  {{Amir}}}, \bibinfo {author} {\bibfnamefont {B.}~\bibnamefont {{Ahmedov}}},\
  and\ \bibinfo {author} {\bibfnamefont {S.~G.}\ \bibnamefont {{Ghosh}}},\
  }\href {https://doi.org/10.1103/PhysRevD.93.104004} {\bibfield  {journal}
  {\bibinfo  {journal} {Phys. Rev. D}\ }\textbf {\bibinfo {volume} {93}},\
  \bibinfo {eid} {104004} (\bibinfo {year} {2016})},\ \Eprint
  {https://arxiv.org/abs/1604.03809} {arXiv:1604.03809 [gr-qc]} \BibitemShut
  {NoStop}%
\bibitem [{\citenamefont {{Mazur}}\ and\ \citenamefont
  {{Mottola}}(2004)}]{Mazur2004}%
  \BibitemOpen
  \bibfield  {author} {\bibinfo {author} {\bibfnamefont {P.~O.}\ \bibnamefont
  {{Mazur}}}\ and\ \bibinfo {author} {\bibfnamefont {E.}~\bibnamefont
  {{Mottola}}},\ }\href {https://doi.org/10.1073/pnas.0402717101} {\bibfield
  {journal} {\bibinfo  {journal} {Proceedings of the National Academy of
  Science}\ }\textbf {\bibinfo {volume} {101}},\ \bibinfo {pages} {9545}
  (\bibinfo {year} {2004})},\ \Eprint {https://arxiv.org/abs/gr-qc/0407075}
  {gr-qc/0407075} \BibitemShut {NoStop}%
\bibitem [{\citenamefont {{Chirenti}}\ and\ \citenamefont
  {{Rezzolla}}(2008)}]{Chirenti2008}%
  \BibitemOpen
  \bibfield  {author} {\bibinfo {author} {\bibfnamefont {C.~B.~M.~H.}\
  \bibnamefont {{Chirenti}}}\ and\ \bibinfo {author} {\bibfnamefont
  {L.}~\bibnamefont {{Rezzolla}}},\ }\href
  {https://doi.org/10.1103/PhysRevD.78.084011} {\bibfield  {journal} {\bibinfo
  {journal} {Phys. Rev. D}\ }\textbf {\bibinfo {volume} {78}},\ \bibinfo {eid}
  {084011} (\bibinfo {year} {2008})},\ \Eprint
  {https://arxiv.org/abs/0808.4080} {arXiv:0808.4080 [gr-qc]} \BibitemShut
  {NoStop}%
\bibitem [{\citenamefont {{Shaikh}}\ \emph {et~al.}(2019)\citenamefont
  {{Shaikh}}, \citenamefont {{Kocherlakota}}, \citenamefont {{Narayan}},\ and\
  \citenamefont {{Joshi}}}]{Shaikh2019}%
  \BibitemOpen
  \bibfield  {author} {\bibinfo {author} {\bibfnamefont {R.}~\bibnamefont
  {{Shaikh}}}, \bibinfo {author} {\bibfnamefont {P.}~\bibnamefont
  {{Kocherlakota}}}, \bibinfo {author} {\bibfnamefont {R.}~\bibnamefont
  {{Narayan}}},\ and\ \bibinfo {author} {\bibfnamefont {P.~S.}\ \bibnamefont
  {{Joshi}}},\ }\href {https://doi.org/10.1093/mnras/sty2624} {\bibfield
  {journal} {\bibinfo  {journal} {Mon. Not. R. Astron. Soc.}\ }\textbf
  {\bibinfo {volume} {482}},\ \bibinfo {pages} {52} (\bibinfo {year} {2019})},\
  \Eprint {https://arxiv.org/abs/1802.08060} {arXiv:1802.08060 [astro-ph.HE]}
  \BibitemShut {NoStop}%
\bibitem [{\citenamefont {{Dey}}\ \emph {et~al.}(2020)\citenamefont {{Dey}},
  \citenamefont {{Shaikh}},\ and\ \citenamefont {{Joshi}}}]{Dey2020}%
  \BibitemOpen
  \bibfield  {author} {\bibinfo {author} {\bibfnamefont {D.}~\bibnamefont
  {{Dey}}}, \bibinfo {author} {\bibfnamefont {R.}~\bibnamefont {{Shaikh}}},\
  and\ \bibinfo {author} {\bibfnamefont {P.~S.}\ \bibnamefont {{Joshi}}},\
  }\href@noop {} {\bibfield  {journal} {\bibinfo  {journal} {arXiv e-prints}\
  ,\ \bibinfo {eid} {arXiv:2009.07487}} (\bibinfo {year} {2020})},\ \Eprint
  {https://arxiv.org/abs/2009.07487} {arXiv:2009.07487 [gr-qc]} \BibitemShut
  {NoStop}%
\bibitem [{\citenamefont {Wald}(1984)}]{Wald84Book}%
  \BibitemOpen
  \bibfield  {author} {\bibinfo {author} {\bibfnamefont {R.~M.}\ \bibnamefont
  {Wald}},\ }\href@noop {} {\emph {\bibinfo {title} {General Relativity}}}\
  (\bibinfo  {publisher} {The University of Chicago Press},\ \bibinfo {address}
  {Chicago},\ \bibinfo {year} {1984})\BibitemShut {NoStop}%
\bibitem [{\citenamefont {Bardeen}(1968)}]{Bardeen68}%
  \BibitemOpen
  \bibfield  {author} {\bibinfo {author} {\bibfnamefont {J.}~\bibnamefont
  {Bardeen}},\ }in\ \href@noop {} {\emph {\bibinfo {booktitle} {Proceedings of
  GR5}}}\ (\bibinfo {organization} {Tbilisi, USSR},\ \bibinfo {year} {1968})\
  p.\ \bibinfo {pages} {174}\BibitemShut {NoStop}%
\bibitem [{\citenamefont {{Hayward}}(2006)}]{Hayward2006PRL}%
  \BibitemOpen
  \bibfield  {author} {\bibinfo {author} {\bibfnamefont {S.~A.}\ \bibnamefont
  {{Hayward}}},\ }\href {https://doi.org/10.1103/PhysRevLett.96.031103}
  {\bibfield  {journal} {\bibinfo  {journal} {Phys. Rev. Lett.}\ }\textbf
  {\bibinfo {volume} {96}},\ \bibinfo {eid} {031103} (\bibinfo {year}
  {2006})},\ \Eprint {https://arxiv.org/abs/gr-qc/0506126} {arXiv:gr-qc/0506126
  [gr-qc]} \BibitemShut {NoStop}%
\bibitem [{\citenamefont {{Frolov}}(2016)}]{Frolov2016}%
  \BibitemOpen
  \bibfield  {author} {\bibinfo {author} {\bibfnamefont {V.~P.}\ \bibnamefont
  {{Frolov}}},\ }\href {https://doi.org/10.1103/PhysRevD.94.104056} {\bibfield
  {journal} {\bibinfo  {journal} {Phys. Rev. D}\ }\textbf {\bibinfo {volume}
  {94}},\ \bibinfo {eid} {104056} (\bibinfo {year} {2016})},\ \Eprint
  {https://arxiv.org/abs/1609.01758} {arXiv:1609.01758 [gr-qc]} \BibitemShut
  {NoStop}%
\bibitem [{\citenamefont {{Kazakov}}\ and\ \citenamefont
  {{Solodukhin}}(1994)}]{Kazakov1994}%
  \BibitemOpen
  \bibfield  {author} {\bibinfo {author} {\bibfnamefont {D.~I.}\ \bibnamefont
  {{Kazakov}}}\ and\ \bibinfo {author} {\bibfnamefont {S.~N.}\ \bibnamefont
  {{Solodukhin}}},\ }\href {https://doi.org/10.1016/S0550-3213(94)80045-6}
  {\bibfield  {journal} {\bibinfo  {journal} {Nuclear Physics B}\ }\textbf
  {\bibinfo {volume} {429}},\ \bibinfo {pages} {153} (\bibinfo {year}
  {1994})},\ \Eprint {https://arxiv.org/abs/hep-th/9310150}
  {arXiv:hep-th/9310150 [hep-th]} \BibitemShut {NoStop}%
\bibitem [{\citenamefont {{Gibbons}}\ and\ \citenamefont
  {{Maeda}}(1988)}]{Gibbons1988}%
  \BibitemOpen
  \bibfield  {author} {\bibinfo {author} {\bibfnamefont {G.~W.}\ \bibnamefont
  {{Gibbons}}}\ and\ \bibinfo {author} {\bibfnamefont {K.-I.}\ \bibnamefont
  {{Maeda}}},\ }\href {https://doi.org/10.1016/0550-3213(88)90006-5} {\bibfield
   {journal} {\bibinfo  {journal} {Nuclear Physics B}\ }\textbf {\bibinfo
  {volume} {298}},\ \bibinfo {pages} {741} (\bibinfo {year}
  {1988})}\BibitemShut {NoStop}%
\bibitem [{\citenamefont {{Garfinkle}}\ \emph {et~al.}(1991)\citenamefont
  {{Garfinkle}}, \citenamefont {{Horowitz}},\ and\ \citenamefont
  {{Strominger}}}]{Garfinkle1991}%
  \BibitemOpen
  \bibfield  {author} {\bibinfo {author} {\bibfnamefont {D.}~\bibnamefont
  {{Garfinkle}}}, \bibinfo {author} {\bibfnamefont {G.~T.}\ \bibnamefont
  {{Horowitz}}},\ and\ \bibinfo {author} {\bibfnamefont {A.}~\bibnamefont
  {{Strominger}}},\ }\href {https://doi.org/10.1103/PhysRevD.43.3140}
  {\bibfield  {journal} {\bibinfo  {journal} {Phys. Rev. D}\ }\textbf {\bibinfo
  {volume} {43}},\ \bibinfo {pages} {3140} (\bibinfo {year}
  {1991})}\BibitemShut {NoStop}%
\bibitem [{\citenamefont {{Garc{\'{\i}}a}}\ \emph {et~al.}(1995)\citenamefont
  {{Garc{\'{\i}}a}}, \citenamefont {{Galtsov}},\ and\ \citenamefont
  {{Kechkin}}}]{Garcia1995}%
  \BibitemOpen
  \bibfield  {author} {\bibinfo {author} {\bibfnamefont {A.}~\bibnamefont
  {{Garc{\'{\i}}a}}}, \bibinfo {author} {\bibfnamefont {D.}~\bibnamefont
  {{Galtsov}}},\ and\ \bibinfo {author} {\bibfnamefont {O.}~\bibnamefont
  {{Kechkin}}},\ }\href {https://doi.org/10.1103/PhysRevLett.74.1276}
  {\bibfield  {journal} {\bibinfo  {journal} {Phys. Rev. Lett.}\ }\textbf
  {\bibinfo {volume} {74}},\ \bibinfo {pages} {1276} (\bibinfo {year}
  {1995})}\BibitemShut {NoStop}%
\bibitem [{\citenamefont {{Kallosh}}\ \emph {et~al.}(1992)\citenamefont
  {{Kallosh}}, \citenamefont {{Linde}}, \citenamefont {{Ort{\'\i}n}},
  \citenamefont {{Peet}},\ and\ \citenamefont {{van Proeyen}}}]{Kallosh1992}%
  \BibitemOpen
  \bibfield  {author} {\bibinfo {author} {\bibfnamefont {R.}~\bibnamefont
  {{Kallosh}}}, \bibinfo {author} {\bibfnamefont {A.}~\bibnamefont {{Linde}}},
  \bibinfo {author} {\bibfnamefont {T.}~\bibnamefont {{Ort{\'\i}n}}}, \bibinfo
  {author} {\bibfnamefont {A.}~\bibnamefont {{Peet}}},\ and\ \bibinfo {author}
  {\bibfnamefont {A.}~\bibnamefont {{van Proeyen}}},\ }\href
  {https://doi.org/10.1103/PhysRevD.46.5278} {\bibfield  {journal} {\bibinfo
  {journal} {Phys. Rev. D}\ }\textbf {\bibinfo {volume} {46}},\ \bibinfo
  {pages} {5278} (\bibinfo {year} {1992})},\ \Eprint
  {https://arxiv.org/abs/hep-th/9205027} {arXiv:hep-th/9205027 [hep-th]}
  \BibitemShut {NoStop}%
\bibitem [{\citenamefont {{Janis}}\ \emph {et~al.}(1968)\citenamefont
  {{Janis}}, \citenamefont {{Newman}},\ and\ \citenamefont
  {{Winicour}}}]{Janis1968}%
  \BibitemOpen
  \bibfield  {author} {\bibinfo {author} {\bibfnamefont {A.~I.}\ \bibnamefont
  {{Janis}}}, \bibinfo {author} {\bibfnamefont {E.~T.}\ \bibnamefont
  {{Newman}}},\ and\ \bibinfo {author} {\bibfnamefont {J.}~\bibnamefont
  {{Winicour}}},\ }\href {https://doi.org/10.1103/PhysRevLett.20.878}
  {\bibfield  {journal} {\bibinfo  {journal} {Phys. Rev. Lett.}\ }\textbf
  {\bibinfo {volume} {20}},\ \bibinfo {pages} {878} (\bibinfo {year}
  {1968})}\BibitemShut {NoStop}%
\bibitem [{\citenamefont {{Kerr}}(1963)}]{Kerr1963}%
  \BibitemOpen
  \bibfield  {author} {\bibinfo {author} {\bibfnamefont {R.~P.}\ \bibnamefont
  {{Kerr}}},\ }\href {https://doi.org/10.1103/PhysRevLett.11.237} {\bibfield
  {journal} {\bibinfo  {journal} {Phys. Rev. Lett.}\ }\textbf {\bibinfo
  {volume} {11}},\ \bibinfo {pages} {237} (\bibinfo {year} {1963})}\BibitemShut
  {NoStop}%
\bibitem [{\citenamefont {{Newman}}\ \emph {et~al.}(1965)\citenamefont
  {{Newman}}, \citenamefont {{Couch}}, \citenamefont {{Chinnapared}},
  \citenamefont {{Exton}}, \citenamefont {{Prakash}},\ and\ \citenamefont
  {{Torrence}}}]{Newman1965}%
  \BibitemOpen
  \bibfield  {author} {\bibinfo {author} {\bibfnamefont {E.~T.}\ \bibnamefont
  {{Newman}}}, \bibinfo {author} {\bibfnamefont {E.}~\bibnamefont {{Couch}}},
  \bibinfo {author} {\bibfnamefont {K.}~\bibnamefont {{Chinnapared}}}, \bibinfo
  {author} {\bibfnamefont {A.}~\bibnamefont {{Exton}}}, \bibinfo {author}
  {\bibfnamefont {A.}~\bibnamefont {{Prakash}}},\ and\ \bibinfo {author}
  {\bibfnamefont {R.}~\bibnamefont {{Torrence}}},\ }\href
  {https://doi.org/10.1063/1.1704351} {\bibfield  {journal} {\bibinfo
  {journal} {Journal of Mathematical Physics}\ }\textbf {\bibinfo {volume}
  {6}},\ \bibinfo {pages} {918} (\bibinfo {year} {1965})}\BibitemShut {NoStop}%
\bibitem [{\citenamefont {{Sen}}(1992)}]{Sen1992}%
  \BibitemOpen
  \bibfield  {author} {\bibinfo {author} {\bibfnamefont {A.}~\bibnamefont
  {{Sen}}},\ }\href {https://doi.org/10.1103/PhysRevLett.69.1006} {\bibfield
  {journal} {\bibinfo  {journal} {Phys. Rev. Lett.}\ }\textbf {\bibinfo
  {volume} {69}},\ \bibinfo {pages} {1006} (\bibinfo {year} {1992})},\ \Eprint
  {https://arxiv.org/abs/hep-th/9204046} {hep-th/9204046} \BibitemShut
  {NoStop}%
\bibitem [{\citenamefont {{Bambi}}\ and\ \citenamefont
  {{Modesto}}(2013)}]{Bambi2013}%
  \BibitemOpen
  \bibfield  {author} {\bibinfo {author} {\bibfnamefont {C.}~\bibnamefont
  {{Bambi}}}\ and\ \bibinfo {author} {\bibfnamefont {L.}~\bibnamefont
  {{Modesto}}},\ }\href {https://doi.org/10.1016/j.physletb.2013.03.025}
  {\bibfield  {journal} {\bibinfo  {journal} {Physics Letters B}\ }\textbf
  {\bibinfo {volume} {721}},\ \bibinfo {pages} {329} (\bibinfo {year}
  {2013})},\ \Eprint {https://arxiv.org/abs/1302.6075} {arXiv:1302.6075
  [gr-qc]} \BibitemShut {NoStop}%
\bibitem [{\citenamefont {{Newman}}\ and\ \citenamefont
  {{Janis}}(1965)}]{Newman1965JMPa}%
  \BibitemOpen
  \bibfield  {author} {\bibinfo {author} {\bibfnamefont {E.~T.}\ \bibnamefont
  {{Newman}}}\ and\ \bibinfo {author} {\bibfnamefont {A.~I.}\ \bibnamefont
  {{Janis}}},\ }\href {https://doi.org/10.1063/1.1704350} {\bibfield  {journal}
  {\bibinfo  {journal} {Journal of Mathematical Physics}\ }\textbf {\bibinfo
  {volume} {6}},\ \bibinfo {pages} {915} (\bibinfo {year} {1965})}\BibitemShut
  {NoStop}%
\bibitem [{\citenamefont {{Carter}}(1968)}]{Carter68}%
  \BibitemOpen
  \bibfield  {author} {\bibinfo {author} {\bibfnamefont {B.}~\bibnamefont
  {{Carter}}},\ }\href {https://doi.org/10.1103/PhysRev.174.1559} {\bibfield
  {journal} {\bibinfo  {journal} {Phys. Rev.}\ }\textbf {\bibinfo {volume}
  {174}},\ \bibinfo {pages} {1559} (\bibinfo {year} {1968})}\BibitemShut
  {NoStop}%
\bibitem [{\citenamefont {{Astorga}}\ \emph {et~al.}(2018)\citenamefont
  {{Astorga}}, \citenamefont {{Salazar}},\ and\ \citenamefont
  {{Zannias}}}]{Astorga17}%
  \BibitemOpen
  \bibfield  {author} {\bibinfo {author} {\bibfnamefont {F.}~\bibnamefont
  {{Astorga}}}, \bibinfo {author} {\bibfnamefont {J.~F.}\ \bibnamefont
  {{Salazar}}},\ and\ \bibinfo {author} {\bibfnamefont {T.}~\bibnamefont
  {{Zannias}}},\ }\href {https://doi.org/10.1088/1402-4896/aacd44} {\bibfield
  {journal} {\bibinfo  {journal} {Phys. Scr.}\ }\textbf {\bibinfo {volume}
  {93}},\ \bibinfo {eid} {085205} (\bibinfo {year} {2018})},\ \Eprint
  {https://arxiv.org/abs/1712.08972} {arXiv:1712.08972 [gr-qc]} \BibitemShut
  {NoStop}%
\bibitem [{\citenamefont {{Yazadjiev}}(2000)}]{Yazadjiev2000}%
  \BibitemOpen
  \bibfield  {author} {\bibinfo {author} {\bibfnamefont {S.}~\bibnamefont
  {{Yazadjiev}}},\ }\href {https://doi.org/10.1023/A:1002080003862} {\bibfield
  {journal} {\bibinfo  {journal} {General Relativity and Gravitation}\ }\textbf
  {\bibinfo {volume} {32}},\ \bibinfo {pages} {2345} (\bibinfo {year}
  {2000})},\ \Eprint {https://arxiv.org/abs/gr-qc/9907092} {arXiv:gr-qc/9907092
  [gr-qc]} \BibitemShut {NoStop}%
\bibitem [{\citenamefont {{Hioki}}\ and\ \citenamefont
  {{Maeda}}(2009)}]{Hioki2009}%
  \BibitemOpen
  \bibfield  {author} {\bibinfo {author} {\bibfnamefont {K.}~\bibnamefont
  {{Hioki}}}\ and\ \bibinfo {author} {\bibfnamefont {K.-I.}\ \bibnamefont
  {{Maeda}}},\ }\href {https://doi.org/10.1103/PhysRevD.80.024042} {\bibfield
  {journal} {\bibinfo  {journal} {Phys. Rev. D}\ }\textbf {\bibinfo {volume}
  {80}},\ \bibinfo {eid} {024042} (\bibinfo {year} {2009})},\ \Eprint
  {https://arxiv.org/abs/0904.3575} {arXiv:0904.3575 [astro-ph.HE]}
  \BibitemShut {NoStop}%
\bibitem [{\citenamefont {{Shaikh}}(2019)}]{Shaikh2019PRD}%
  \BibitemOpen
  \bibfield  {author} {\bibinfo {author} {\bibfnamefont {R.}~\bibnamefont
  {{Shaikh}}},\ }\href {https://doi.org/10.1103/PhysRevD.100.024028} {\bibfield
   {journal} {\bibinfo  {journal} {Phys. Rev. D}\ }\textbf {\bibinfo {volume}
  {100}},\ \bibinfo {eid} {024028} (\bibinfo {year} {2019})},\ \Eprint
  {https://arxiv.org/abs/1904.08322} {arXiv:1904.08322 [gr-qc]} \BibitemShut
  {NoStop}%
\bibitem [{\citenamefont {{Bardeen}}\ \emph {et~al.}(1972)\citenamefont
  {{Bardeen}}, \citenamefont {{Press}},\ and\ \citenamefont
  {{Teukolsky}}}]{Bardeen72}%
  \BibitemOpen
  \bibfield  {author} {\bibinfo {author} {\bibfnamefont {J.~M.}\ \bibnamefont
  {{Bardeen}}}, \bibinfo {author} {\bibfnamefont {W.~H.}\ \bibnamefont
  {{Press}}},\ and\ \bibinfo {author} {\bibfnamefont {S.~A.}\ \bibnamefont
  {{Teukolsky}}},\ }\href {https://doi.org/10.1086/151796} {\bibfield
  {journal} {\bibinfo  {journal} {Astrophys. J.}\ }\textbf {\bibinfo {volume}
  {178}},\ \bibinfo {pages} {347} (\bibinfo {year} {1972})}\BibitemShut
  {NoStop}%
\bibitem [{\citenamefont {{Abdujabbarov}}\ \emph {et~al.}(2015)\citenamefont
  {{Abdujabbarov}}, \citenamefont {{Rezzolla}},\ and\ \citenamefont
  {{Ahmedov}}}]{Abdujabbarov2015}%
  \BibitemOpen
  \bibfield  {author} {\bibinfo {author} {\bibfnamefont {A.~A.}\ \bibnamefont
  {{Abdujabbarov}}}, \bibinfo {author} {\bibfnamefont {L.}~\bibnamefont
  {{Rezzolla}}},\ and\ \bibinfo {author} {\bibfnamefont {B.~J.}\ \bibnamefont
  {{Ahmedov}}},\ }\href {https://doi.org/10.1093/mnras/stv2079} {\bibfield
  {journal} {\bibinfo  {journal} {Mon. Not. R. Astron. Soc.}\ }\textbf
  {\bibinfo {volume} {454}},\ \bibinfo {pages} {2423} (\bibinfo {year}
  {2015})},\ \Eprint {https://arxiv.org/abs/1503.09054} {arXiv:1503.09054
  [gr-qc]} \BibitemShut {NoStop}%
\bibitem [{\citenamefont {{Johannsen}}\ and\ \citenamefont
  {{Psaltis}}(2010)}]{Johannsen2010}%
  \BibitemOpen
  \bibfield  {author} {\bibinfo {author} {\bibfnamefont {T.}~\bibnamefont
  {{Johannsen}}}\ and\ \bibinfo {author} {\bibfnamefont {D.}~\bibnamefont
  {{Psaltis}}},\ }\href {https://doi.org/10.1088/0004-637X/718/1/446}
  {\bibfield  {journal} {\bibinfo  {journal} {Astrophys. J.}\ }\textbf
  {\bibinfo {volume} {718}},\ \bibinfo {pages} {446} (\bibinfo {year}
  {2010})},\ \Eprint {https://arxiv.org/abs/1005.1931} {arXiv:1005.1931
  [astro-ph.HE]} \BibitemShut {NoStop}%
\bibitem [{\citenamefont {{Magueijo}}(2003)}]{Magueijo03}%
  \BibitemOpen
  \bibfield  {author} {\bibinfo {author} {\bibfnamefont {J.}~\bibnamefont
  {{Magueijo}}},\ }\href {https://doi.org/10.1088/0034-4885/66/11/R04}
  {\bibfield  {journal} {\bibinfo  {journal} {Reports on Progress in Physics}\
  }\textbf {\bibinfo {volume} {66}},\ \bibinfo {pages} {2025} (\bibinfo {year}
  {2003})},\ \Eprint {https://arxiv.org/abs/astro-ph/0305457}
  {arXiv:astro-ph/0305457 [astro-ph]} \BibitemShut {NoStop}%
\bibitem [{\citenamefont {{Rodrigues}}\ and\ \citenamefont
  {{Silva}}(2018)}]{Rodrigues2018}%
  \BibitemOpen
  \bibfield  {author} {\bibinfo {author} {\bibfnamefont {M.~E.}\ \bibnamefont
  {{Rodrigues}}}\ and\ \bibinfo {author} {\bibfnamefont {M.~V. d.~S.}\
  \bibnamefont {{Silva}}},\ }\href
  {https://doi.org/10.1088/1475-7516/2018/06/025} {\bibfield  {journal}
  {\bibinfo  {journal} {JCAP}\ }\textbf {\bibinfo {volume} {2018}}\bibfield
  {number} {\bibinfo  {number} { (6)},\ \bibinfo {eid} {025}},\ }\Eprint
  {https://arxiv.org/abs/1802.05095} {arXiv:1802.05095 [gr-qc]} \BibitemShut
  {NoStop}%
\bibitem [{\citenamefont {{Liebling}}\ and\ \citenamefont
  {{Palenzuela}}(2012)}]{Liebling2012}%
  \BibitemOpen
  \bibfield  {author} {\bibinfo {author} {\bibfnamefont {S.~L.}\ \bibnamefont
  {{Liebling}}}\ and\ \bibinfo {author} {\bibfnamefont {C.}~\bibnamefont
  {{Palenzuela}}},\ }\href {https://doi.org/10.12942/lrr-2012-6} {\bibfield
  {journal} {\bibinfo  {journal} {Living Reviews in Relativity}\ }\textbf
  {\bibinfo {volume} {15}},\ \bibinfo {eid} {6} (\bibinfo {year} {2012})},\
  \Eprint {https://arxiv.org/abs/1202.5809} {arXiv:1202.5809 [gr-qc]}
  \BibitemShut {NoStop}%
\bibitem [{\citenamefont {{Meliani}}\ \emph {et~al.}(2016)\citenamefont
  {{Meliani}}, \citenamefont {{Grandcl{\'e}ment}}, \citenamefont {{Casse}},
  \citenamefont {{Vincent}}, \citenamefont {{Straub}},\ and\ \citenamefont
  {{Dauvergne}}}]{Meliani2016}%
  \BibitemOpen
  \bibfield  {author} {\bibinfo {author} {\bibfnamefont {Z.}~\bibnamefont
  {{Meliani}}}, \bibinfo {author} {\bibfnamefont {P.}~\bibnamefont
  {{Grandcl{\'e}ment}}}, \bibinfo {author} {\bibfnamefont {F.}~\bibnamefont
  {{Casse}}}, \bibinfo {author} {\bibfnamefont {F.~H.}\ \bibnamefont
  {{Vincent}}}, \bibinfo {author} {\bibfnamefont {O.}~\bibnamefont
  {{Straub}}},\ and\ \bibinfo {author} {\bibfnamefont {F.}~\bibnamefont
  {{Dauvergne}}},\ }\href {https://doi.org/10.1088/0264-9381/33/15/155010}
  {\bibfield  {journal} {\bibinfo  {journal} {Classical and Quantum Gravity}\
  }\textbf {\bibinfo {volume} {33}},\ \bibinfo {eid} {155010} (\bibinfo {year}
  {2016})}\BibitemShut {NoStop}%
\bibitem [{\citenamefont {{Hirschmann}}\ \emph {et~al.}(2018)\citenamefont
  {{Hirschmann}}, \citenamefont {{Lehner}}, \citenamefont {{Liebling}},\ and\
  \citenamefont {{Palenzuela}}}]{Hirschmann2018}%
  \BibitemOpen
  \bibfield  {author} {\bibinfo {author} {\bibfnamefont {E.~W.}\ \bibnamefont
  {{Hirschmann}}}, \bibinfo {author} {\bibfnamefont {L.}~\bibnamefont
  {{Lehner}}}, \bibinfo {author} {\bibfnamefont {S.~L.}\ \bibnamefont
  {{Liebling}}},\ and\ \bibinfo {author} {\bibfnamefont {C.}~\bibnamefont
  {{Palenzuela}}},\ }\href {https://doi.org/10.1103/PhysRevD.97.064032}
  {\bibfield  {journal} {\bibinfo  {journal} {Phys. Rev. D}\ }\textbf {\bibinfo
  {volume} {97}},\ \bibinfo {eid} {064032} (\bibinfo {year} {2018})},\ \Eprint
  {https://arxiv.org/abs/1706.09875} {arXiv:1706.09875 [gr-qc]} \BibitemShut
  {NoStop}%
\bibitem [{\citenamefont {Wiltshire}\ \emph {et~al.}(2009)\citenamefont
  {Wiltshire}, \citenamefont {Visser},\ and\ \citenamefont
  {Scott}}]{Wiltshire2009}%
  \BibitemOpen
  \bibfield  {author} {\bibinfo {author} {\bibfnamefont {D.~L.}\ \bibnamefont
  {Wiltshire}}, \bibinfo {author} {\bibfnamefont {M.}~\bibnamefont {Visser}},\
  and\ \bibinfo {author} {\bibfnamefont {S.~M.}\ \bibnamefont {Scott}},\
  }\href@noop {} {\emph {\bibinfo {title} {{The Kerr spacetime: Rotating black
  holes in general relativity}}}}\ (\bibinfo  {publisher} {Cambridge University
  Press},\ \bibinfo {year} {2009})\BibitemShut {NoStop}%
\bibitem [{\citenamefont {{Held}}\ \emph {et~al.}(2019)\citenamefont {{Held}},
  \citenamefont {{Gold}},\ and\ \citenamefont {{Eichhorn}}}]{Held2019}%
  \BibitemOpen
  \bibfield  {author} {\bibinfo {author} {\bibfnamefont {A.}~\bibnamefont
  {{Held}}}, \bibinfo {author} {\bibfnamefont {R.}~\bibnamefont {{Gold}}},\
  and\ \bibinfo {author} {\bibfnamefont {A.}~\bibnamefont {{Eichhorn}}},\
  }\href {https://doi.org/10.1088/1475-7516/2019/06/029} {\bibfield  {journal}
  {\bibinfo  {journal} {JCAP}\ }\textbf {\bibinfo {volume} {2019}}\bibfield
  {number} {\bibinfo  {number} { (6)},\ \bibinfo {eid} {029}},\ }\Eprint
  {https://arxiv.org/abs/1904.07133} {arXiv:1904.07133 [gr-qc]} \BibitemShut
  {NoStop}%
\bibitem [{\citenamefont {{Bambi}}(2014)}]{Bambi2014}%
  \BibitemOpen
  \bibfield  {author} {\bibinfo {author} {\bibfnamefont {C.}~\bibnamefont
  {{Bambi}}},\ }\href {https://doi.org/10.1016/j.physletb.2014.01.037}
  {\bibfield  {journal} {\bibinfo  {journal} {Physics Letters B}\ }\textbf
  {\bibinfo {volume} {730}},\ \bibinfo {pages} {59} (\bibinfo {year} {2014})},\
  \Eprint {https://arxiv.org/abs/1401.4640} {arXiv:1401.4640 [gr-qc]}
  \BibitemShut {NoStop}%
\bibitem [{\citenamefont {{Barausse}}\ \emph
  {et~al.}(2016{\natexlab{b}})\citenamefont {{Barausse}}, \citenamefont
  {{Yunes}},\ and\ \citenamefont {{Chamberlain}}}]{Barausse2016b}%
  \BibitemOpen
  \bibfield  {author} {\bibinfo {author} {\bibfnamefont {E.}~\bibnamefont
  {{Barausse}}}, \bibinfo {author} {\bibfnamefont {N.}~\bibnamefont
  {{Yunes}}},\ and\ \bibinfo {author} {\bibfnamefont {K.}~\bibnamefont
  {{Chamberlain}}},\ }\href {https://doi.org/10.1103/PhysRevLett.116.241104}
  {\bibfield  {journal} {\bibinfo  {journal} {Phys. Rev. Lett.}\ }\textbf
  {\bibinfo {volume} {116}},\ \bibinfo {eid} {241104} (\bibinfo {year}
  {2016}{\natexlab{b}})},\ \Eprint {https://arxiv.org/abs/1603.04075}
  {arXiv:1603.04075 [gr-qc]} \BibitemShut {NoStop}%
\bibitem [{\citenamefont {{Konoplya}}\ and\ \citenamefont
  {{Zhidenko}}(2016)}]{Konoplya2016c}%
  \BibitemOpen
  \bibfield  {author} {\bibinfo {author} {\bibfnamefont {R.}~\bibnamefont
  {{Konoplya}}}\ and\ \bibinfo {author} {\bibfnamefont {A.}~\bibnamefont
  {{Zhidenko}}},\ }\href {https://doi.org/10.1016/j.physletb.2016.03.044}
  {\bibfield  {journal} {\bibinfo  {journal} {Physics Letters B}\ }\textbf
  {\bibinfo {volume} {756}},\ \bibinfo {pages} {350} (\bibinfo {year}
  {2016})},\ \Eprint {https://arxiv.org/abs/1602.04738} {arXiv:1602.04738
  [gr-qc]} \BibitemShut {NoStop}%
\bibitem [{\citenamefont {{Juli{\'e}}}(2018{\natexlab{a}})}]{Felix-Louis2018}%
  \BibitemOpen
  \bibfield  {author} {\bibinfo {author} {\bibfnamefont {F.-L.}\ \bibnamefont
  {{Juli{\'e}}}},\ }\href {https://doi.org/10.1088/1475-7516/2018/01/026}
  {\bibfield  {journal} {\bibinfo  {journal} {JCAP}\ }\textbf {\bibinfo
  {volume} {2018}}\bibfield  {number} {\bibinfo  {number} { (1)},\ \bibinfo
  {eid} {026}},\ }\Eprint {https://arxiv.org/abs/1711.10769} {arXiv:1711.10769
  [gr-qc]} \BibitemShut {NoStop}%
\bibitem [{\citenamefont {{Juli{\'e}}}(2018{\natexlab{b}})}]{Felix-Louis2018b}%
  \BibitemOpen
  \bibfield  {author} {\bibinfo {author} {\bibfnamefont {F.-L.}\ \bibnamefont
  {{Juli{\'e}}}},\ }\href {https://doi.org/10.1088/1475-7516/2018/10/033}
  {\bibfield  {journal} {\bibinfo  {journal} {JCAP}\ }\textbf {\bibinfo
  {volume} {2018}}\bibfield  {number} {\bibinfo  {number} { (10)},\ \bibinfo
  {eid} {033}},\ }\Eprint {https://arxiv.org/abs/1809.05041} {arXiv:1809.05041
  [gr-qc]} \BibitemShut {NoStop}%
\bibitem [{\citenamefont {{Siahaan}}(2020)}]{Siahaan2020}%
  \BibitemOpen
  \bibfield  {author} {\bibinfo {author} {\bibfnamefont {H.~M.}\ \bibnamefont
  {{Siahaan}}},\ }\href {https://doi.org/10.1103/PhysRevD.101.064036}
  {\bibfield  {journal} {\bibinfo  {journal} {Phys. Rev. D}\ }\textbf {\bibinfo
  {volume} {101}},\ \bibinfo {eid} {064036} (\bibinfo {year} {2020})},\ \Eprint
  {https://arxiv.org/abs/1907.02158} {arXiv:1907.02158 [gr-qc]} \BibitemShut
  {NoStop}%
\bibitem [{\citenamefont {{V{\"o}lkel}}\ \emph {et~al.}(2020)\citenamefont
  {{V{\"o}lkel}}, \citenamefont {{Barausse}}, \citenamefont {{Franchini}},\
  and\ \citenamefont {{Broderick}}}]{Voelkel2020}%
  \BibitemOpen
  \bibfield  {author} {\bibinfo {author} {\bibfnamefont {S.~H.}\ \bibnamefont
  {{V{\"o}lkel}}}, \bibinfo {author} {\bibfnamefont {E.}~\bibnamefont
  {{Barausse}}}, \bibinfo {author} {\bibfnamefont {N.}~\bibnamefont
  {{Franchini}}},\ and\ \bibinfo {author} {\bibfnamefont {A.~E.}\ \bibnamefont
  {{Broderick}}},\ }\href@noop {} {\bibfield  {journal} {\bibinfo  {journal}
  {arXiv e-prints}\ } (\bibinfo {year} {2020})},\ \Eprint
  {https://arxiv.org/abs/2011.06812} {arXiv:2011.06812 [gr-qc]} \BibitemShut
  {NoStop}%
\bibitem [{\citenamefont {{Greisen}}(2003)}]{Greisen2003}%
  \BibitemOpen
  \bibfield  {author} {\bibinfo {author} {\bibfnamefont {E.~W.}\ \bibnamefont
  {{Greisen}}},\ }\bibinfo {title} {{AIPS, the VLA, and the VLBA}},\ in\ \href
  {https://doi.org/10.1007/0-306-48080-8\_7} {\emph {\bibinfo {booktitle}
  {Information Handling in Astronomy - Historical Vistas}}},\ Vol.\ \bibinfo
  {volume} {285},\ \bibinfo {editor} {edited by\ \bibinfo {editor}
  {\bibfnamefont {A.}~\bibnamefont {{Heck}}}}\ (\bibinfo {year} {2003})\ p.\
  \bibinfo {pages} {109}\BibitemShut {NoStop}%
\bibitem [{\citenamefont {{Kettenis}}\ \emph {et~al.}(2006)\citenamefont
  {{Kettenis}}, \citenamefont {{van Langevelde}}, \citenamefont {{Reynolds}},\
  and\ \citenamefont {{Cotton}}}]{Kettenis2006}%
  \BibitemOpen
  \bibfield  {author} {\bibinfo {author} {\bibfnamefont {M.}~\bibnamefont
  {{Kettenis}}}, \bibinfo {author} {\bibfnamefont {H.~J.}\ \bibnamefont {{van
  Langevelde}}}, \bibinfo {author} {\bibfnamefont {C.}~\bibnamefont
  {{Reynolds}}},\ and\ \bibinfo {author} {\bibfnamefont {B.}~\bibnamefont
  {{Cotton}}},\ }in\ \href@noop {} {\emph {\bibinfo {booktitle} {Astronomical
  Data Analysis Software and Systems XV}}},\ \bibinfo {series} {Astronomical
  Society of the Pacific Conference Series}, Vol.\ \bibinfo {volume} {351},\
  \bibinfo {editor} {edited by\ \bibinfo {editor} {\bibfnamefont
  {C.}~\bibnamefont {{Gabriel}}}, \bibinfo {editor} {\bibfnamefont
  {C.}~\bibnamefont {{Arviset}}}, \bibinfo {editor} {\bibfnamefont
  {D.}~\bibnamefont {{Ponz}}},\ and\ \bibinfo {editor} {\bibfnamefont
  {S.}~\bibnamefont {{Enrique}}}}\ (\bibinfo {year} {2006})\ p.\ \bibinfo
  {pages} {497}\BibitemShut {NoStop}%
\bibitem [{\citenamefont {{Tange}}(2011)}]{Tange2011}%
  \BibitemOpen
  \bibfield  {author} {\bibinfo {author} {\bibfnamefont {O.}~\bibnamefont
  {{Tange}}},\ }\href@noop {} {\bibfield  {journal} {\bibinfo  {journal}
  {login: The USENIX Magazine}\ }\textbf {\bibinfo {volume} {36}},\ \bibinfo
  {pages} {42} (\bibinfo {year} {2011})}\BibitemShut {NoStop}%
\bibitem [{\citenamefont {{Chael}}\ \emph {et~al.}(2016)\citenamefont
  {{Chael}}, \citenamefont {{Johnson}}, \citenamefont {{Narayan}},
  \citenamefont {{Doeleman}}, \citenamefont {{Wardle}},\ and\ \citenamefont
  {{Bouman}}}]{Chael2016}%
  \BibitemOpen
  \bibfield  {author} {\bibinfo {author} {\bibfnamefont {A.~A.}\ \bibnamefont
  {{Chael}}}, \bibinfo {author} {\bibfnamefont {M.~D.}\ \bibnamefont
  {{Johnson}}}, \bibinfo {author} {\bibfnamefont {R.}~\bibnamefont
  {{Narayan}}}, \bibinfo {author} {\bibfnamefont {S.~S.}\ \bibnamefont
  {{Doeleman}}}, \bibinfo {author} {\bibfnamefont {J.~F.~C.}\ \bibnamefont
  {{Wardle}}},\ and\ \bibinfo {author} {\bibfnamefont {K.~L.}\ \bibnamefont
  {{Bouman}}},\ }\href {https://doi.org/10.3847/0004-637X/829/1/11} {\bibfield
  {journal} {\bibinfo  {journal} {Astrophys. J.}\ }\textbf {\bibinfo {volume}
  {829}},\ \bibinfo {eid} {11} (\bibinfo {year} {2016})},\ \Eprint
  {https://arxiv.org/abs/1605.06156} {arXiv:1605.06156 [astro-ph.IM]}
  \BibitemShut {NoStop}%
\bibitem [{\citenamefont {{Shepherd}}(2011)}]{Shepherd2011}%
  \BibitemOpen
  \bibfield  {author} {\bibinfo {author} {\bibfnamefont {M.}~\bibnamefont
  {{Shepherd}}},\ }\href@noop {} {\bibinfo {title} {{Difmap: Synthesis Imaging
  of Visibility Data}}} (\bibinfo {year} {2011}),\ \Eprint
  {https://arxiv.org/abs/1103.001} {ascl:1103.001} \BibitemShut {NoStop}%
\bibitem [{\citenamefont {van~der Walt}\ \emph {et~al.}(2011)\citenamefont
  {van~der Walt}, \citenamefont {Colbert},\ and\ \citenamefont
  {Varoquaux}}]{vanderWalt2011}%
  \BibitemOpen
  \bibfield  {author} {\bibinfo {author} {\bibfnamefont {S.}~\bibnamefont
  {van~der Walt}}, \bibinfo {author} {\bibfnamefont {S.~C.}\ \bibnamefont
  {Colbert}},\ and\ \bibinfo {author} {\bibfnamefont {G.}~\bibnamefont
  {Varoquaux}},\ }\href {https://doi.org/10.1109/MCSE.2011.37} {\bibfield
  {journal} {\bibinfo  {journal} {Computing in Science \& Engineering}\
  }\textbf {\bibinfo {volume} {13}},\ \bibinfo {pages} {22} (\bibinfo {year}
  {2011})},\ \Eprint
  {https://arxiv.org/abs/http://aip.scitation.org/doi/pdf/10.1109/MCSE.2011.37}
  {http://aip.scitation.org/doi/pdf/10.1109/MCSE.2011.37} \BibitemShut
  {NoStop}%
\bibitem [{\citenamefont {{Jones}}\ \emph {et~al.}(2001)\citenamefont {{Jones}}
  \emph {et~al.}}]{Jones2001}%
  \BibitemOpen
  \bibfield  {author} {\bibinfo {author} {\bibfnamefont {E.}~\bibnamefont
  {{Jones}}} \emph {et~al.},\ }\bibfield  {journal} {\bibinfo  {journal}
  {SciPy: Open Source Scientific Tools for Python}\ }\href
  {https://doi.org/http://www.scipy.org/} {http://www.scipy.org/} (\bibinfo
  {year} {2001})\BibitemShut {NoStop}%
\bibitem [{\citenamefont {{McKinney}}(2010)}]{McKinney2010Pandas}%
  \BibitemOpen
  \bibfield  {author} {\bibinfo {author} {\bibfnamefont {W.}~\bibnamefont
  {{McKinney}}},\ }in\ \href@noop {} {\emph {\bibinfo {booktitle} {Proc. IX
  Python in Science Conf.}}},\ \bibinfo {editor} {edited by\ \bibinfo {editor}
  {\bibfnamefont {S.}~\bibnamefont {{van der Walt}}}\ and\ \bibinfo {editor}
  {\bibfnamefont {J.}~\bibnamefont {{Millman}}}}\ (\bibinfo {year} {2010})\
  p.~\bibinfo {pages} {51}\BibitemShut {NoStop}%
\bibitem [{\citenamefont {{Astropy Collaboration}}\ \emph
  {et~al.}(2013)\citenamefont {{Astropy Collaboration}}, \citenamefont
  {{Robitaille}}, \citenamefont {{Tollerud}} \emph {et~al.}}]{Astropy2013}%
  \BibitemOpen
  \bibfield  {author} {\bibinfo {author} {\bibnamefont {{Astropy
  Collaboration}}}, \bibinfo {author} {\bibfnamefont {T.~P.}\ \bibnamefont
  {{Robitaille}}}, \bibinfo {author} {\bibfnamefont {E.~J.}\ \bibnamefont
  {{Tollerud}}}, \emph {et~al.},\ }\href
  {https://doi.org/10.1051/0004-6361/201322068} {\bibfield  {journal} {\bibinfo
   {journal} {Astron. Astrophys.}\ }\textbf {\bibinfo {volume} {558}},\
  \bibinfo {eid} {A33} (\bibinfo {year} {2013})},\ \Eprint
  {https://arxiv.org/abs/1307.6212} {arXiv:1307.6212 [astro-ph.IM]}
  \BibitemShut {NoStop}%
\bibitem [{\citenamefont {{Astropy Collaboration}}\ \emph
  {et~al.}(2018)\citenamefont {{Astropy Collaboration}}, \citenamefont
  {{Price-Whelan}}, \citenamefont {{Sip{\H{o}}cz}} \emph
  {et~al.}}]{Astropy2018}%
  \BibitemOpen
  \bibfield  {author} {\bibinfo {author} {\bibnamefont {{Astropy
  Collaboration}}}, \bibinfo {author} {\bibfnamefont {A.~M.}\ \bibnamefont
  {{Price-Whelan}}}, \bibinfo {author} {\bibfnamefont {B.~M.}\ \bibnamefont
  {{Sip{\H{o}}cz}}}, \emph {et~al.},\ }\href
  {https://doi.org/10.3847/1538-3881/aabc4f} {\bibfield  {journal} {\bibinfo
  {journal} {Astron. J.}\ }\textbf {\bibinfo {volume} {156}},\ \bibinfo {eid}
  {123} (\bibinfo {year} {2018})},\ \Eprint {https://arxiv.org/abs/1801.02634}
  {arXiv:1801.02634 [astro-ph.IM]} \BibitemShut {NoStop}%
\bibitem [{\citenamefont {{Kluyver}}\ \emph {et~al.}(2016)\citenamefont
  {{Kluyver}} \emph {et~al.}}]{Kluyver2016}%
  \BibitemOpen
  \bibfield  {author} {\bibinfo {author} {\bibfnamefont {T.}~\bibnamefont
  {{Kluyver}}} \emph {et~al.},\ }in\ \href@noop {} {\emph {\bibinfo {booktitle}
  {Positioning and Power in Academic Publishing: Players, Agents and
  Agendas}}},\ \bibinfo {editor} {edited by\ \bibinfo {editor} {\bibfnamefont
  {F.}~\bibnamefont {{Loizides}}}\ and\ \bibinfo {editor} {\bibfnamefont
  {B.}~\bibnamefont {{Schmidt}}}}\ (\bibinfo  {publisher} {IOS Press},\
  \bibinfo {year} {2016})\ p.~\bibinfo {pages} {87}\BibitemShut {NoStop}%
\bibitem [{\citenamefont {Hunter}(2007)}]{Hunter2007}%
  \BibitemOpen
  \bibfield  {author} {\bibinfo {author} {\bibfnamefont {J.~D.}\ \bibnamefont
  {Hunter}},\ }\href {https://doi.org/10.1109/MCSE.2007.55} {\bibfield
  {journal} {\bibinfo  {journal} {Computing In Science \& Engineering}\
  }\textbf {\bibinfo {volume} {9}},\ \bibinfo {pages} {90} (\bibinfo {year}
  {2007})}\BibitemShut {NoStop}%
\bibitem [{\citenamefont {{Broderick}}\ \emph {et~al.}(2020)\citenamefont
  {{Broderick}} \emph {et~al.}}]{Broderick2020}%
  \BibitemOpen
  \bibfield  {author} {\bibinfo {author} {\bibfnamefont {A.~E.}\ \bibnamefont
  {{Broderick}}} \emph {et~al.},\ }\href
  {https://doi.org/10.3847/1538-4357/ab91a4} {\bibfield  {journal} {\bibinfo
  {journal} {Astrophys. J.}\ }\textbf {\bibinfo {volume} {897}},\ \bibinfo
  {eid} {139} (\bibinfo {year} {2020})}\BibitemShut {NoStop}%
\bibitem [{\citenamefont {{Pesce}}(2021)}]{Pesce2021}%
  \BibitemOpen
  \bibfield  {author} {\bibinfo {author} {\bibfnamefont {D.~W.}\ \bibnamefont
  {{Pesce}}},\ }\href {https://doi.org/10.3847/1538-3881/abe3f8} {\bibfield
  {journal} {\bibinfo  {journal} {Astron. J.}\ }\textbf {\bibinfo {volume}
  {161}},\ \bibinfo {eid} {178} (\bibinfo {year} {2021})},\ \Eprint
  {https://arxiv.org/abs/2102.03328} {arXiv:2102.03328 [astro-ph.IM]}
  \BibitemShut {NoStop}%
\bibitem [{\citenamefont {{Mart{\'\i}-Vidal}}\ \emph
  {et~al.}(2021)\citenamefont {{Mart{\'\i}-Vidal}}, \citenamefont {{Mus}},
  \citenamefont {{Janssen}}, \citenamefont {{de Vicente}},\ and\ \citenamefont
  {{Gonz{\'a}lez}}}]{MartiVidal2021}%
  \BibitemOpen
  \bibfield  {author} {\bibinfo {author} {\bibfnamefont {I.}~\bibnamefont
  {{Mart{\'\i}-Vidal}}}, \bibinfo {author} {\bibfnamefont {A.}~\bibnamefont
  {{Mus}}}, \bibinfo {author} {\bibfnamefont {M.}~\bibnamefont {{Janssen}}},
  \bibinfo {author} {\bibfnamefont {P.}~\bibnamefont {{de Vicente}}},\ and\
  \bibinfo {author} {\bibfnamefont {J.}~\bibnamefont {{Gonz{\'a}lez}}},\ }\href
  {https://doi.org/10.1051/0004-6361/202039527} {\bibfield  {journal} {\bibinfo
   {journal} {Astron. Astrophys.}\ }\textbf {\bibinfo {volume} {646}},\
  \bibinfo {eid} {A52} (\bibinfo {year} {2021})},\ \Eprint
  {https://arxiv.org/abs/2012.05581} {arXiv:2012.05581 [astro-ph.IM]}
  \BibitemShut {NoStop}%
\bibitem [{\citenamefont {{Park}}\ \emph {et~al.}(2021)\citenamefont {{Park}},
  \citenamefont {{Byun}}, \citenamefont {{Asada}},\ and\ \citenamefont
  {{Yun}}}]{Park2021}%
  \BibitemOpen
  \bibfield  {author} {\bibinfo {author} {\bibfnamefont {J.}~\bibnamefont
  {{Park}}}, \bibinfo {author} {\bibfnamefont {D.-Y.}\ \bibnamefont {{Byun}}},
  \bibinfo {author} {\bibfnamefont {K.}~\bibnamefont {{Asada}}},\ and\ \bibinfo
  {author} {\bibfnamefont {Y.}~\bibnamefont {{Yun}}},\ }\href
  {https://doi.org/10.3847/1538-4357/abcc6e} {\bibfield  {journal} {\bibinfo
  {journal} {Astrophys. J.}\ }\textbf {\bibinfo {volume} {906}},\ \bibinfo
  {eid} {85} (\bibinfo {year} {2021})},\ \Eprint
  {https://arxiv.org/abs/2011.09713} {arXiv:2011.09713 [astro-ph.IM]}
  \BibitemShut {NoStop}%
\end{thebibliography}%


\begin{appendix}

\section{Distortion parameters} \label{sec:Distortion_Parameters}

Since the boundary of the shadow region is a closed curve as discussed
above, one can define various characteristic features for a quantitative
comparison \cite{Hioki2009, Abdujabbarov2015}. Out of the many possible
measures of distortion of this curve from a perfect circle discussed in
Ref. \cite{Abdujabbarov2015}, we use here the simplest one which was
originally introduced in Ref. \cite{Hioki2009}, namely
\begin{equation}
\delta_{\text{sh}} = \frac{\alpha_{l, \text{c}} - \alpha_{l}}{r_{\text{sh, c}}}\,, 
\end{equation}
where $r_{\text{sh, c}}$ is the radius of the circumcircle passing
through the two points (since the images here are symmetric about the
$\alpha$-axis) with coordinates $(\alpha_{\text{r}}, 0)$ and
$(\alpha_{\text{t}}, \beta_{\text{t}})$, which are the rightmost and
topmost points of the shadow curve, and is given as \cite{Hioki2009},
\begin{equation}
r_{\text{sh, c}} = \frac{(\alpha_{\text{t}} - \alpha_{\text{r}})^2 +
  \beta_{\text{t}}^2}{2|\alpha_{\text{t}} - \alpha_{\text{r}}|}\,,
\end{equation}
with $(\alpha_{l}, 0)$ and $(\alpha_{l, \text{c}}, 0)$ the leftmost points of
the shadow curve and of the circumcircle respectively (see Fig. 3 of
\cite{Abdujabbarov2016}).

\begin{figure}
\centering
\includegraphics[width=\columnwidth]{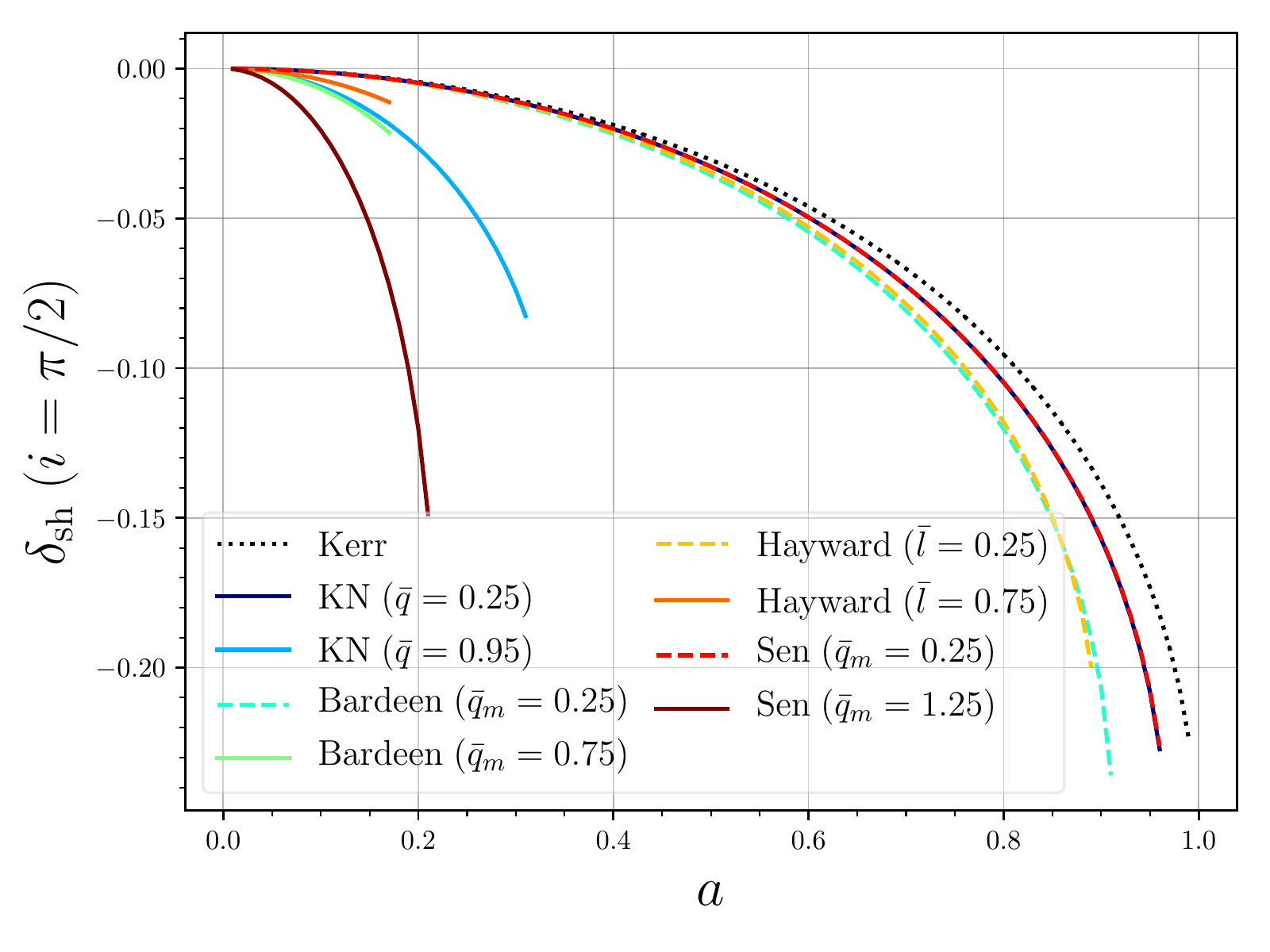}
\caption{Distortion parameter $\delta_{\text{sh}}$ for a number of
  stationary black holes observed on the equatorial plane ($i=\pi/2$)
  with dimensionless spin $a$. Because for observers viewing the black
  hole from inclinations increasingly close to the pole, the shadow
  boundary appears increasingly circular, the distortions reported can be
  taken as upper limits.}
\label{fig:Distortion_Parameter}
\end{figure}

In Fig. \ref{fig:Distortion_Parameter} we display the distortion
parameter $\delta_{\text{sh}}$ for the shadow curves of various rotating
black holes, for an equatorial observer, as an additional simple
comparable characteristic. We note also that the deviation of
$\delta_{\text{sh}}$ from zero is insignificant for observer viewing
angles that are close to the pole of the black hole, as anticipated (not
displayed here).

As a concluding remark we note that the EHT bounds on the size of the
shadow of M87*, as discussed above and displayed in Eq. \ref{eq:EHT-Constraint}, do not
impose straightforward bounds on its shape. In particular, we can see
from Fig. \ref{fig:Distortion_Parameter} that the rotating Bardeen black
hole with $\bar{q}_{m} = 0.25$ for high spins can be more distorted from
a circle than a Kerr black hole but still be compatible with the EHT
measurement (see Fig. \ref{fig:Spherical_Shadow_Radius_Constraints}). On the other hand, even though
we are able to exclude Sen black holes with large electromagnetic charges
(see, \eg the Sen curve for $\bar{q}_{m}=1.25$ in the right panel of
Fig. \ref{fig:Spherical_Shadow_Radius_Constraints}) as viable models for M87*, its shadow is less
distorted from a circle than that of an extremal Kerr black hole (see
Fig. \ref{fig:Distortion_Parameter}). In other words, the examples
just made highlight the importance of using the appropriate bounds on a
sufficiently robust quantity when using the EHT measurement to test
theories of gravity. Failing to do so may lead to incorrect bounds on the
black-hole properties. For instance, Ref. \cite{Kumar2019b} is able to
set bounds on the parameter space of the uncharged, rotating Hayward
black hole by imposing bounds on the maximum distortion of the shape of
its shadow boundaries, albeit using a different measure for the
distortion from a circle [see Eq. (58) there], whereas we have shown that
this is not possible, upon using the bounds $4.31\,M - 6.08\,M$ for the
size of their shadows (\cf right panel of Fig. \ref{fig:Spherical_Shadow_Radius_Constraints}).

\end{appendix}

\end{document}